\documentclass{egpubl}
\usepackage{egsr18}

\usepackage{booktabs} % For formal tables

\usepackage{import}

\usepackage{t1enc,dfadobe}

\usepackage{cite}

\usepackage{amsmath}
\usepackage{caption}
\usepackage{subcaption}

\usepackage{amssymb}
\usepackage{caption} 
\usepackage{subcaption}
\usepackage{tabularx} % in the preamble
\usepackage{tabulary}
\usepackage{amsmath}
\usepackage{multirow}

\usepackage{enumerate}
\usepackage{array}
\usepackage{makecell}
\usepackage{multirow}
\usepackage{algorithm, algpseudocode}
\usepackage{algorithmicx}
\usepackage{rotating}

\title[Conformal Mesh Parameterization Using Discrete Calabi Flow]%
{Conformal Mesh Parameterization Using Discrete Calabi Flow}

\author[Hui Zhao \& Xuan Li \& Huabin Ge \& Xianfeng Gu \& Na Lei] 
{\parbox{\textwidth}
	{
		\centering  Hui Zhao\thanks{e-mail:alanzhaohui@qq.com} $^{1}$,
	            	Xuan Li\thanks{e-mail:li-xuan13@mails.tsinghua.edu.cn}$^{1,2}$,  
	            	Huabin Ge\thanks{e-mail:hbge@bjtu.edu.cn}$^{4}$,
	                Na Lei\thanks{e-mail:nalei@outlook.com}$^3$
	          and	Xianfeng Gu\thanks{e-mail:gu@cs.stonybrook.edu}$^2$
	}
	\\
	{\parbox{\textwidth}
		    {\centering 
		    	$^1$Tsinghua University, China\\ 
		    	$^2$State University of New York at Stony Brook, USA\\
		    	$^3$Dalian University of Technology, China \\
		        $^4$Beijing Jiaotong University, China 		
	        }
	}
}

\begin{document}
	
     \teaser{
	   \includegraphics[width=\linewidth]{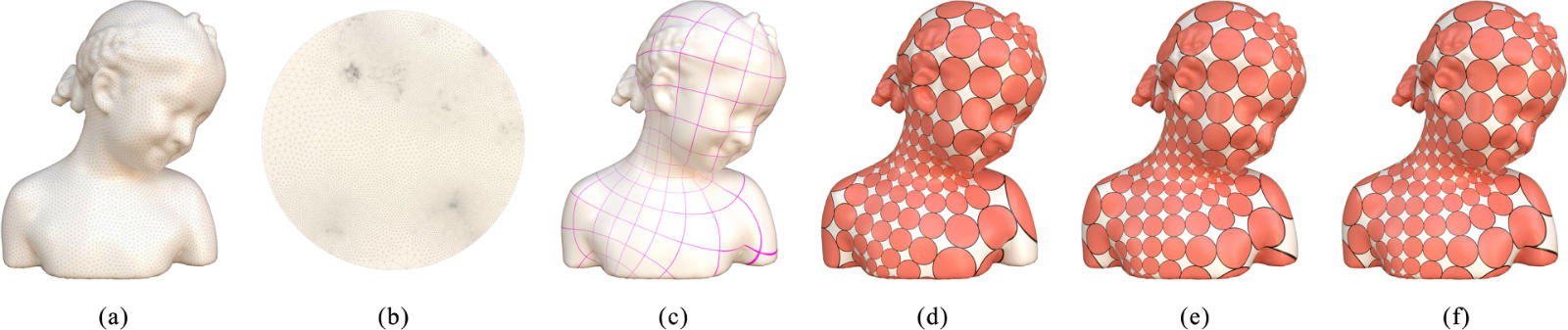}
        \centering
        \caption{(a) original mesh; (b) the parameterization with Calabi flow; (c), (d) the texturing with Calabi flow; (e) the texturing with CETM; (f) the texturing with Ricci flow.}
	    \label{fig:teaser}
	  }
	
	\maketitle

	%-------------------------------------------------------------------------
\begin{abstract}

In this paper, we introduce discrete Calabi flow to the graphics research community and present a novel conformal mesh  parameterization algorithm. 
Calabi energy has a succinct and explicit format. Its corresponding flow is conformal and convergent under certain conditions.
Our method is based on the Calabi energy and Calabi flow with solid  theoretical  and mathematical   base.  
We demonstrate our approach on dozens of models and compare it with other related flow based methods, such as the well-known  Ricci flow and CETM. 
Our experiments show that the  performance of our algorithm is comparably the same with other  methods. 
The discrete Calabi flow in our method provides another perspective on conformal flow and conformal parameterization.

\begin{classification} % according to http:http://www.acm.org/about/class/1998
	\CCScat{Computer Graphics}{I.3.5}{Computational Geometry and Object Modeling}{Curve, surface, solid, and object representations}
\end{classification}

\end{abstract}	
\section{Introduction}
 In this paper, we present a novel conformal flow-based method for conformal mapping. Our method is based on discrete Calabi energy and Calabi flow \cite{chen2008calabi,ge2012combinatorial,ge20162}. Discrete Calabi flow is inspired by discrete Ricci flow \cite{chow2003combinatorial, luo2003yamabe,Jin2007DiscreteSR},   it    is also a conformal flow which preserves the angles. Conformal parameterization can keep the shape of the original mesh and is especially useful in all kinds of applications. 

Mesh mapping and parameterization are  crucial operations in computer graphics modeling. 
Researchers designed a lot of different algorithms in past twenty years.
One   important application of mesh parameterization is texturing which assigns a  2D  image onto a  3D  mesh surface, another one is remeshing.

Given a 3D mesh, the parameterization looks for a corresponding 2D flat mesh. The perfect mapping is an isometric one that  can only exist on the developable surfaces.
Therefore in practice, we try to preserve the area or angle. They are called authalic (area-preserving) mapping, conformal (angle-preserving) mapping, isometric (length-preserving)
or some combination of them.  

The algorithms proposed in \cite{hormann2000mips, fu2015computing} can be designed on the discrete triangle mesh directly. The optimal mapping \cite{desbrun2002intrinsic,levy2002least,liu2008local} results from defining and minimizing an energy related to the mesh triangles. Other methodologies are based on the smooth surface mapping theories and then derive their corresponding discrete counterpart \cite{Jin2007DiscreteSR,gu2003global}.

\begin{figure} [th!] 
	\centering   
	\begin{subfigure}[b]{0.15\textwidth}
		\includegraphics[width=\textwidth]{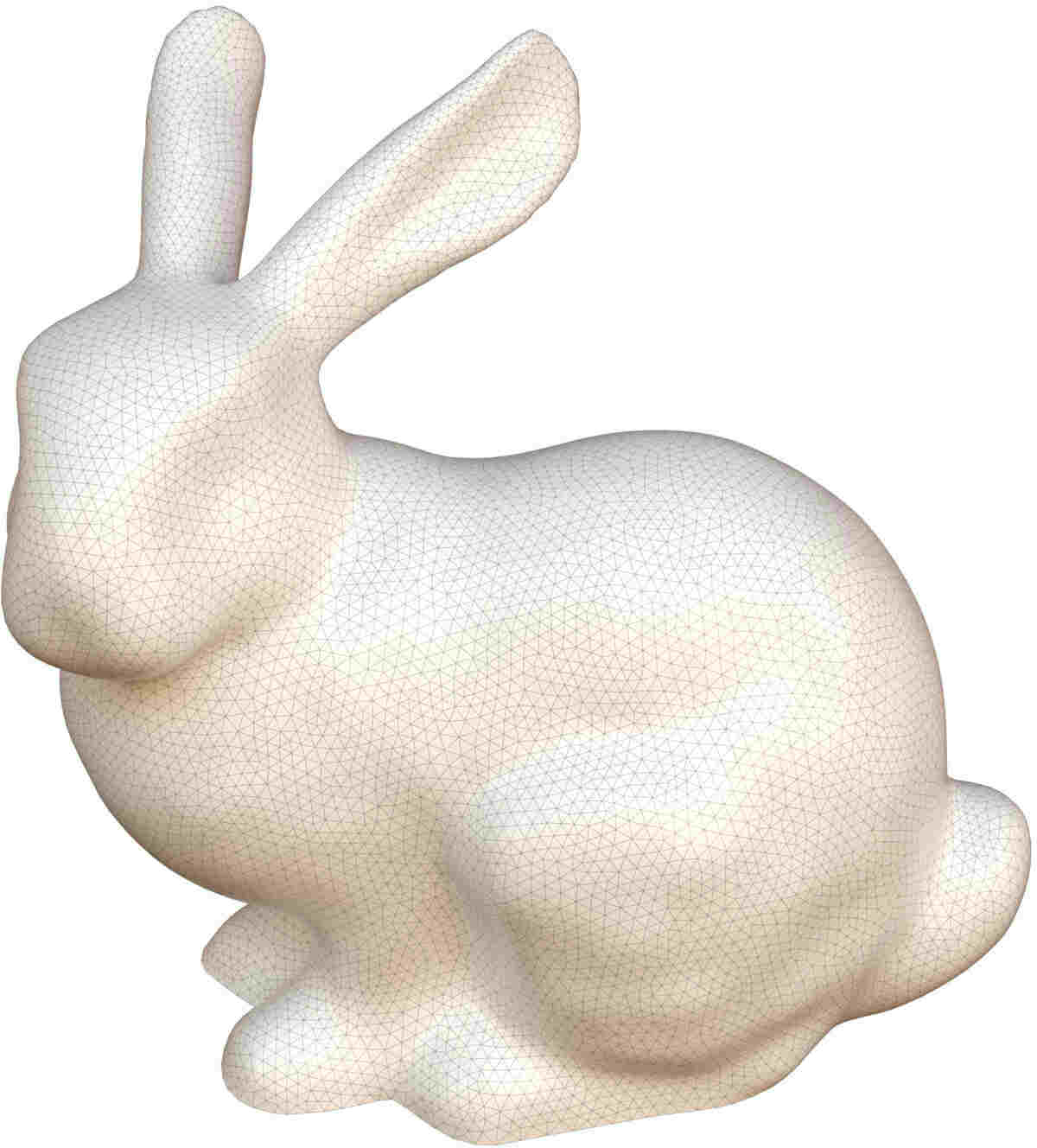}
		\caption{ }   \label{fig:bunny0}
	\end{subfigure}
	\begin{subfigure}[b]{0.15\textwidth}
		\includegraphics[width=\textwidth]{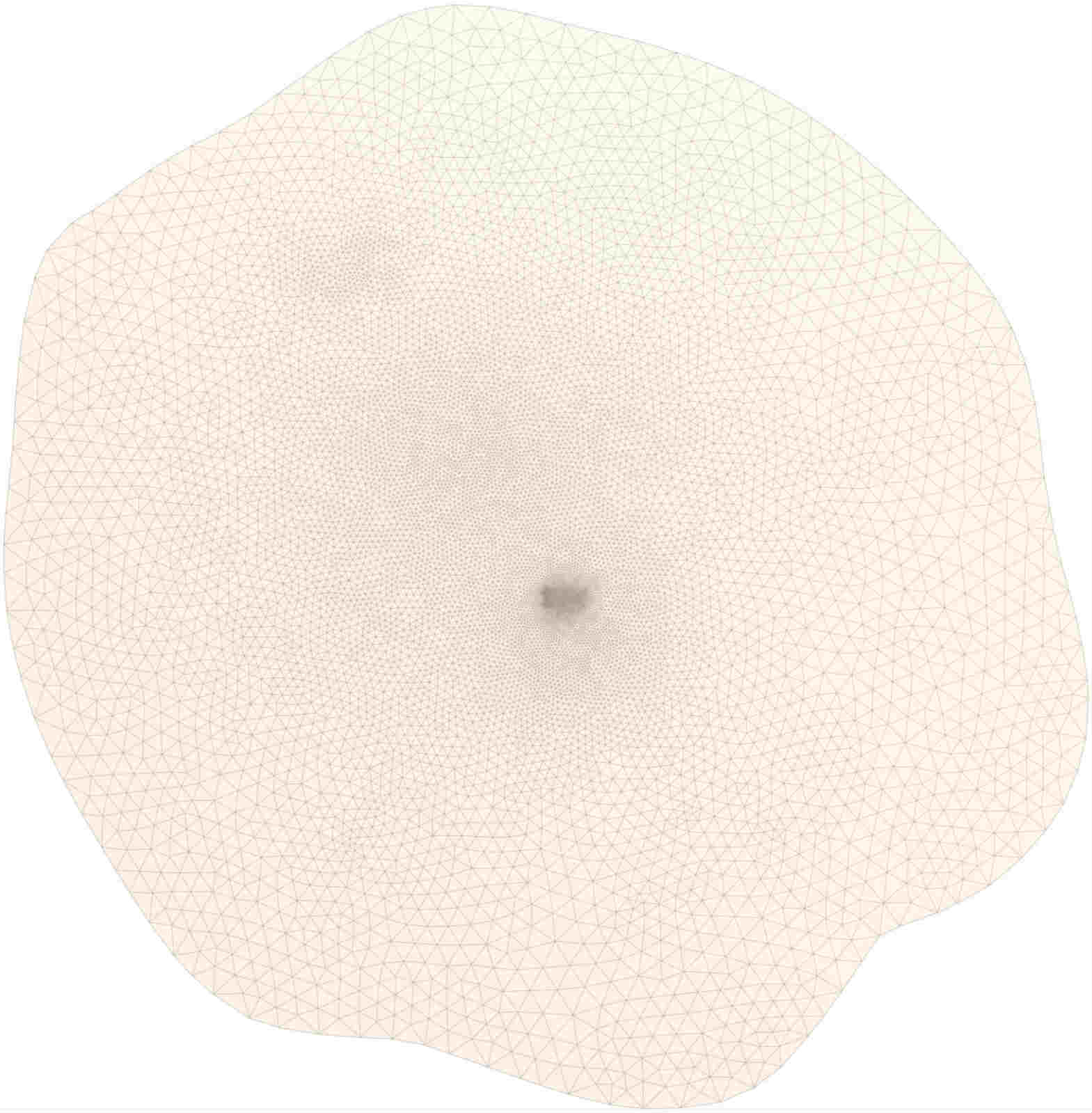}
		\caption{ }   
	\end{subfigure}   
	\begin{subfigure}[b]{0.15\textwidth}
		\includegraphics[width=\textwidth]{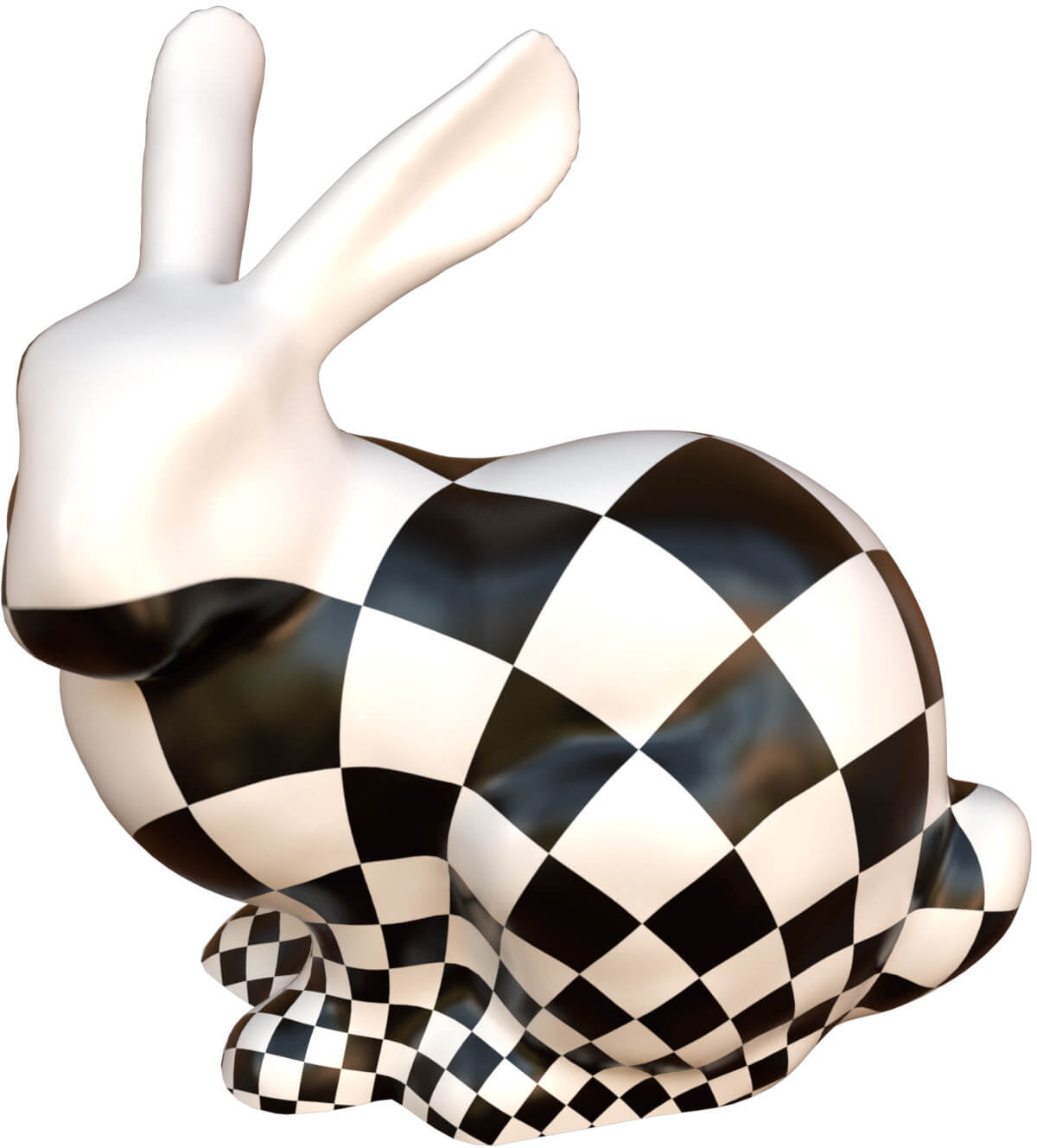}
		\caption{ }     
	\end{subfigure}  
	
	\begin{subfigure}[b]{0.15\textwidth}
		\includegraphics[width=\textwidth]{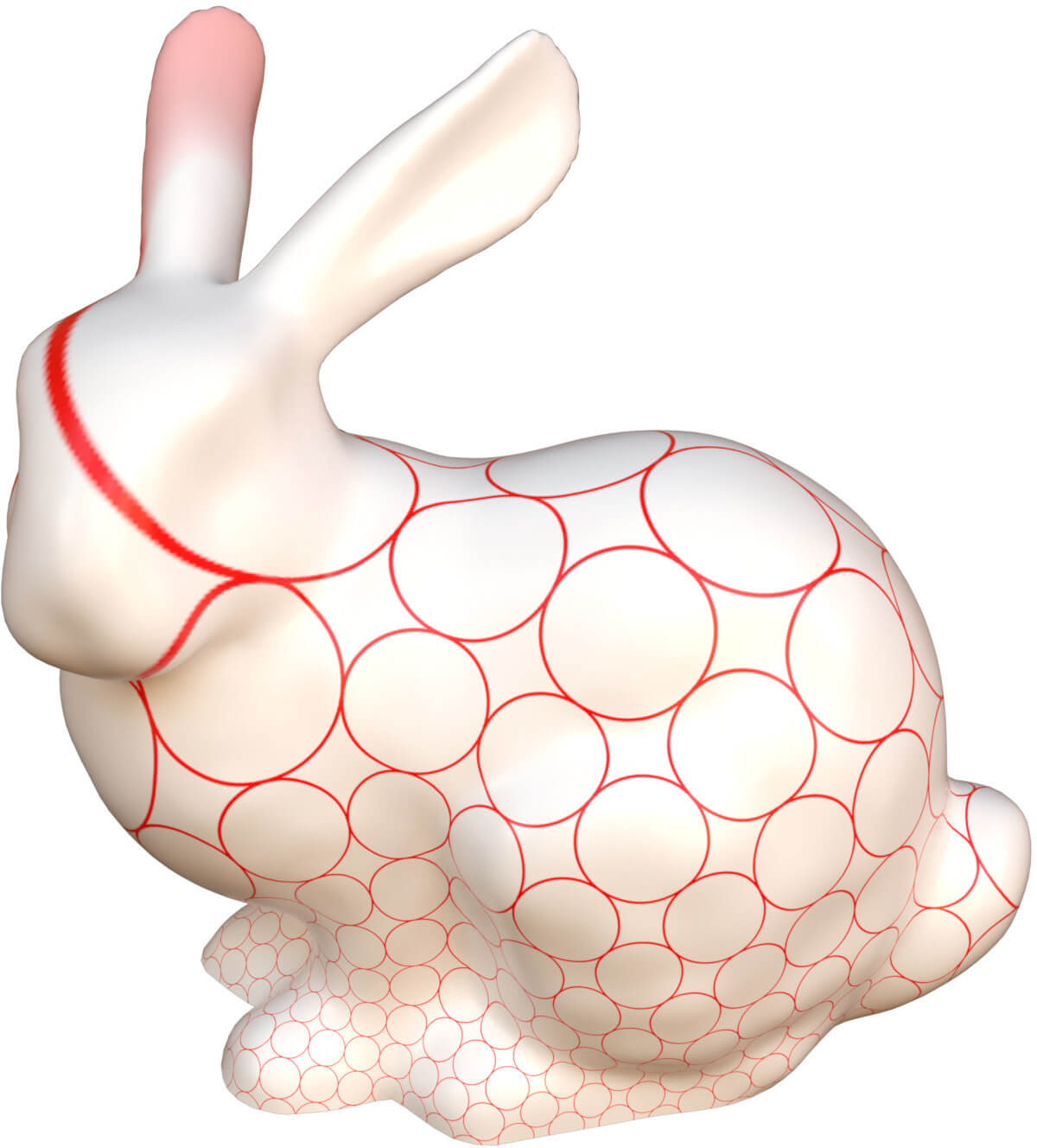}
		\caption{ }   
	\end{subfigure}
	\begin{subfigure}[b]{0.15\textwidth}
		\includegraphics[width=\textwidth]{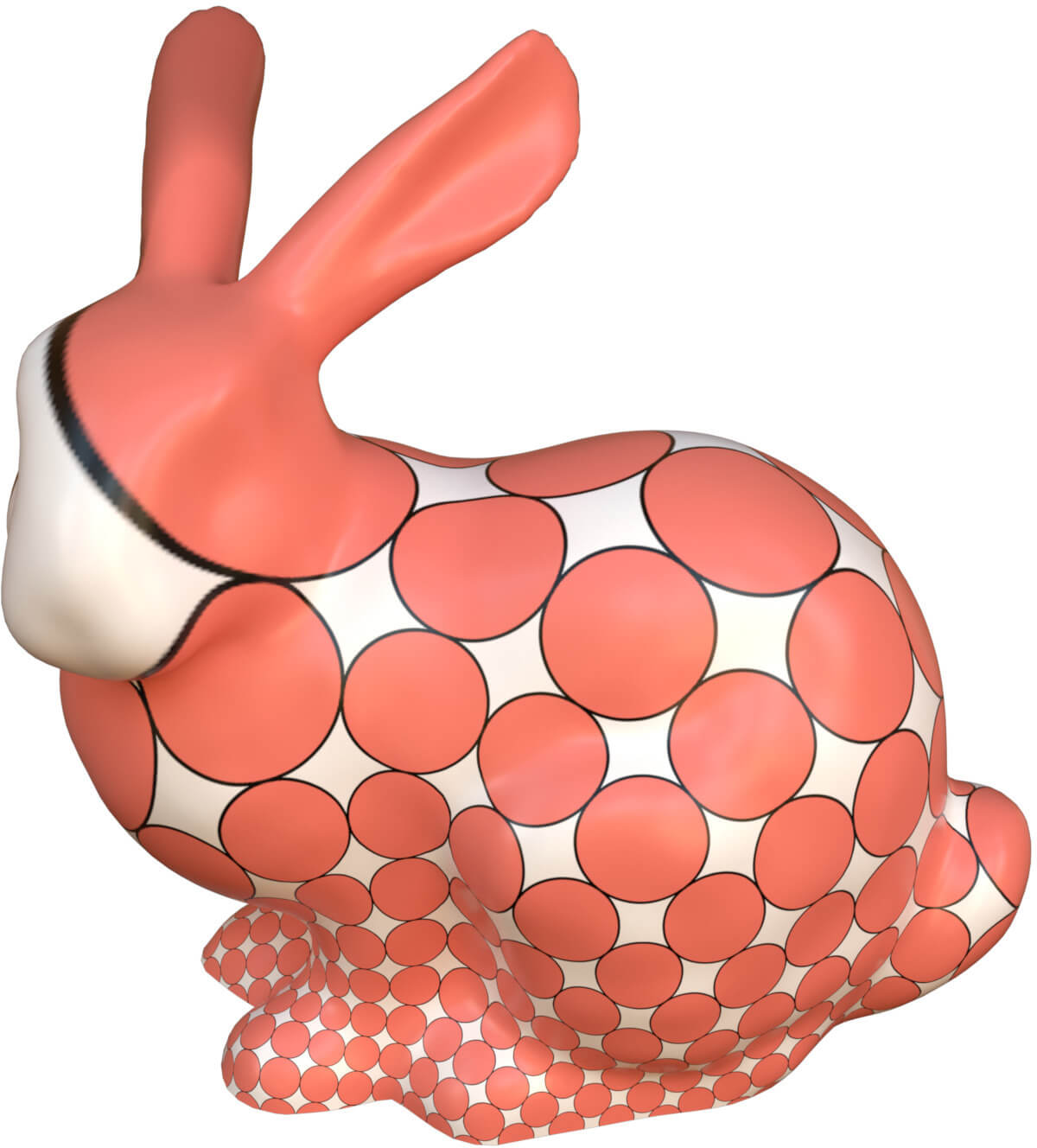}
		\caption{ }   
	\end{subfigure}   
	\begin{subfigure}[b]{0.15\textwidth}
		\includegraphics[width=\textwidth]{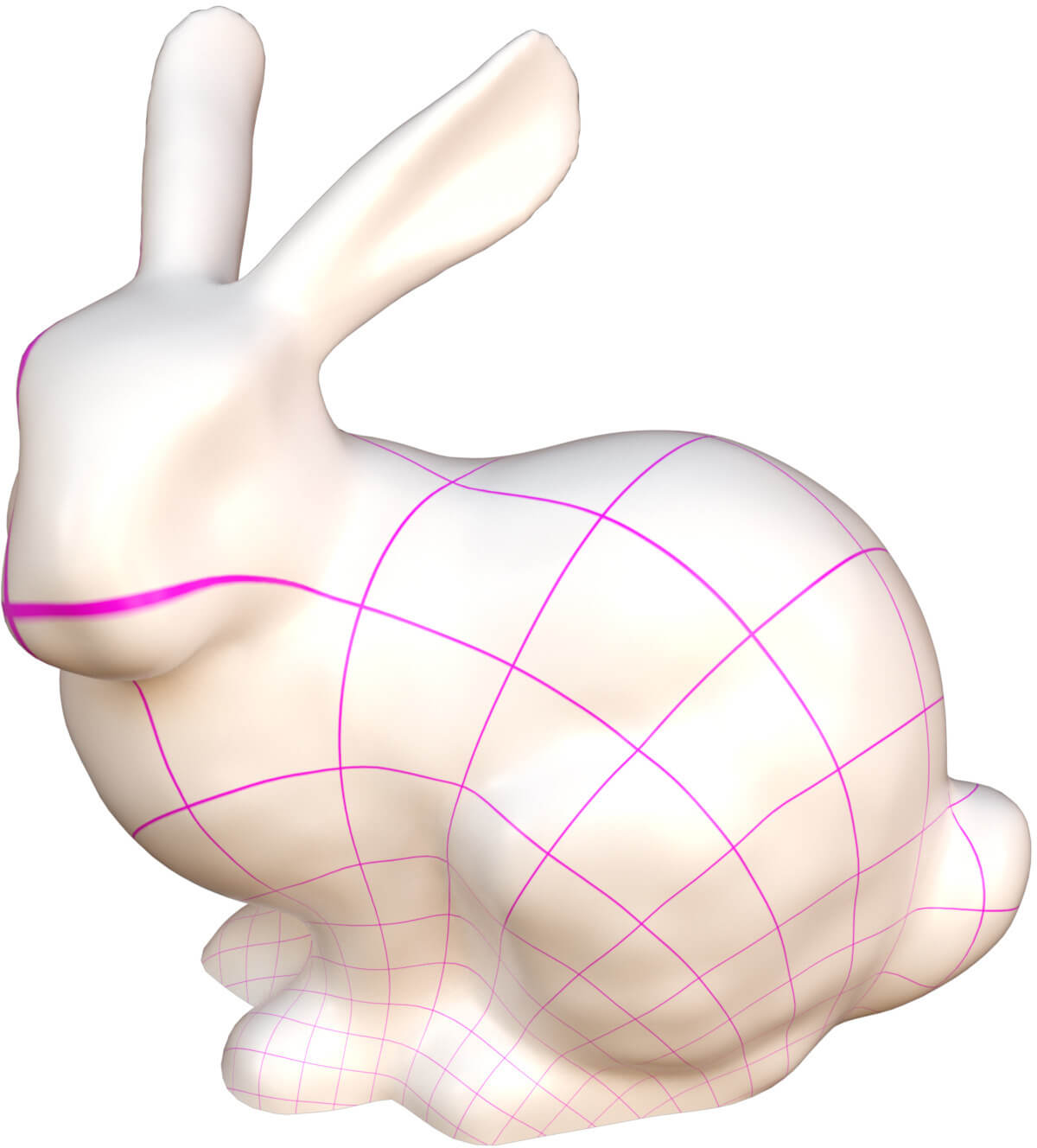}
		\caption{ }     
	\end{subfigure}     
	\caption{(a) the bunny; (b) its  parameterization with Calabi flow;  (c), (d), (e) rendered  with different textures.}\label{fig:calabiBunny}
\end{figure}

Flow-based algorithms do not work on the  positions directly, instead they evolve  the surface metric into a flat one. The final parametrization is obtained by embedding the surface of the flat metric to the 2D plane.
In the Figure \ref{fig:calabiBunny}, we show that our Calabi-flow-based conformal parameterization. The angles are preserved very well in several corresponding  rendering results.

\textbf{Contribution.} We design a different conformal parameterization algorithm based on discrete Calabi flow which is derived from the Calabi energy. In our algorithm, we use a new dual-laplacian operator to obtain the optimal solution.  To the best of our knowledge, this is the first time that the discrete Calabi flow and dual-laplacian operator are introduced to the graphics literature. To summarize, our algorithm is a new perspective method to gain parameterization. The energy expression is simple and very easy to understand: it is squared difference between current curvature vector and target curvature. This is one of main advantages towards Ricci flow and CETM.

\section{Related Works}

Due to the abundance of literature on mesh parameterization. Here, we focus on approaches that are the most relevant to ours. We refer the reader to some excellent surveys \cite{Sheffer2006MeshPM,Hormann2007MeshPT} for more complete information.

\textbf{Tutte's embedding.}
The algorithms \cite{Floater2003OnetoonePL,Desbrun2002IntrinsicPO,Weber2014LocallyIP,cohen2006designing} based on Tutte's embedding of planar graphs are fundamental ones. 
They map 3D disk-topology meshes onto Euclidean flat plane. \cite{Gortler2006DiscreteOO} generalized them to handle genus-one meshes by integrating harmonic one-forms on the torus. 
Euclidean orbifolds are used to process sphere-topology meshes in \cite{Aigerman2015OrbifoldTE} which achieves flat surfaces with cone singularities.
Recently orbifolds are also extended to hyperbolic space in\cite{Aigerman2016HyperbolicOT}. Hyperbolic orbifolds is able to handle wider variety of cone arrangements and topologies than Euclidean orbifolds. 
The advantages of all of these  methods are derived from that Tutte's theory are guaranteed to be bijective. 

\textbf{Injective parameterization.}
Besides bijective, the injective and distortion bounded algorithms \cite{myles2013controlled,myles2012global,myles2014robust} are also sought after. The algorithms presented in \cite{hormann2000mips,Sheffer2005ABFFA,schuller2013locally,Aigerman2014LiftedBF,Weber2014LocallyIP,Fu2015ComputingLI} are locally injective, and the ones in \cite{Lipman2012BoundedDM,Campen2015QuantizedGP,Smith2015BijectivePW} are globally injective.

\textbf{Conformal parameterization.}
Based on conformal geometry theory, discrete conformal mapping definition are proposed, and  Ricci flow \cite{Jin2007DiscreteSR,Jin2008DiscreteSR,Jin2008VariationalMO}, circle packing \cite{stephenson2005introduction}, circle patterns \cite{kharevych2006discrete}, conformal equivalence
of triangle meshes (CETM) \cite{springborn2008conformal} and conformal flattening \cite{ben2008conformal} are presented. The relationships and comparisons among these methods are discussed in \cite{Zhang2014TheUD,Zhang2015SurveyOD}. All of these algorithms achieve a discrete flat metric under certain flows. 
They iteratively update the edge lengths which are conformal to the original mesh in each step. Ricci flow can also work under hyperbolic background geometry \cite{jin2006computing,Yang2009GeneralizedDR,shi2013hyperbolic}. However these flow-based methods are not guaranteed to be injective.

Another approach for conformal parameterization is based on conformal structure in Riemann surface theory. In the seminar paper \cite{gu2003global}, the discrete holomorphic differentials   are defined, and the conformal mapping are achieved by computing discrete conformal structures.

\textbf{Area-preserving.}
Another kind of parameterizations is area-preserving.  Recently discreate optimal mass transport theory are designed \cite{gu2013variational} and applied to obtain the map which can preserve the local triangle areas \cite{su2016area,zhao2013area,su2013area}.  The conformal and area-preserving approaches can also be mixed or interpolated by polar factorization method \cite{yu2017surface} to obtain the parameterization between them.
 
\section{Calabi Energy and  Calabi Flow}

A manifolds $M^n$ is a topological space which is locally Euclidean of dimension $n$. It is covered by a series coordinate charts $\{U_{\alpha},\phi_{\alpha}\}$ which are $C^{\infty}$ compatible \cite{lee2003manifolds}. A Riemannian metric tensor \textbf{g} on the manifold is a Euclidean inner product defined on the tangent space $T_p(S)$ of each point $p$ of $S$ \cite{riemanngeometry}. 

For a 2-dimensional surface $S$, we usually embed it in $\mathbb{R}^3$, and equip each point with a local chart: $$\textbf{r}: D \rightarrow \mathbb{R}^3 $$ where $D\in \mathbb{R}^2$ and  $\textbf{r}$ is smooth. $\textbf{r}$ is called a parameterization of $S$. At each point, let $\textbf{r}_i = \partial \textbf{r} / \partial u^i , i = 1, 2$ be the tangent vectors along the isoparametric curves. They are the basis of tangent space at that point. The length of a general tangent vector $d\textbf{r} = \textbf{r}_1 du_1 + \textbf{r}_2 du_2$ can be computed by:
\begin{equation}
ds^2 = \left<d\textbf{r},d\textbf{r}\right> = 
\left(\begin{matrix}
du_1 & du_2
\end{matrix}\right)
\left(\begin{matrix}
g_{11} & g_{12}\\
g_{21} & g_{22}
\end{matrix}\right)
\left(\begin{matrix}
du_1 \\ du_2
\end{matrix}\right)
\end{equation}
where $\left<-,-\right>$ is the inner product in $\mathbb{R}^3$, and $g_{ij} = \left<r_i, r_j\right>$. In this case, the matrix $\textbf{g} = (g_{ij})$ is the Riemannian metric tensor on $S$. And we denote the inner product induced by $\textbf{g}$ as $\left<-,-\right>_{\textbf{g}}$.

The angle between two tangent vectors can be measured by $\textbf{g}$. Suppose $\delta \textbf{r} = \textbf{r}_1\delta u_1 + \textbf{r}_2\delta u_2$ is another tangent vector, the angle between $d\textbf{r}$ and $\delta \textbf{r}$ measure by $\textbf{g}$ is defined as:
\begin{equation}\label{angle}
\theta_{\textbf{g}} = \cos^{-1} \frac{\left<d\textbf{r},\delta \textbf{r}\right>_\textbf{g}}{\sqrt{\left<d\textbf{r},d\textbf{r}\right>_\textbf{g}}\sqrt{\left<\delta\textbf{r},\delta \textbf{r}\right>_\textbf{g}}}
\end{equation}

Suppose $\lambda: S \rightarrow \mathbb{R}$ is a real function defined on the surface. Define another Riemannian metric 
\begin{equation}
\bar{\textbf{g}} = e^{2\lambda}\textbf{g}
\end{equation}
then we have
\begin{equation}
\left<d\textbf{r},\delta \textbf{r}\right>_{\bar{\textbf{g}}} = e^{2\lambda}\left<d\textbf{r},\delta \textbf{r}\right>_{{\textbf{g}}}
\end{equation}
According to Eq. \ref{angle}, we obtain $\theta_{\textbf{g}} = \theta_{\bar{\textbf{g}}}$. So we say $\textbf{g}$ and $\bar{\textbf{g}}$ is conformally equvalent and $e^{2\lambda}$ is conformal factor between $\bar{\textbf{g}}$ and $\textbf{g}$. 

Any Riemannian metrics of 2-dimensional surface are \textbf{locally} conformally equivalent to the Euclidean flat metric \cite{chern1995isothermal}. That is, we can always choose a special parameters, such that the metric is represented as:
\begin{equation}
ds^2 = e^{2\lambda}(du_1^2 + du_2^2)
\end{equation}
Such kinds of parameterizations are called the isothemal coordinates of the surface.

Under isothermal coordinates, the Gaussian curvature is represented as:
\begin{equation}
K = -e^{-2\lambda}\nabla^2 \lambda
\end{equation}
where $\nabla^2$ is the normal Laplace operator:
\begin{equation}
\nabla^2 = \frac{\partial^2}{\partial u_1^2} + \frac{\partial^2}{\partial u_2^2}
\end{equation}
and $e^{-2\lambda}\nabla^2$ is called the Laplace-Beltrami operator. Here we denote it as $\Delta$

 %For details of defination, we refer the reader to \cite{tian2012kahler}.
2-dimensional Calabi flow was studied in \cite{chen2008calabi}. Suppose $S$ is a smooth surface with a Riemann metric $\textbf{g}$, 
Calabi introduced   the so-called Calabi energy, which is defined as:
\begin{equation}
\Phi(\textbf{g}) = \int_M K^2dA
\end{equation}
where $dA$ is the area element of $S$. 

The Calabi flow on $S$ is defined as:  %, $[\textbf{g}_0]$ is the conformal class of $\textbf{g}_0$. 
 \begin{equation}
 \frac{dg_{ij}}{dt} = (\Delta K)g_{ij} %\textbf{g}\in[\textbf{g}_0]
 \end{equation}
 where $R$ is the scalar curvature induced by metric $\textbf{g}$ and $\Delta$ is the Laplace-Beltrami operator.
 
With isothermal coordinates, we have $\textbf{g}=e^{2\lambda}\textbf{g}_0$, then the Calabi flow becomes:
 \begin{equation}\label{smoothcalabiflow}
 \frac{d\lambda}{dt} = \Delta K 
 \end{equation}
 
It is proved that the above Calabi flow is convergent under certain conditions \cite{calabi1982kahler}.

\section{Discrete Metric and Conformal Class}

In practice, smooth surfaces are often approximated by simplicial complexes, that is, piecewise linear triangle meshes. Concepts in the continuous setting can be generalized to the discrete setting. In this paper, a triangle mesh is denoted as $M$, which is associate with a vertex set $V$, a edge set $E$ and a face set $F$. $v_i$ represents a certain vertex, $e_{ij}$ represents the edge between vertices $v_i$ and $v_j$, and $f_{ijk}$ represents the face formed by $v_i$, $v_j$ and $v_k$.

A Riemannian metric on a piecewise linear discrete mesh $M = (V,E,F)$ is defined as a positive scalar function on edges: \[l: E\rightarrow\mathbb{R}^+\] such that for each triangle face $f_{ijk}$, edge lengths $\{l_{ij}, l_{jk}, l_{ki}\}$ satisfy the triangle inequality: $$l_{ij} + l_{jk} > l_{ki}$$ $$l_{ij} + l_{ki} > l_{jk}$$  $$ l_{jk} + l_{ki} > l_{ij}$$
The discrete metric determines corner angles $\{\theta_i^{jk}, \theta_j^{ki}, \theta_k^{ij}\}$ of triangles by cosine laws in Euclidean background geometries.

In discrete setting, we have infinite ways to assign edge lengths to make triangle inequalities be satisfied. However, it is difficult to settle down what the conformal deformation means. Inspired by the property that conformal map sends infinitesimal circles to infinitesimal circles, Thurston introduced circle packing on weighted meshes in \cite{thurston}. 
Inspired by conformal factors, Luo introduced another metric in \cite{luo2003yamabe} called Yamabe flow metric. 
And as a generalization of Thurston's circle packing metric, inverse distance metric was introduced in \cite{Yang2009GeneralizedDR}. These two metrics induced different definitions of conformal class. Both are approximations of definition of conformal deformation in smooth settings. In \cite{Zhang2014TheUD} , a unified framework for circle packing was introduced and the metrics above were all included.

\textbf{Thurston's Circle Packing Metric}
The circle packing metric was firstly defined on weighted meshes in \cite{thurston}. A weighted mesh $(M, \Gamma, \Phi)$ with a circle packing metric is a mesh with a function $\Gamma$ assigning a radius $r_i$ to each vertex $v_i$: $$\Gamma : V\rightarrow \mathbb{R}^+$$ and a function assigning a weight $\Phi_{ij}$ to each edge: $$\Phi: E \rightarrow [0,\frac{\pi}{2}]$$ 
Under this setting, we can determine edge lengths using different cosine laws in different background geometry:

    \begin{equation} 
    l_{ij}^2 = r_i^2 + r_j^2 + 2r_ir_j\cos\Phi_{ij}  
    \end{equation}

\begin{figure}[ht]
	\centering
	\begin{subfigure}[b]{0.22\textwidth}
		\includegraphics[width=\textwidth]{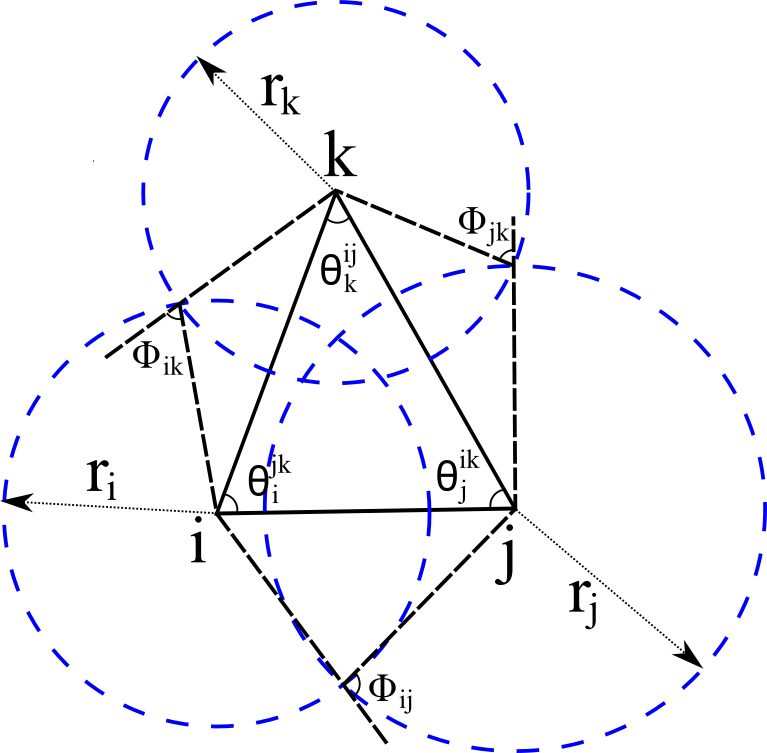}
		\caption{}   \label{fig:circlePacking}
	\end{subfigure}
	\begin{subfigure}[b]{0.22\textwidth}
		\includegraphics[width=\textwidth]{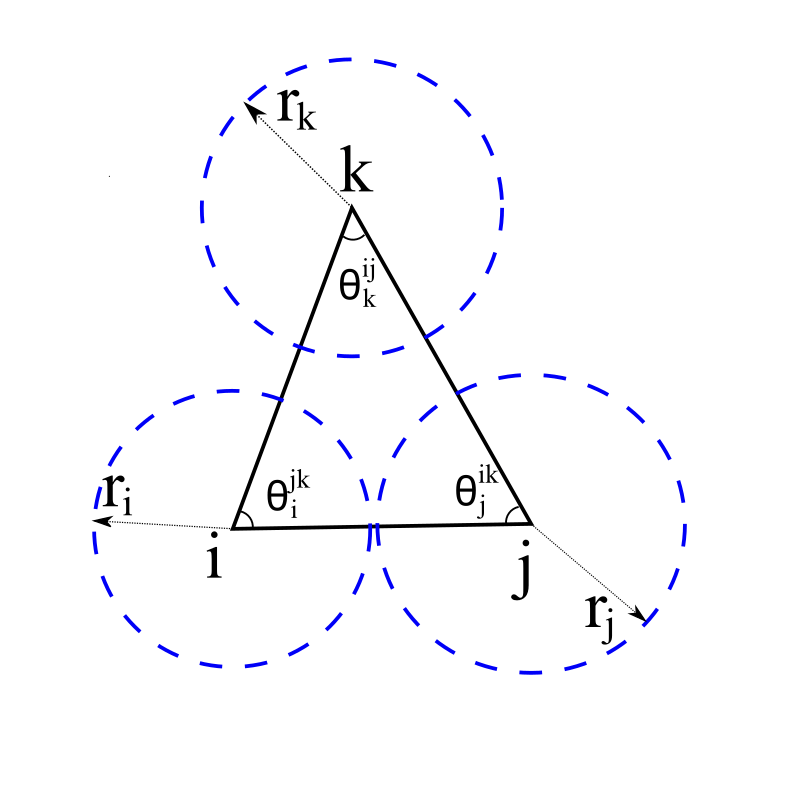}
		\caption{}   
	\end{subfigure}   
    \caption{The two kinds of metrics. (a) Thurston's cirlce packing, (b) inversive distance circle packing}
    \label{fig:metrics}
\end{figure}

Given a mesh $M$, we say two circle metrics $(\Gamma_1, \Phi_1)$ and $(\Gamma_2, \Phi_2)$ are conformally equivalent if $\Phi_1 = \Phi_2$, they are in the same conformal class.

\textbf{Inversive Distance Circle Packing Metric}
This kind of metric is first introduced in \cite{bowers2004uniformizing}. It is a generalization of Thurston's circle packing. We define a function on edges $I: E\rightarrow \mathbb{R}$, which is called inversive distance function. Edge lengths is determined as follows:

	\begin{equation}
	l_{ij}^2 = r_i^2 + r_j^2 + 2r_ir_j I_{ij}   
	\end{equation}

Given a mesh $M$, we say two circle metrics $(\Gamma_1, I_1)$ and $(\Gamma_2, I_2)$ are conformally equivalent if $I_1 = I_2$, they are in the same conformal class.
This kind of metric can approximate initial edge lengths very well, so it is more practical than Thurston's circle packing.

\textbf{Geometric Interpretation}
Two kinds of circle packing metrics can be illustrated as Fig. \ref{fig:metrics}. Each vertex $v_i$ has a circle with radius $r_i$ centering at it.  For Thurston's circle packing metric, two adjacent circles intersect with angle $\Phi_{ij}$. And edge length is the distance between two circle centers. And for inverse distance circle metric, circles need not intersect with each other, and $\cos\Phi_{ij}$ is replaced by inverse distance $I_{ij}$.  

\section{Discrete Calabi flow}

Discrete Calabi flow  \cite{ge2012combinatorial} defined on triangular meshes is a counterpart of smooth Calabi flow on smooth surfaces.
Given weighted mesh $(M, \Gamma, \Phi)$ with circle packing metric, we set:
\begin{equation}
u_i =  \log{r_i}.
\end{equation}
We define discrete Calabi flow as:
\begin{equation}\label{combinatorial calabi}
\frac{d\textbf{u}}{dt} = \Delta_{dual} \mathbf{K},
\end{equation}
where $\Delta_{dual}$ is a new kind of Laplacian operator, we call it discrete \textbf{dual-Laplacian operator} \cite{ge2012combinatorial}, and $\mathbf{K}$ is the well-known traditional Gaussian curvature which is computed as the angle deficit, i.e. $2\pi$ minus angle sum for an inner vertex and $\pi$ minus angel sum for a boundary vertex.

We can also modify discrete Calabi flow into a form with prescribed curvature:
\begin{equation} \label{prescribedcalabi}
\frac{d\textbf{u}}{dt} = \Delta_{dual} (\textbf{K} - \bar{\textbf{K}})
\end{equation}
and $\bar{\textbf{K}} = (\bar{K_1}, \bar{K_2}, ..., \bar{K_N})$ is the prescribed curvature vector. 

To compare, we notice that the Ricci flow is the following:
\begin{equation} \label{prescribedricci}
\frac{d\textbf{u}}{dt} =  \textbf{K} - \bar{\textbf{K}}
\end{equation}

The discrete Calabi energy is defined as the following. And the Calabi flow is the negative gradient flow of it.
\begin{equation}
\mathbf{C}(u)  = \sum_{v_i\in V}(\bar{K_i} - K_i)^2
\end{equation} 

With a little calculation, we can find that
\begin{equation}
\nabla_u \mathbf{C}= \Delta_{dual}(\bar{\textbf{K}} - \textbf{K})
\end{equation}

If Calabi flow converges, Calabi energy will arrive at its critical point where $$\Delta_{dual}(\bar{\textbf{K}} - \textbf{K}) = 0$$
then we can obtain prescribed curvatures.

We use  the method of gradient descending to solve  the optimization problem of Calabi energy. The scheme of conjugate gradient descending and accelerated gradient descending can also be applied.

Intuitively, we can treate $\frac{\partial K_j}{\partial u_i}(\bar{K}_j - K_j)$ as the descending direction of $K_j$ along axis $u_i$. So the gradient of Calabi energy can be seen as the average of adjecent descending directions.

\textbf{Dual-laplacian operator.}
The dual-Laplacian operator $\Delta_{dual}$ is different from the well-known cotangent Laplacian operator $\Delta_{cot}$.
It is a special type of the discrete Laplacian, and comes from the dual structure of the circle packings. It has been discussed and studied by Glickenstein in  \cite{glickenstein2005geometric,glickenstein2005combinatorial} systematically.
It is  defined to be $\Delta_{dual} = - L_{dual}^T$, where

\begin{align}
L_{dual} = \nabla_u K = (L_{ij})_{N\times N} &= \frac{\partial(K_1, ..., K_N)}{\partial(u_1,...,u_N)}\\
\nonumber
&=\begin{pmatrix}
    \frac{\partial K_1}{\partial u_1} & \cdot & \cdot & \cdot & \frac{\partial K_1}{\partial u_N} \\
    \cdot & \cdot & \cdot & \cdot & \cdot\\
    \cdot & \cdot & \cdot & \cdot & \cdot\\
    \frac{\partial K_N}{\partial u_1} & \cdot & \cdot & \cdot & \frac{\partial K_N}{\partial u_N}
  \end{pmatrix}
\end{align}

Both $\Delta_{dual}$ and $\Delta_{cot}$ operate on the column vector functions which are defined on mesh vertices with a matrix multiplication.

We need to derive the explicit form of $L = \nabla_u \mathbf{K}$. The calculation is direct. We write $i \sim j$ in the following if $v_i$ and $v_j$ are adjacent.

If $i \sim j$, 
\begin{equation}
\label{eqn:Ldual}
(L_{dual})_{ij} = (L_{dual})_{ji} = \frac{\partial K_i}{\partial u_j} = - \sum_{f_{ijk}}\frac{\partial \theta_i^{jk}}{\partial u_j},
\end{equation}

If $i = j$, according to Gauss-Bonnet Theorem, we have
\begin{equation}
(L_{dual})_{ii} = \frac{\partial K_i}{\partial u_i} =  - \sum_{j \sim i}\frac{\partial K_j}{\partial u_i}, 
\end{equation}

Otherwise, 
\begin{equation}
(L_{dual})_{ij} = 0.
\end{equation}

So the calculation of dual Laplacian boils down to the calculation of $(L_{dual})_{ij}$. This quantity can be associated to edge $e_{ij}$. We call it edge weight, and denote it as $(w_{d})_{ij}$, which is only determined by its two adjecent faces.

\begin{figure}[ht]
	\centering
		\includegraphics[width=0.5\textwidth]{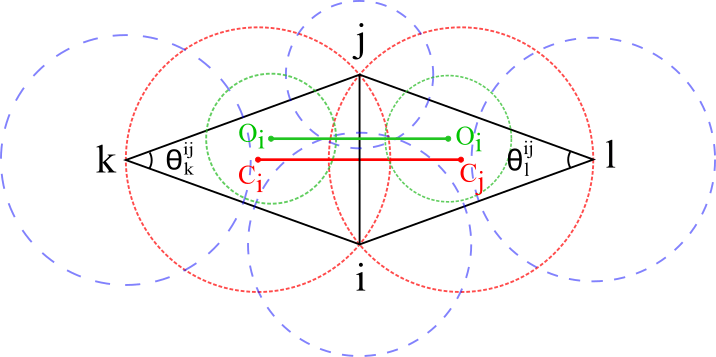}
	\caption{Geometric interpretation of dual Laplacian and cotangent Laplacian}
    \label{fig:laplacian}
\end{figure}

For the equation \ref{eqn:Ldual},  there is a formula with nice geometric interpretation provided in \cite{ge2012combinatorial}.
As shown in Fig. \ref{fig:laplacian}, the face  $f_{ijk}$ and face $f_{ijl}$ are two adjecent faces at edge $e_{ij}$.
On each vertex of the face, there is a circle with the radius value of its corresponding metric.
 For each face of the triangle  we can assign a circle orthogonal to three vertex circles simultaneously. This circle is called the power circle of the face and its center is called power center. 
 It can be shown that the line segment connecting two power centers $O_i$ and $O_j$ in this figure is orthogonal to edge $e_{ij}$, and we call this segment as dual edge of $e_{ij}$. Denote the length of $e_{ij}$ as $l_{ij}$ and the length of its dual edge as $l^o_{ij} = |O_i O_j|$.
 
Finlay we have a very nice formula for edge weight $(w_{d})_{ij}$:
\begin{equation}
(L_{dual})_{ij} = (w_{d})_{ij} = - \frac{\partial (\theta_i^{jk}+ \theta_i^{jl})}{\partial u_j} =  \frac{l^o_{ij}}{l_{ij}},
\end{equation}
While the formula of cotangent-Laplacian operator is :
\begin{equation}
\Delta_{cot} = -L^T_{cot}.
\end{equation}

The operator $L_{cot}$ is compute as follows:

If $i \sim j$, 
\begin{equation}
(L_{cot})_{(ij)} = (L_{cot})_{ji} = \frac{1}{2}\sum_{f_{ijk}} \cot {\theta_k^{ij}},
\end{equation}

If $i = j$, 
\begin{equation}
(L_{cot})_{ii} = - \sum_{j \sim i}(L_{cot})_{ij},
\end{equation}

Otherwise, 
\begin{equation}
(L_{cot})_{ij} = 0.
\end{equation}

The computation of  the cotangent Laplacian are also   related to the corresponding edge weights. And it  has an geometric interpretation too.  As shown in figure. \ref{fig:laplacian},  there is a circumcircle for every triangle, the line segment between two circumcenter $C_i$ and $C_j$ is also called dual edge of $e_{ij}$. Denote the length of $e_{ij}$ as $l_{ij}$ and length of its dual edge as $l^c_{ij}$, then we have 
\begin{equation}
(L_{cot})_{ij} = (w_c)_{ij} = \frac{\cot \theta_k^{ij} + \cot \theta_l^{ij}}{2} = \frac{l^c_{ij}}{l_{ij}}
\end{equation}

Actually, the circumcenter can be also treated as some kind of power center. In this case, three vertex circles shrink to a point and the orthogonal circle of them is exactly the same as the circumcicle. Therefore these two kinds of laplacian operators have unified forms.

\begin{figure} [h!] 
	\centering   
	\begin{subfigure}[b]{0.15\textwidth}
		\includegraphics[width=\textwidth]{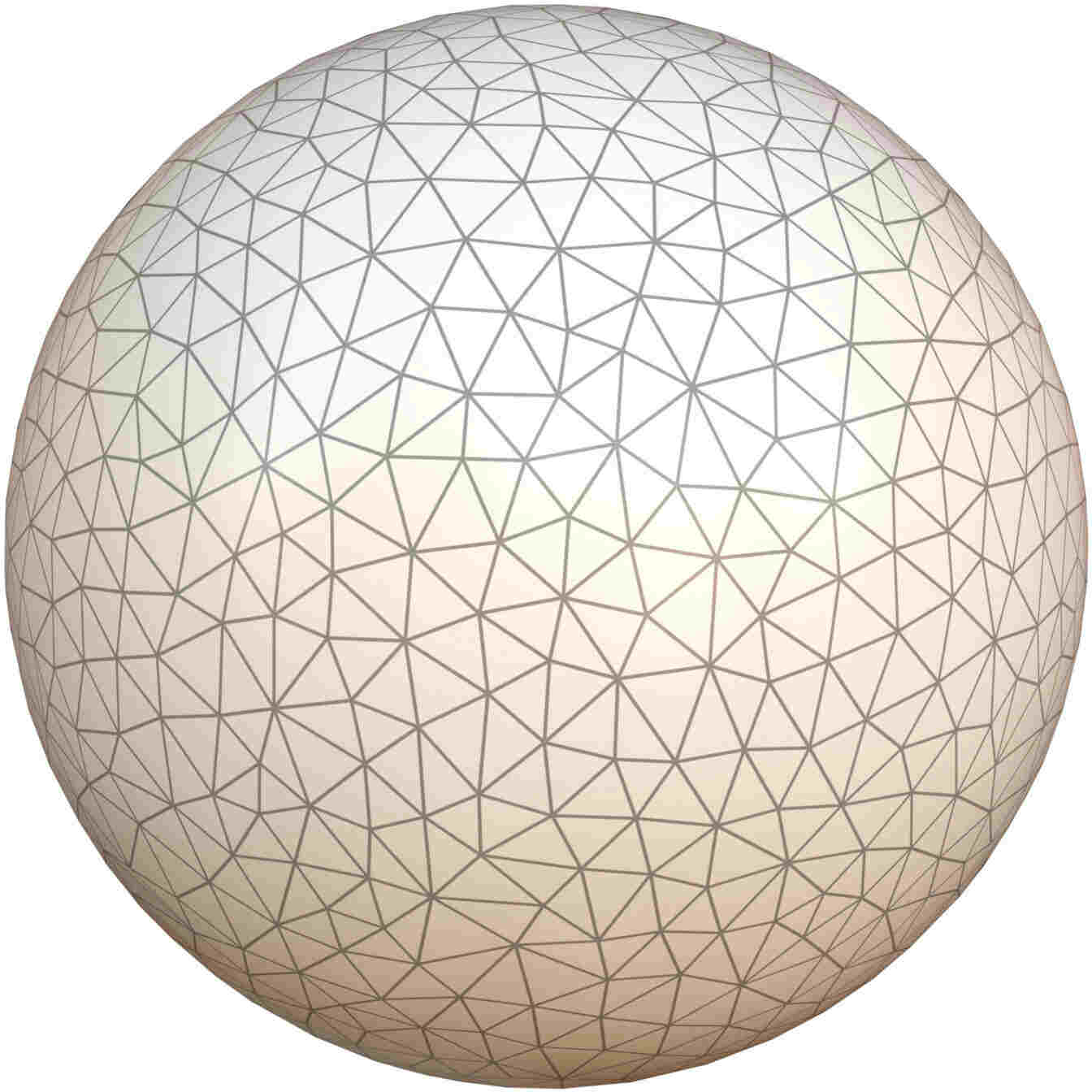}
		\caption{ }   \label{fig:bunny0}
	\end{subfigure}
	\begin{subfigure}[b]{0.15\textwidth}
		\includegraphics[width=\textwidth]{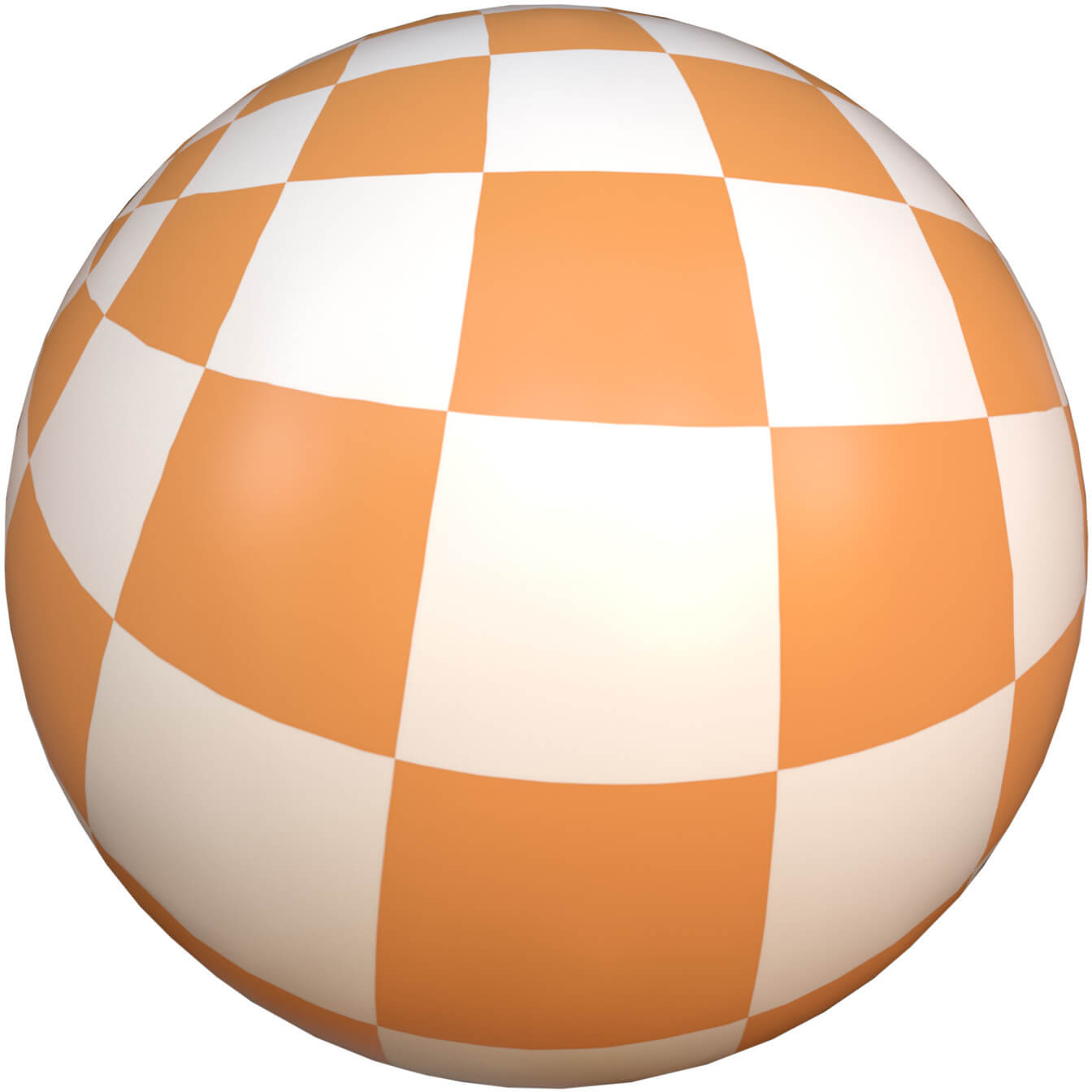}
		\caption{ }   
	\end{subfigure}   
	\begin{subfigure}[b]{0.15\textwidth}
		\includegraphics[width=\textwidth]{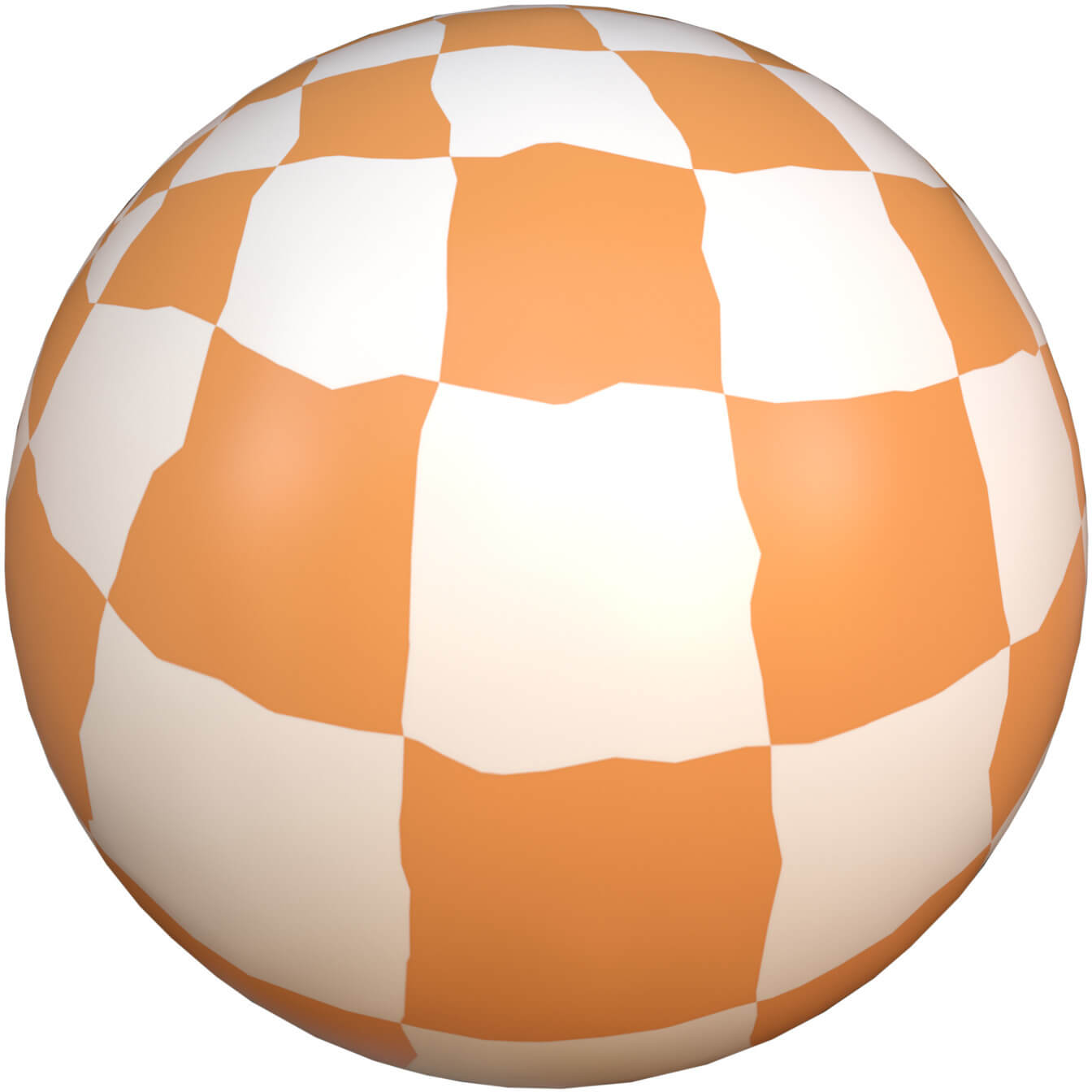}
		\caption{ }     
	\end{subfigure}  
	
	\begin{subfigure}[b]{0.15\textwidth}
		\includegraphics[width=\textwidth]{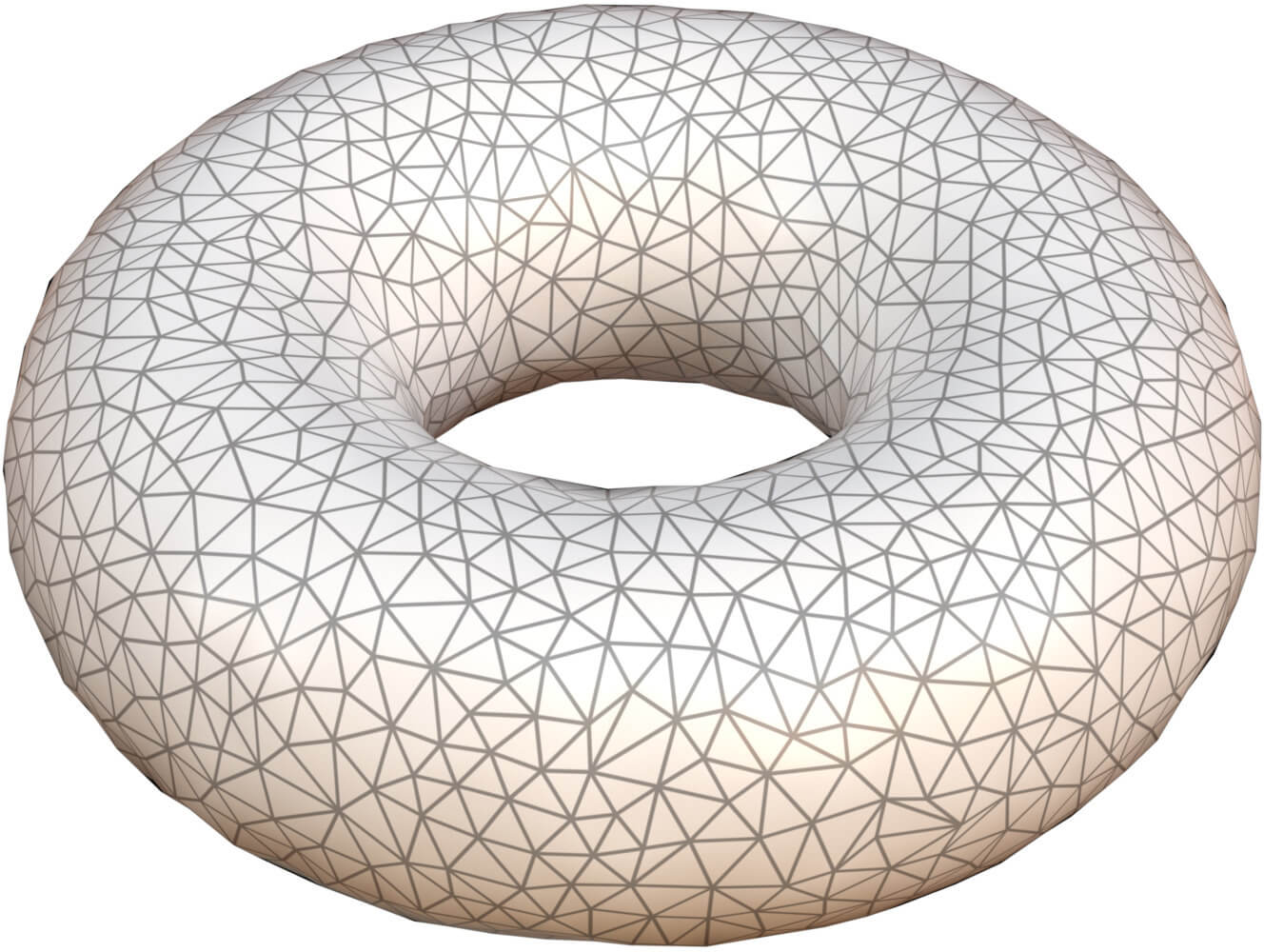}
		\caption{ }   
	\end{subfigure}
	\begin{subfigure}[b]{0.15\textwidth}
		\includegraphics[width=\textwidth]{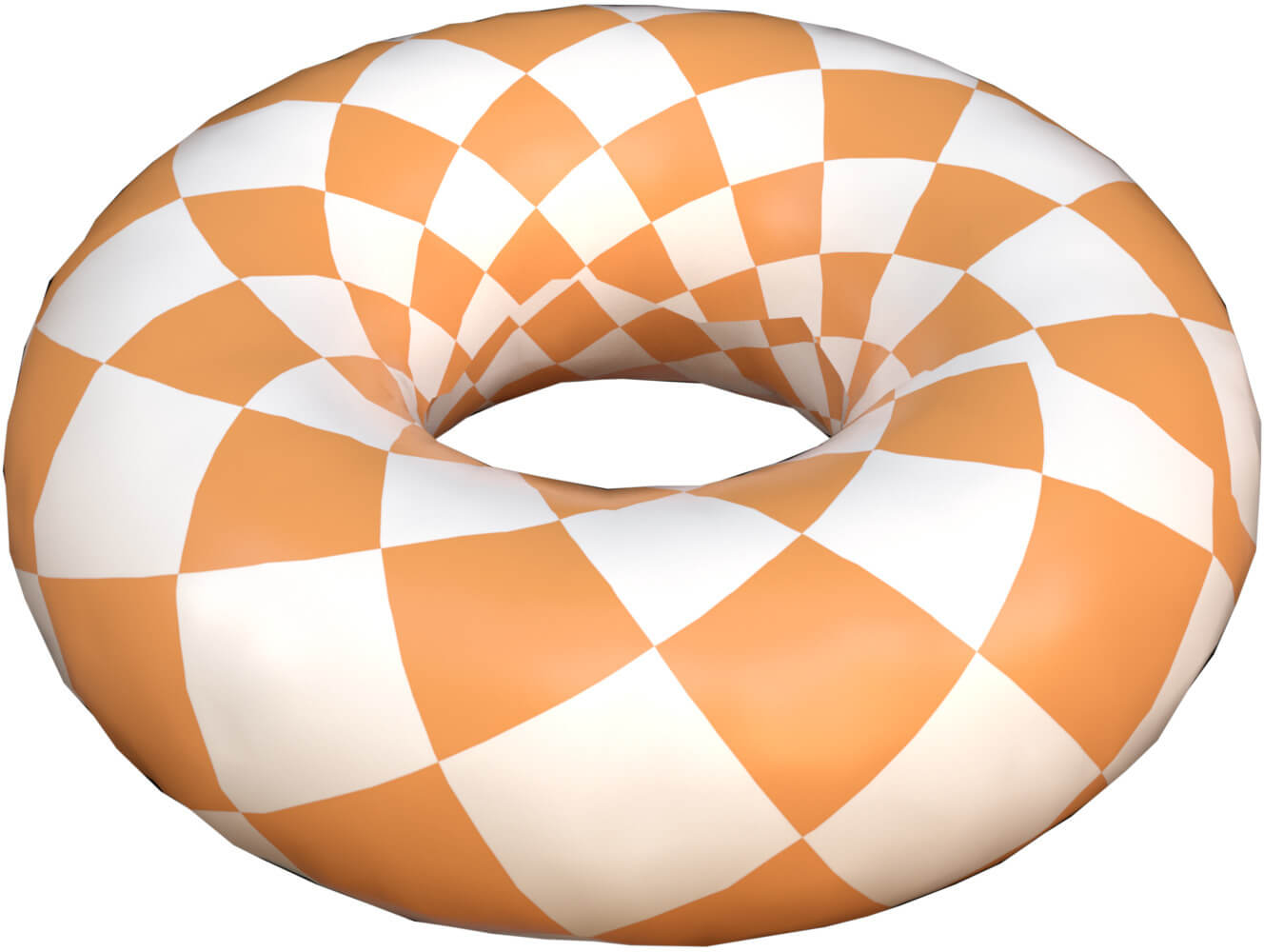}
		\caption{ }   
	\end{subfigure}   
	\begin{subfigure}[b]{0.15\textwidth}
		\includegraphics[width=\textwidth]{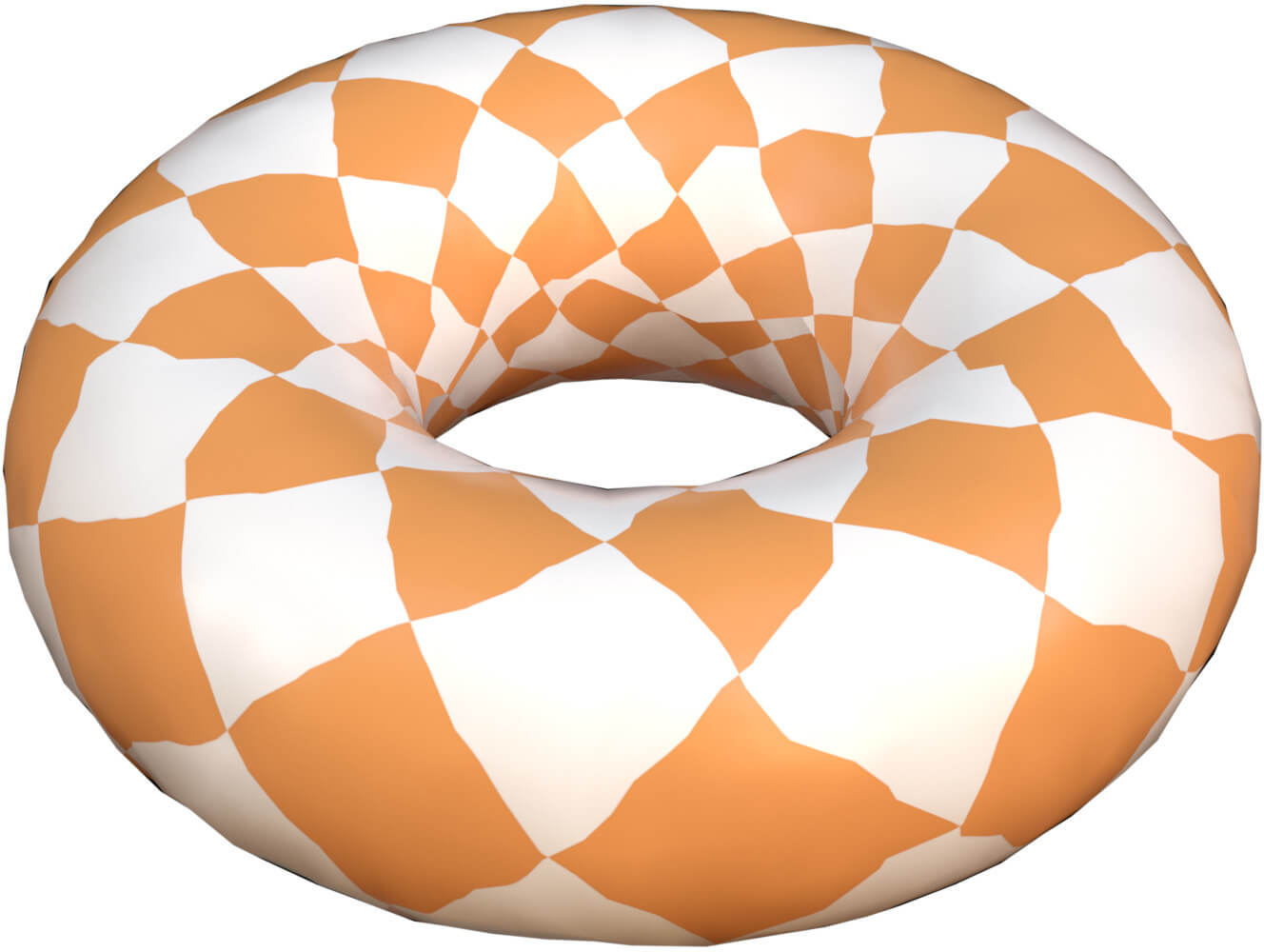}
		\caption{ }     
	\end{subfigure}     
	\caption{(a), (d) the original meshes; (b), (e) with inverse distance circle packing metric; (c), (f) with Thurston's circle packing metric.}\label{fig:initialMetric}
\end{figure} 

\begin{algorithm}
	\caption{Compute Initial Inversive Distance Circle Packing Metric}
	\label{init_metric}
	\begin{algorithmic}[1]
		\For{$e_{ij}\in E$}
		\State $d_{ij} \leftarrow$ the original edge length of $e_{ij}$ in $\mathbb{R}^3$
		\EndFor
		
		\For{$f_{ijk}\in F$}
		\State $r_i^{jk} \leftarrow \frac{d_{ki} + d_{ij} - d_{jk}}{2}$
		\EndFor
		
		\For{$v_i\in V$}
		\State Dertermine vertex radius by: $r_i = \min_{f_{ijk}} r_i^{jk}$
		\EndFor
		
		\For{$e_{ij}\in E}$
		\State compute edge metric weight by $I_{ij} \leftarrow \frac{d_{ij}^2 - r_i^2 - r_j^2}{2r_ir_j}$
		\EndFor
	\end{algorithmic}
\end{algorithm}

 As same with discrete Ricci flow, 
the solution exists and converges if and only if it satisfies Thurston's circle packing condition which is explained in detail in \cite{ge2012combinatorial}.
The convergent analysis of discrete Calabi flow for circle packing metric and inverse distance metric  are discussed    in \cite{ge2012combinatorial} and  \cite{ge2017deformation2,ge2017deformation3}    respectively.

\textbf{Initial metric.}
The Calabi flow starts with an initial metric, there are some choices, such as Thurston's circle packing, inverse distance circle packing, and so on \cite{springborn2008conformal,Zhang2014TheUD}. The desired metric should  approximate original edge lengths in $\mathbb{R}^3$ as much as possible. Inverse distance circle packing metric is equal to the original edge length, and the Thurston's one can only approximate them. In our experiments,  inversive distance circle packing metric do well in terms of this aspect. Therefore we use the inversive distance circle packing metric in our experiments. We follow the method in \cite{Yang2009GeneralizedDR} to compute  the initiate inverse distance circle packing metric.  The detail of the algorithms is shown in the  Alg. \ref{init_metric}.

In the figure \ref{fig:initialMetric}, we show the comparison of Thurston's circle packing metric and inverse distance circle packing metric, we observe that inverse distance metric has better conformal results when meshes are coarse.

Calabi energy is minimized and we use the gradient decedent algorithm  to obtain the optimal solution.
The whole procedure of the algorithm is shown in Alg. \ref{fix_k}.

\begin{algorithm}
	\caption{Calabi Flow}
	\label{fix_k}
	\begin{algorithmic}[1] 
		\State Compute an initial circle packing metric. 
		\State Set target curvatures of each vertex.
		\While{$\max_i |K_i - \bar{K}_i| < \epsilon$}
		\State Calculate curvatures according current metric.
		\State Calculate dual laplacian $L$.
		\State Compute the updating direction $d\textbf{u}\leftarrow L^T(\bar{\textbf{K}} - \textbf{K})$.
		\State Update conformal factors of each vertex by $\textbf{u} \leftarrow \textbf{u} + \delta d\textbf{u}$.
		\EndWhile \Comment{This procedure is the optimization of Calabi energy.}
		\State Embed the mesh to Euclidean plane.
	\end{algorithmic}
\end{algorithm}

\textbf{Embedding.}
After the Calabi flow converges, we obtain a flat metric.  The 2D vertex positions which are compatible to the metric need to be calculated.
We choose a triangle face as root and embed it onto Euclidean plane, then we use breadth-first method to embed other triangle faces. The details of our algorithm is shown in  Alg. \ref{embed}.

\begin{algorithm}
	\caption{Embed the Mesh with Flat Metric to Euclidean Plane.}
	\label{embed}
	\begin{algorithmic}[1]
		\State Choose a root face and then embed it.
		\For{$f_{ijk}$ in the sequence of breadth-first search of all faces}
		\If{all vertices of $f_{ijk}$ have been embedded}
		\State \Return
		\Else
		\State Compute intersections of circles $C(p_i, l_{ik})$ and $C(p_j, l_{jk})$.
		\State Choose the intersection which preserve the orientation of $f_{ijk}$ as the embedding of $v_k$. \Comment{We assume $v_i$ and $v_j$ have been embeded.}
		\EndIf
		\EndFor
	\end{algorithmic}
\end{algorithm}

\section{Experiments}

\begin{figure} [th!] 
	\centering   
	\begin{subfigure}[b]{0.14\textwidth}
		\includegraphics[width=\textwidth]{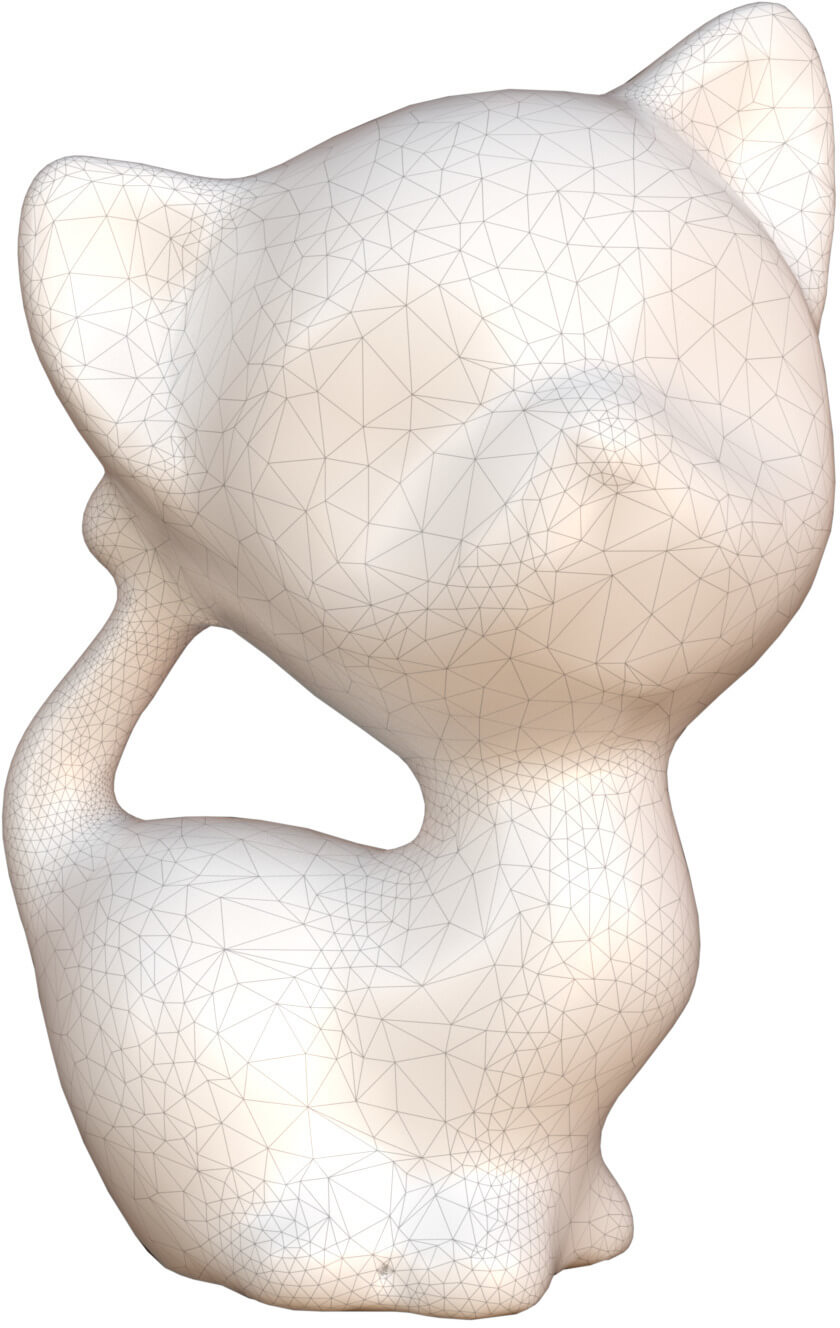}
		\caption{ }   \label{fig:bunny0}
	\end{subfigure}
	\begin{subfigure}[b]{0.17\textwidth}
		\includegraphics[width=\textwidth]{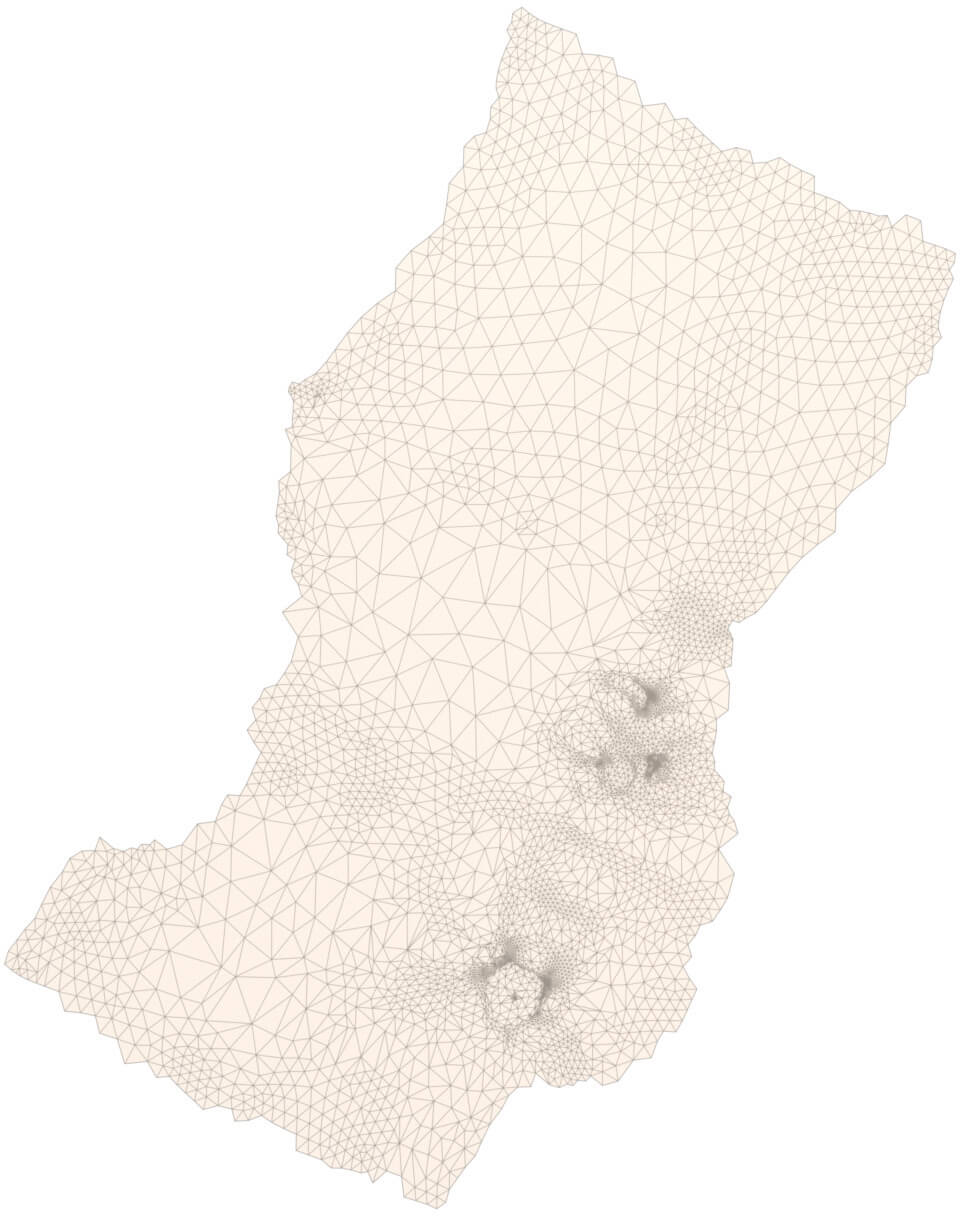}
		\caption{ }   
	\end{subfigure}   
	\begin{subfigure}[b]{0.14\textwidth}
		\includegraphics[width=\textwidth]{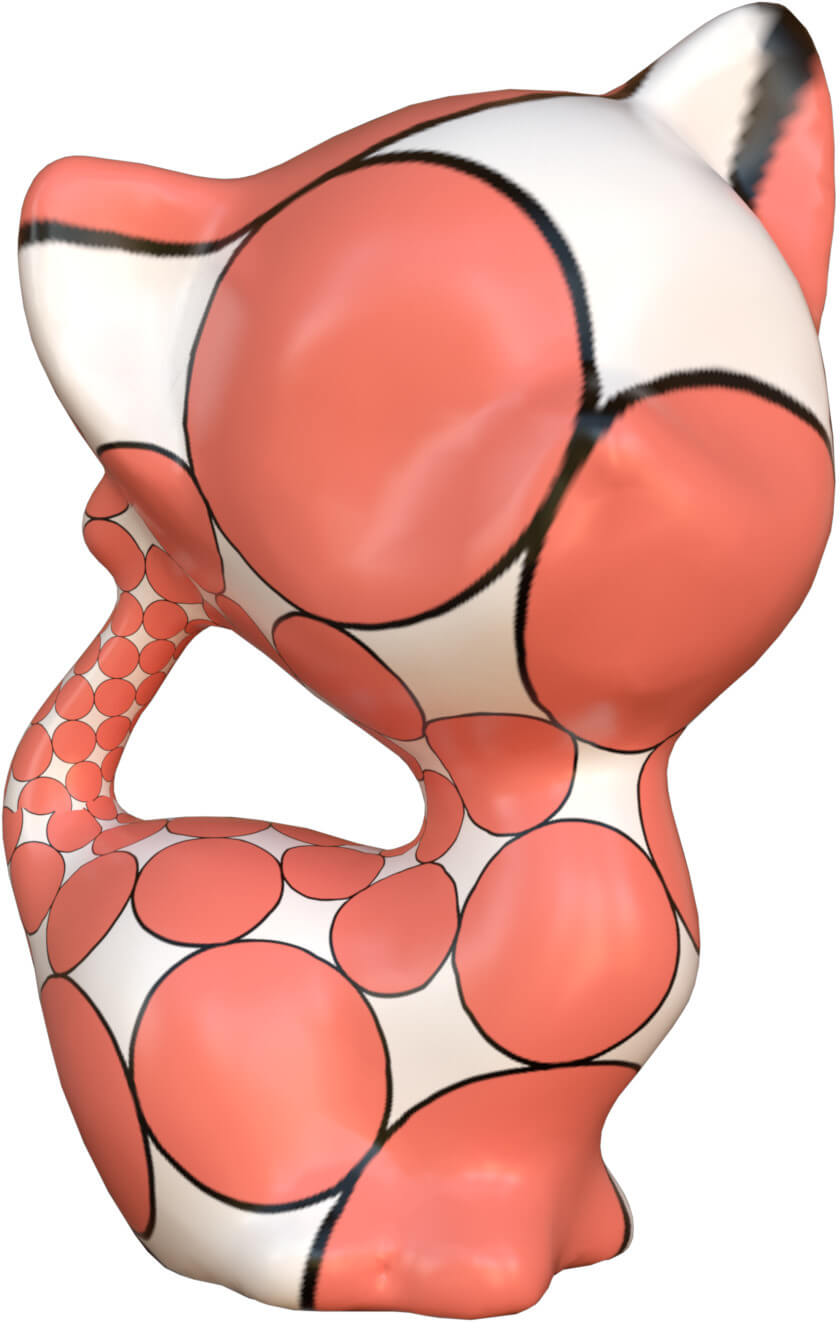}
		\caption{ }     
	\end{subfigure}  
	
	\begin{subfigure}[b]{0.17\textwidth}
		\includegraphics[width=\textwidth]{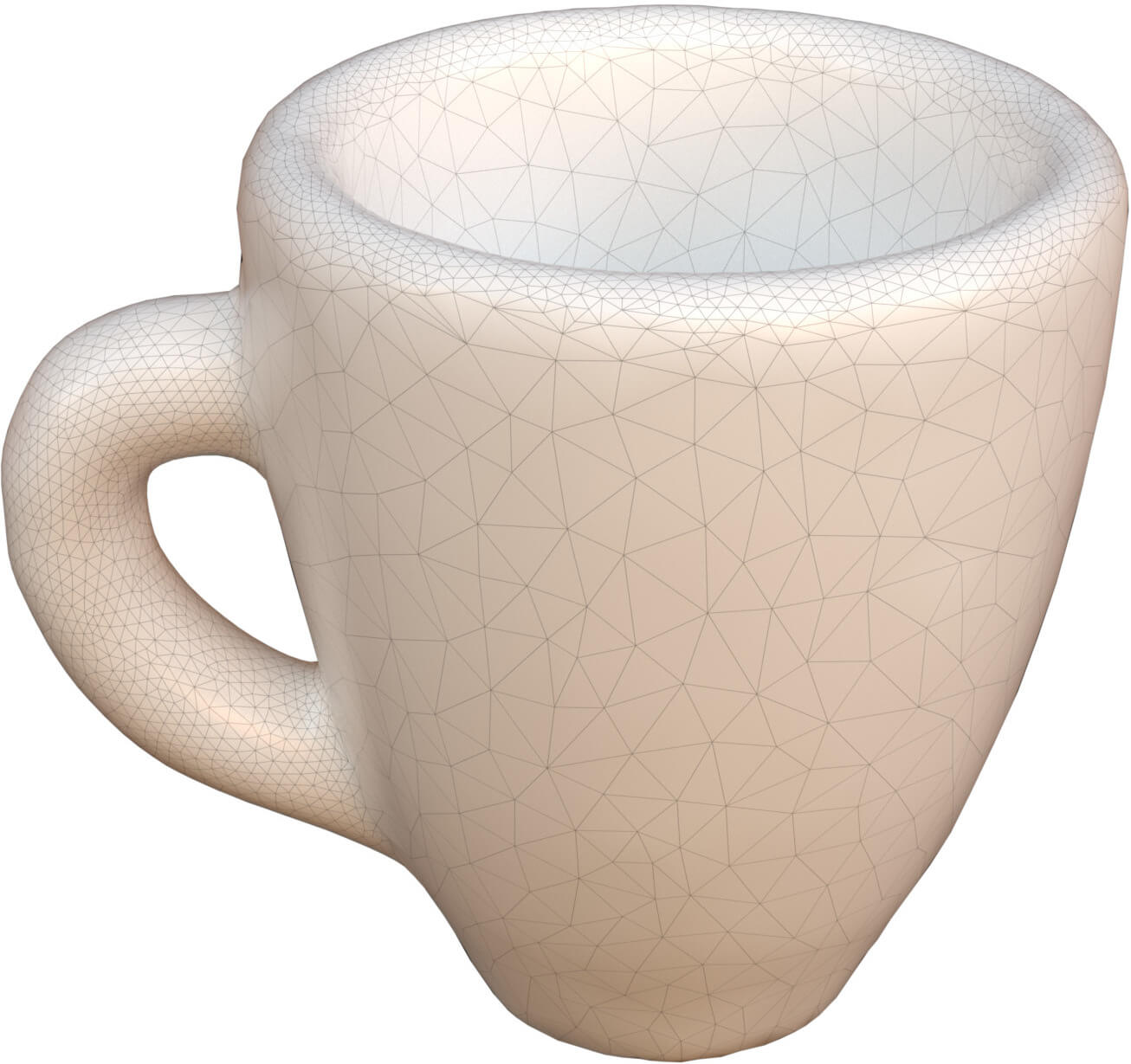}
		\caption{ }   
	\end{subfigure}
	\begin{subfigure}[b]{0.11\textwidth}
		\includegraphics[width=\textwidth]{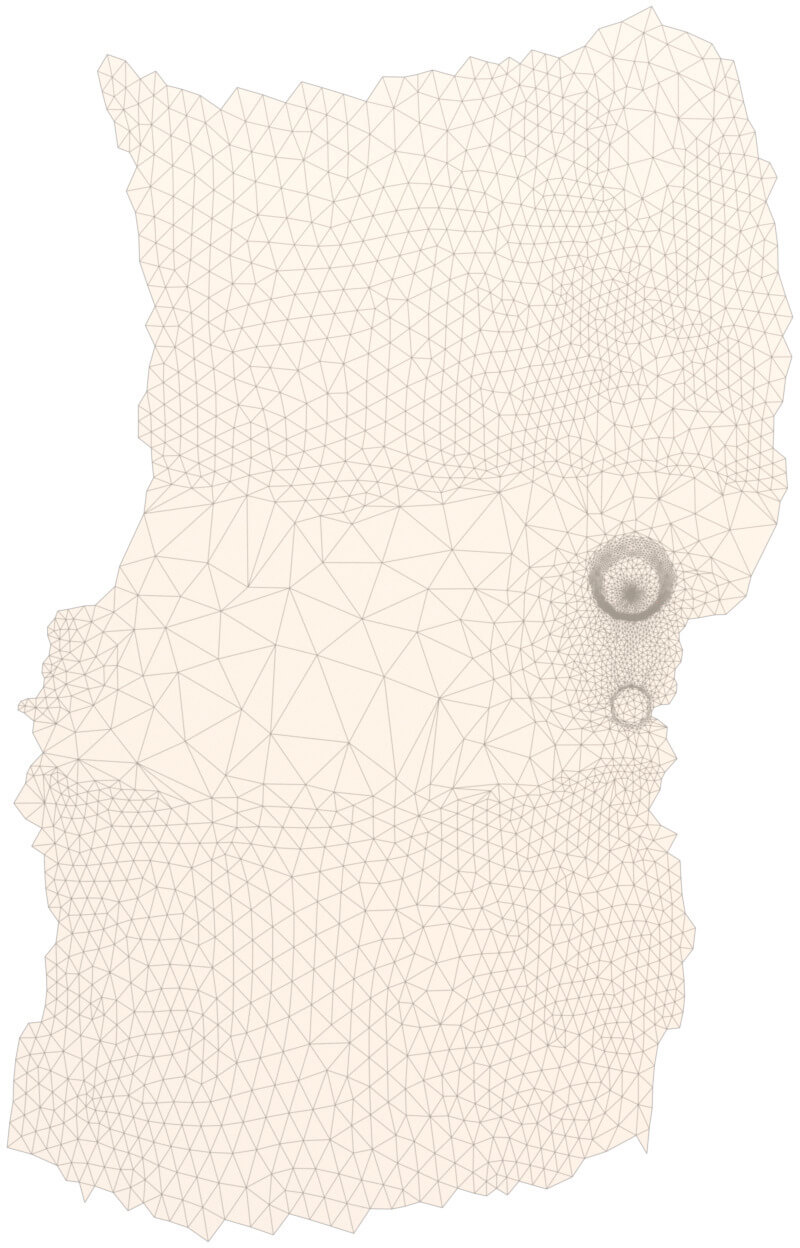}
		\caption{ }   
	\end{subfigure}   
	\begin{subfigure}[b]{0.17\textwidth}
		\includegraphics[width=\textwidth]{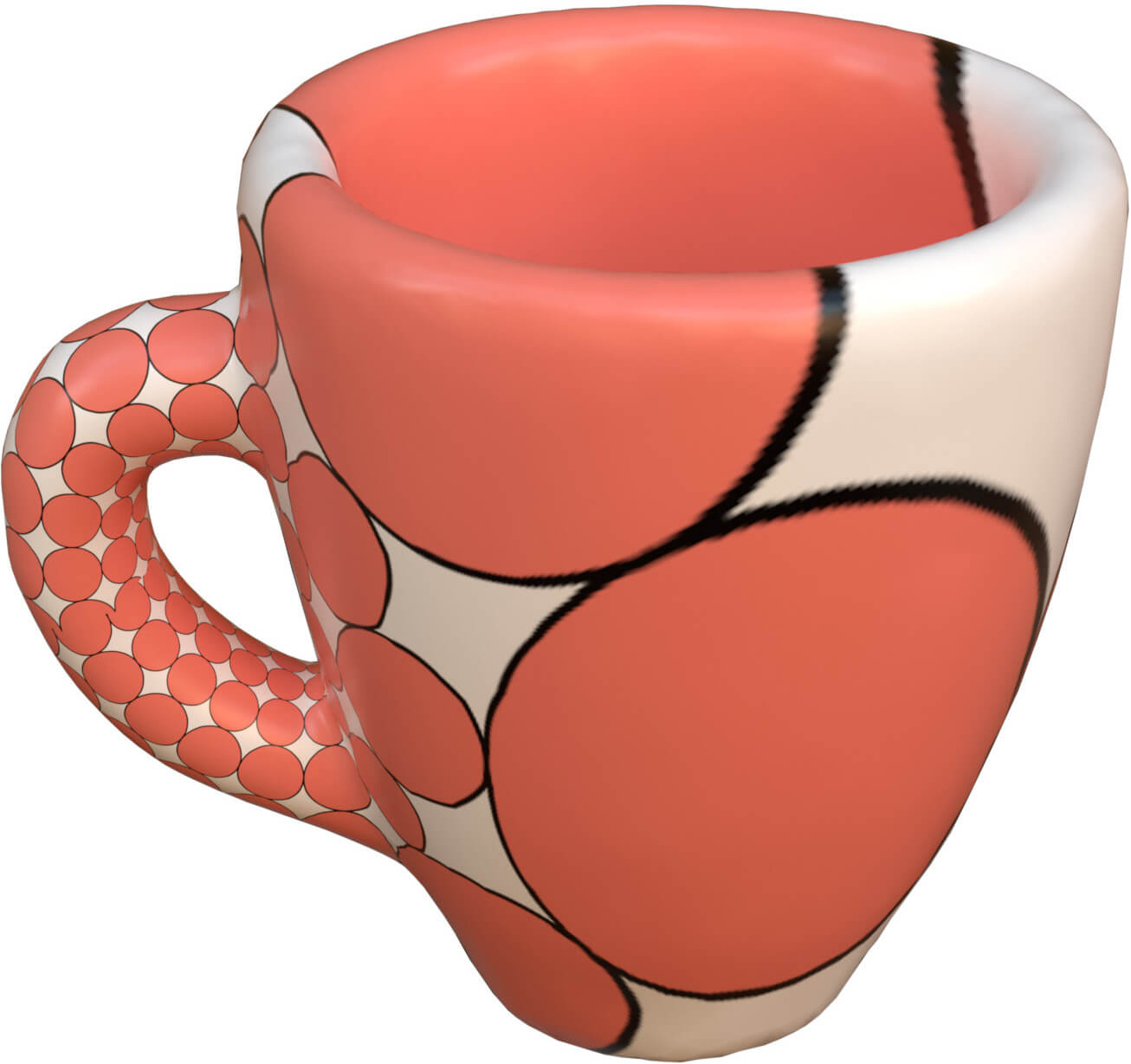}
		\caption{ }     
	\end{subfigure}     
	\caption{The meshes of genus one and their Calabi flow based parameterizations.}\label{fig:calabiTorus}
\end{figure}

According to Gauss-Bonnet theorem, only genus one closed surfaces admit euclidean conformal structure, which means a flat metric without any singularity. In this setting, we set all target curvatures be zero, and run flow on it directly. When the flow converges, we slice the mesh into a disk and then the mesh can be embedded into Euclidean plane. In figure \ref{fig:calabiTorus}, we show the parameterizations of the genus one meshes.

For the closed meshes with genus but one, we need cut them into disk-type topology. 
We  design experiments with three different kinds of  boundary settings: fixing target boundary curvatures, mapping boundary to circle, and free boundary.

 \begin{figure} [h!] 
 	\centering   
 	\begin{subfigure}[b]{0.13\textwidth}
 		\includegraphics[width=\textwidth]{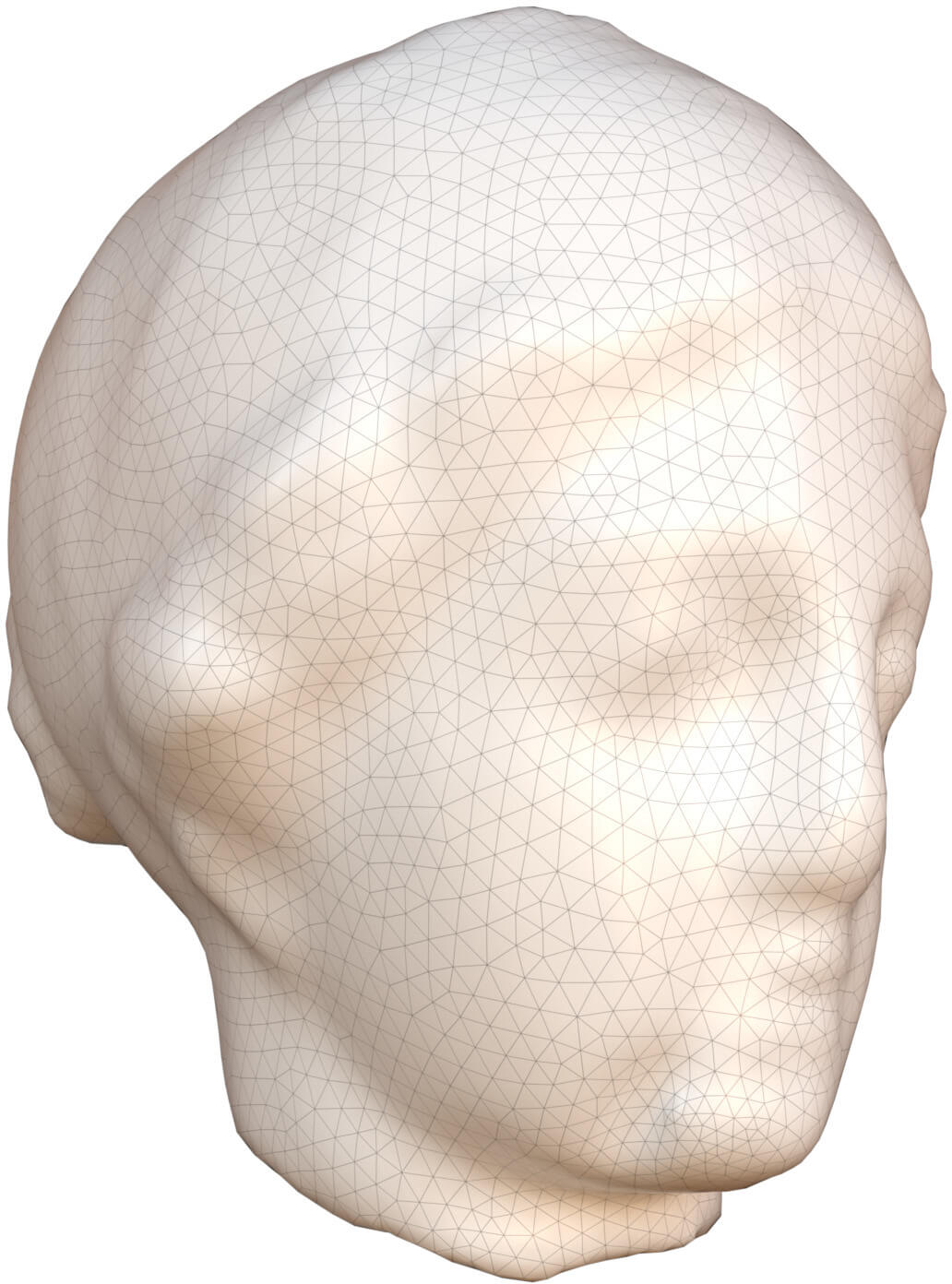}
 		\caption{ }   \label{fig:bunny0}
 	\end{subfigure}
 	\begin{subfigure}[b]{0.19\textwidth}
 		\includegraphics[width=\textwidth]{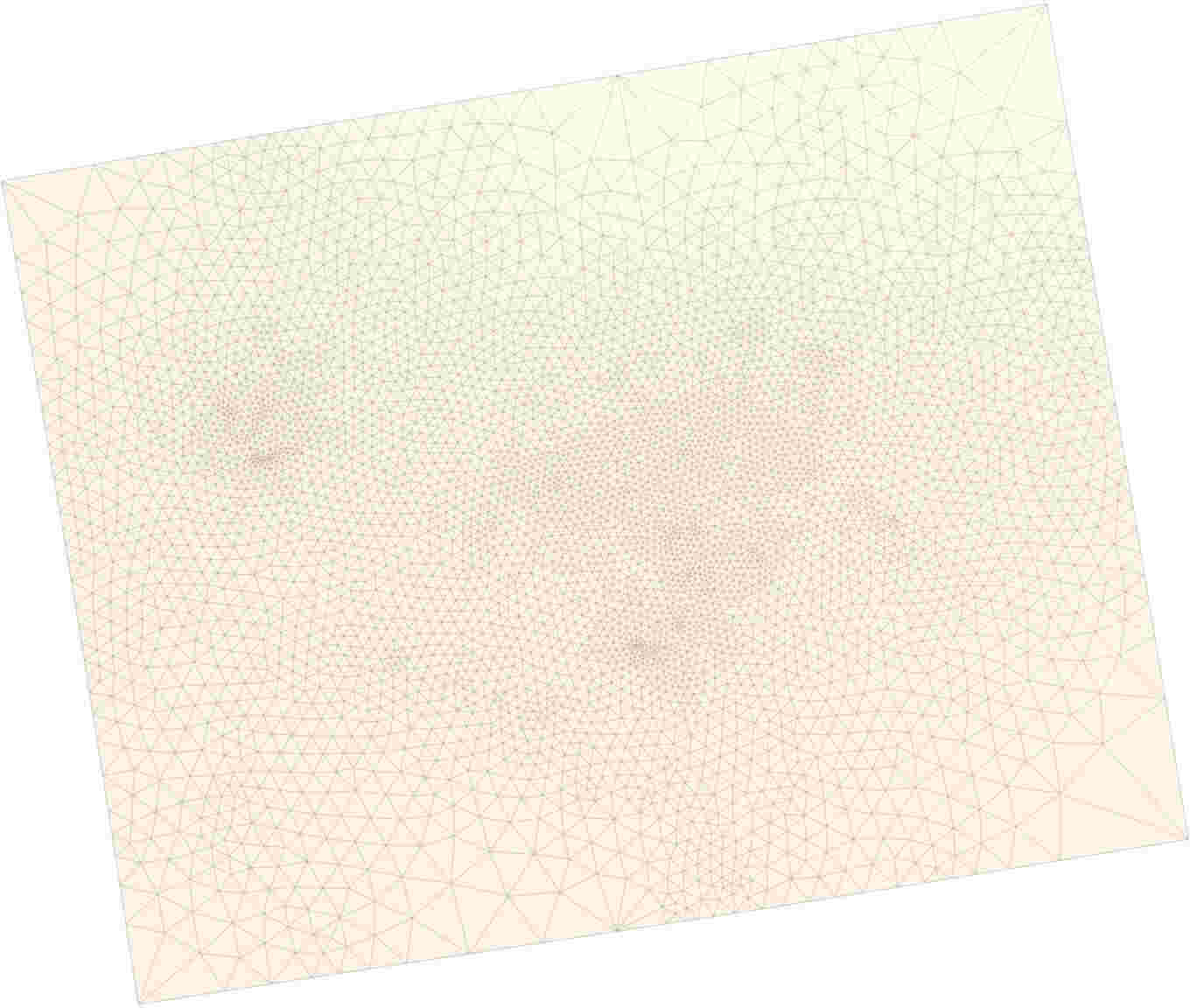}
 		\caption{ }   
 	\end{subfigure}   
 	\begin{subfigure}[b]{0.13\textwidth}
 		\includegraphics[width=\textwidth]{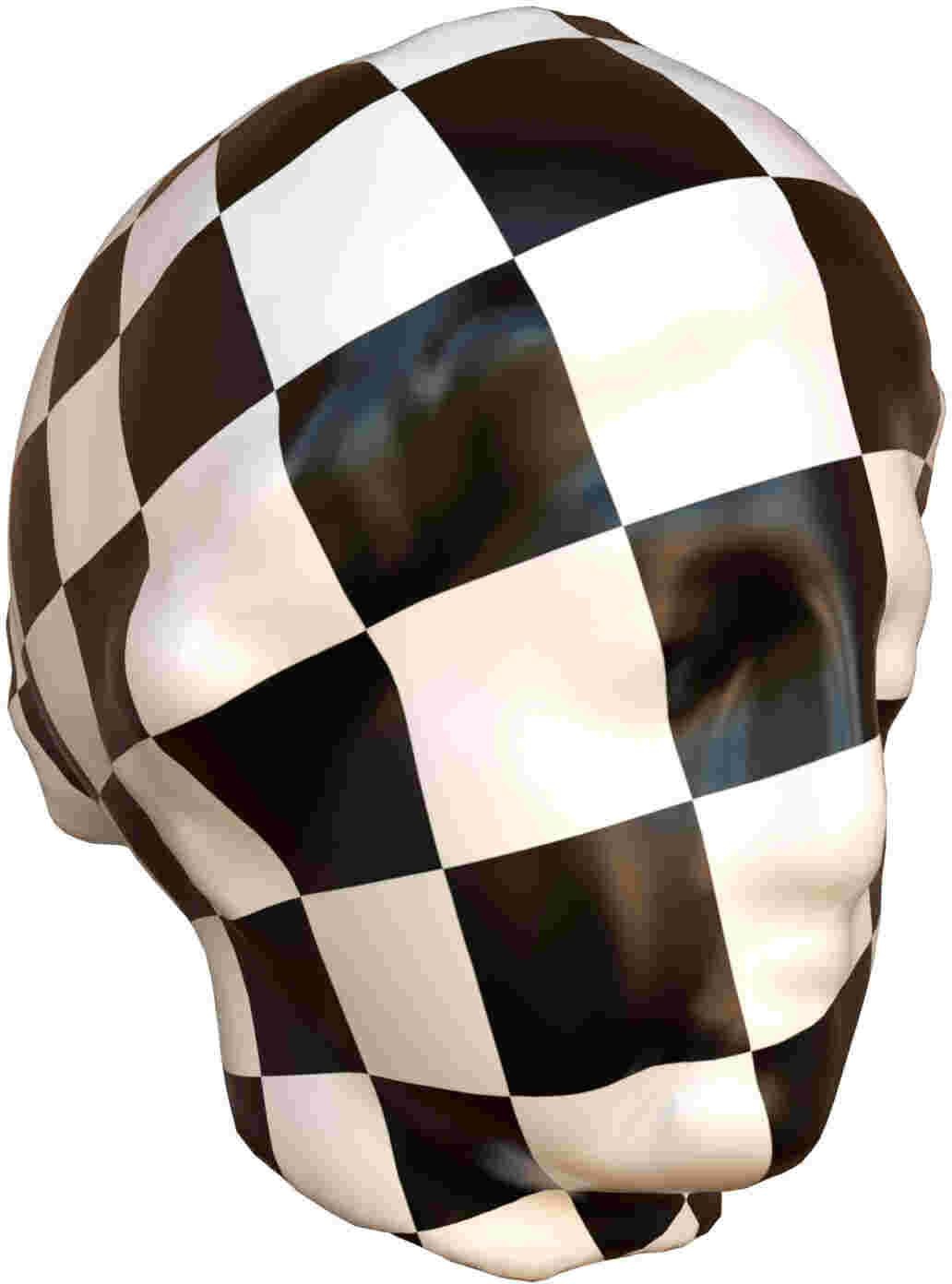}
 		\caption{ }     
 	\end{subfigure}  
 	
 	\begin{subfigure}[b]{0.11\textwidth}
 		\includegraphics[width=\textwidth]{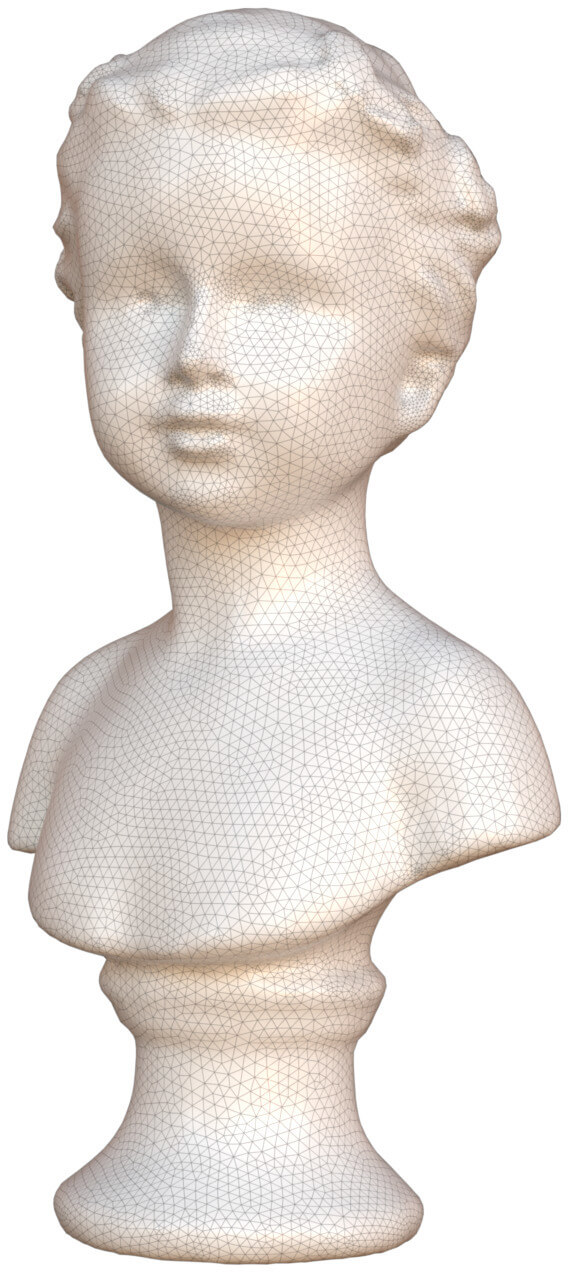}
 		\caption{ }   
 	\end{subfigure}
 	\begin{subfigure}[b]{0.21\textwidth}
 		\includegraphics[width=\textwidth]{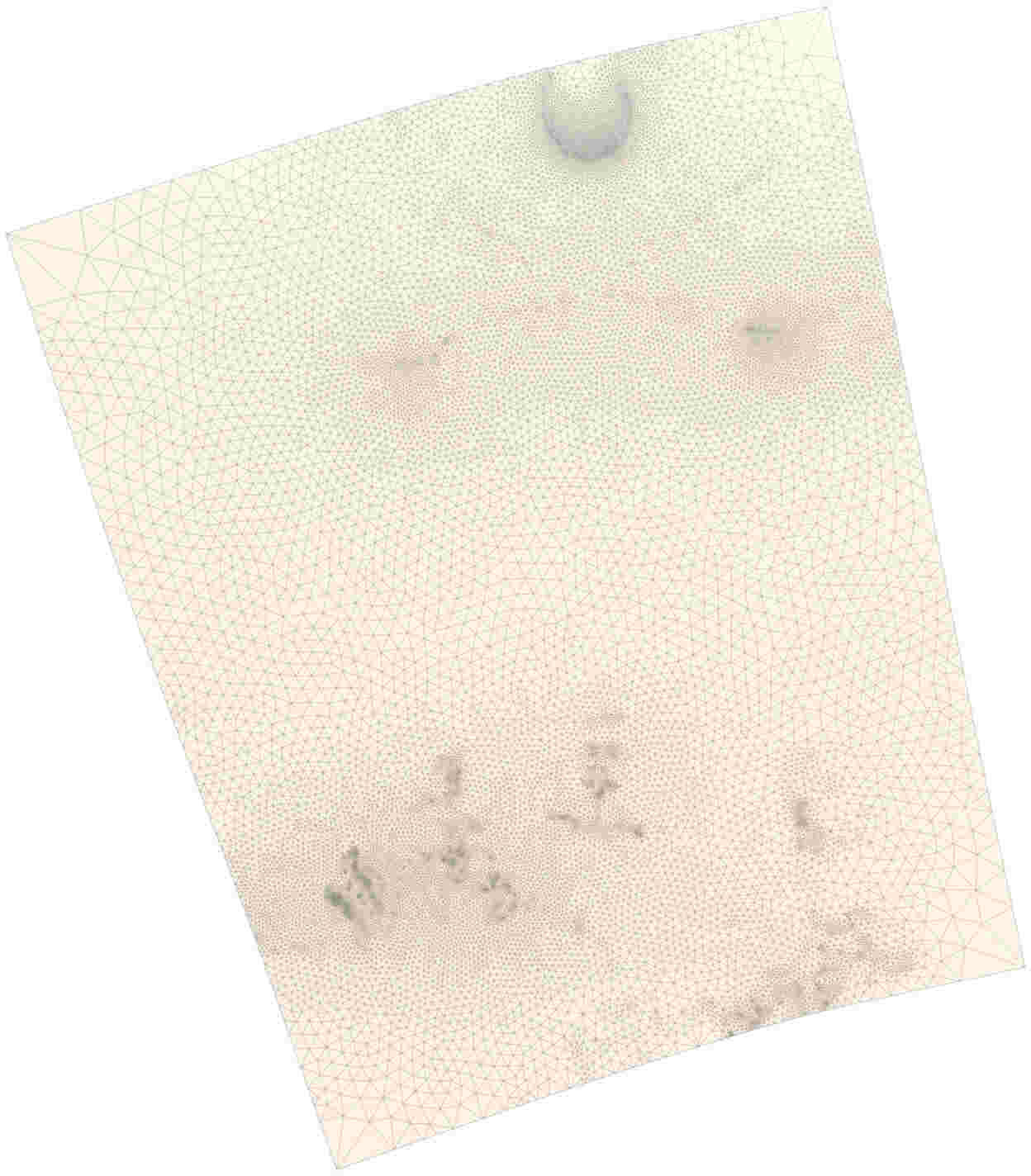}
 		\caption{ }   
 	\end{subfigure}   
 	\begin{subfigure}[b]{0.11\textwidth}
 		\includegraphics[width=\textwidth]{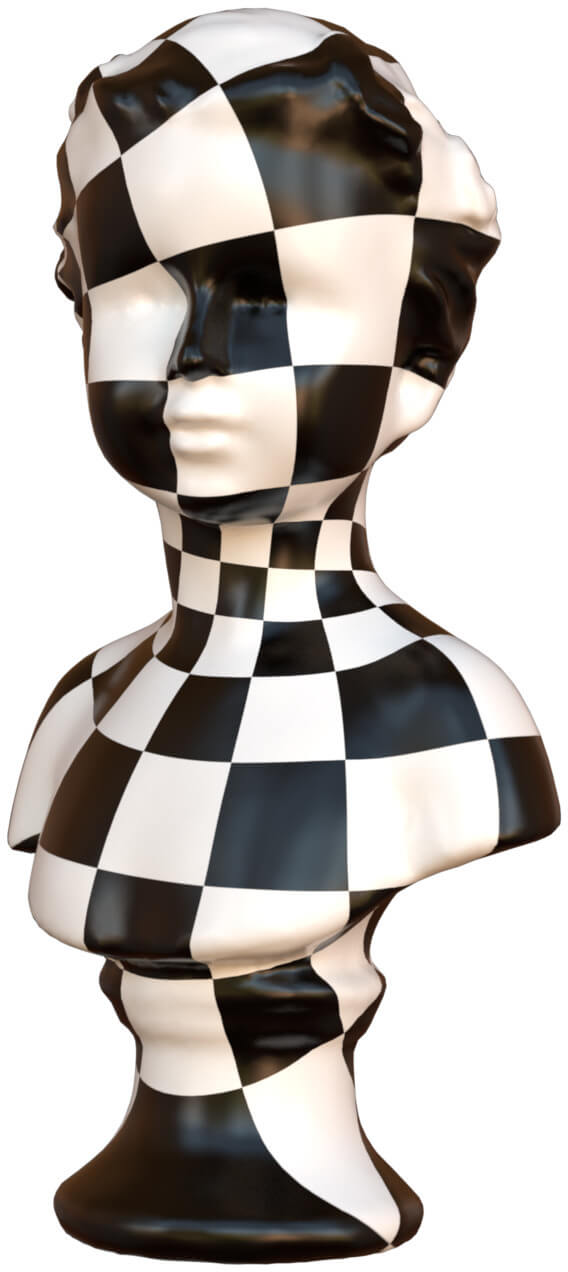}
 		\caption{ }     
 	\end{subfigure}     
 	\caption{The Calabi flow based parametrizations with   fixed rectangle boundary.}\label{fig:calabiRectangle}
 \end{figure} 

\textbf{Fixed Boundary.} The result of fixing boundary is a polygon with corner angles being the $\pi$ minus target curvature $\bar{K}_i$.  
In the step of setting target curvatures, we set all target curvatures of each interior vertices zero, and set all target curvatures of boundary vertices zero except corner vertices. We should guarantee that the the sum of target curvatures of these corner vertices equals to $2\pi$, make it admissible with Gauss-Bonnet Theorem.

In the figure \ref{fig:calabiRectangle}, we show two parametrization results  with   fixed rectangle boundaries.

\textbf{Circular Boundary.} If the target boundary is a circle, we cannot simply set all boundary target curvatures be $\frac{2\pi}{m}$, where $m$ is the number of vertices of the boundary. In fact, if the boundary is on a circle, the curvatures of each boundary vertex should satisfy the following conditions. And we must update these target curvatures in each iteration.

\begin{equation} \label{c1}
\frac{K_i}{l_{i-1,i} + l_{i, i+1}} \equiv c,  \forall v_i \in \partial M
\end{equation}
where $l_{ij}$ is edge length under the target metric. 
And 
\begin{equation}\label{c2}
\sum_{v_i \in \partial M} K_i = 2\pi
\end{equation}

\begin{figure} [h!] 
	\centering   
	\begin{subfigure}[b]{0.13\textwidth}
		\includegraphics[width=\textwidth]{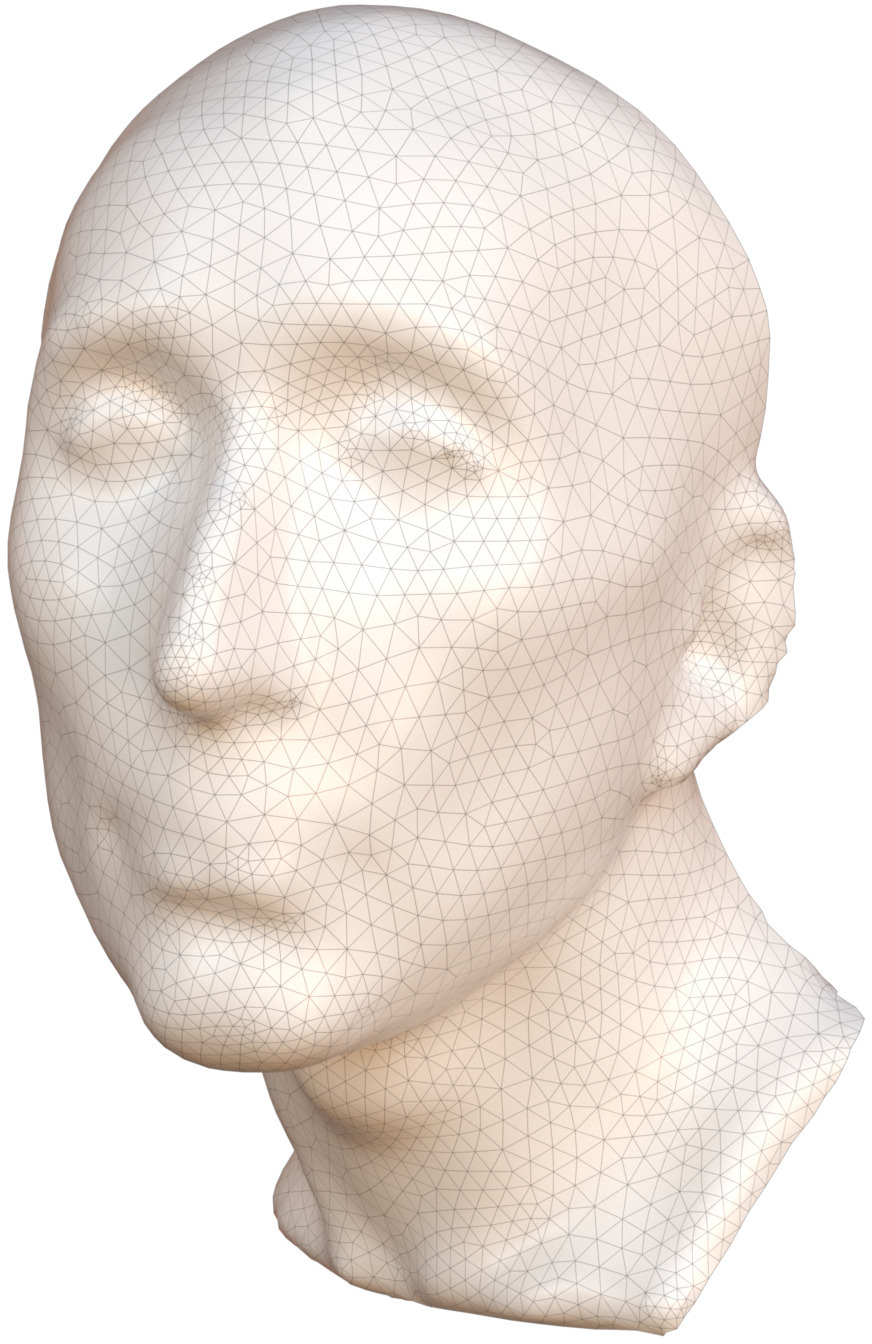}
		\caption{ }   \label{fig:bunny0}
	\end{subfigure}
	\begin{subfigure}[b]{0.19\textwidth}
		\includegraphics[width=\textwidth]{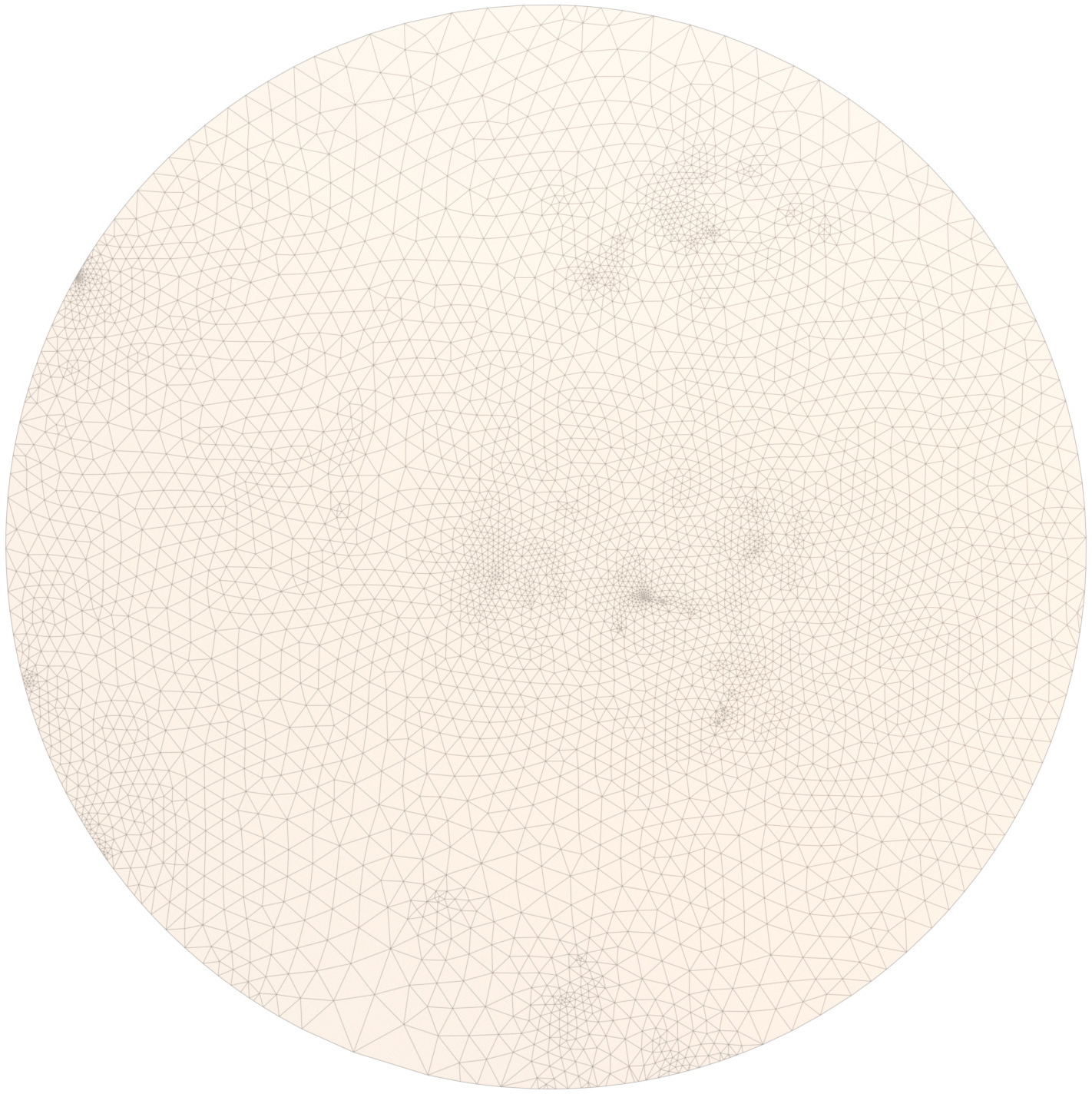}
		\caption{ }   
	\end{subfigure}   
	\begin{subfigure}[b]{0.13\textwidth}
		\includegraphics[width=\textwidth]{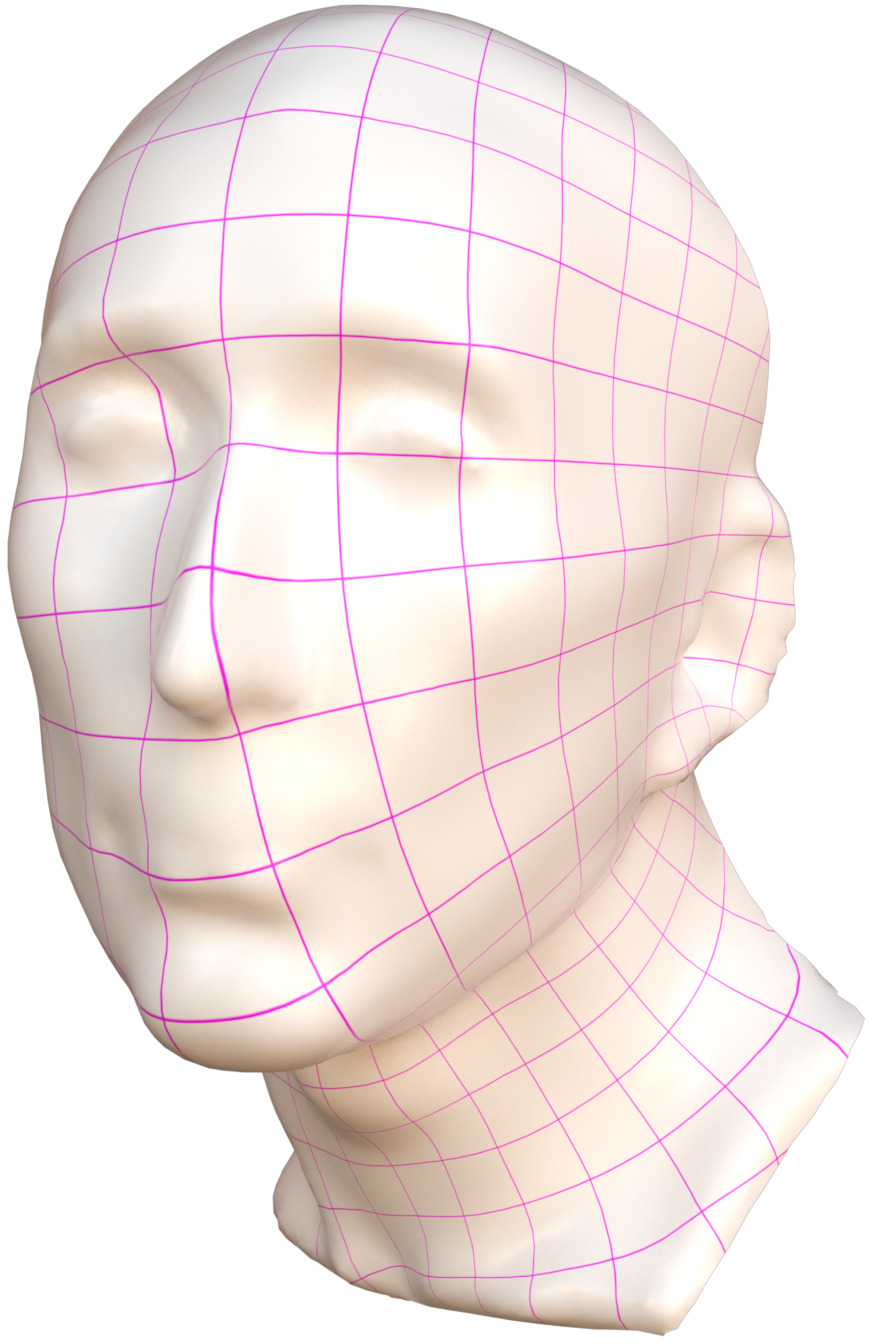}
		\caption{ }     
	\end{subfigure}  
	
	\begin{subfigure}[b]{0.15\textwidth}
		\includegraphics[width=\textwidth]{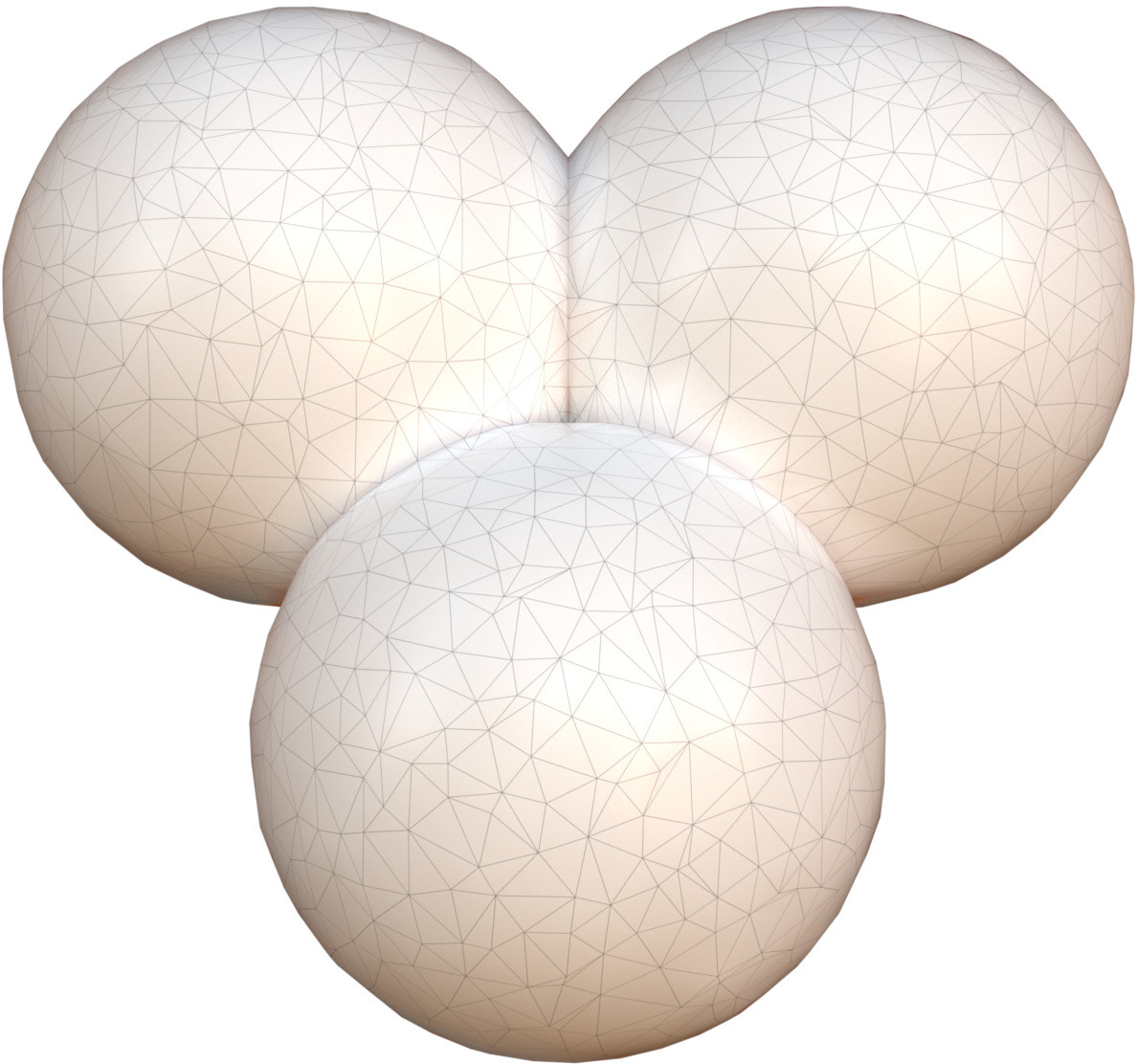}
		\caption{ }   
	\end{subfigure}
	\begin{subfigure}[b]{0.15\textwidth}
		\includegraphics[width=\textwidth]{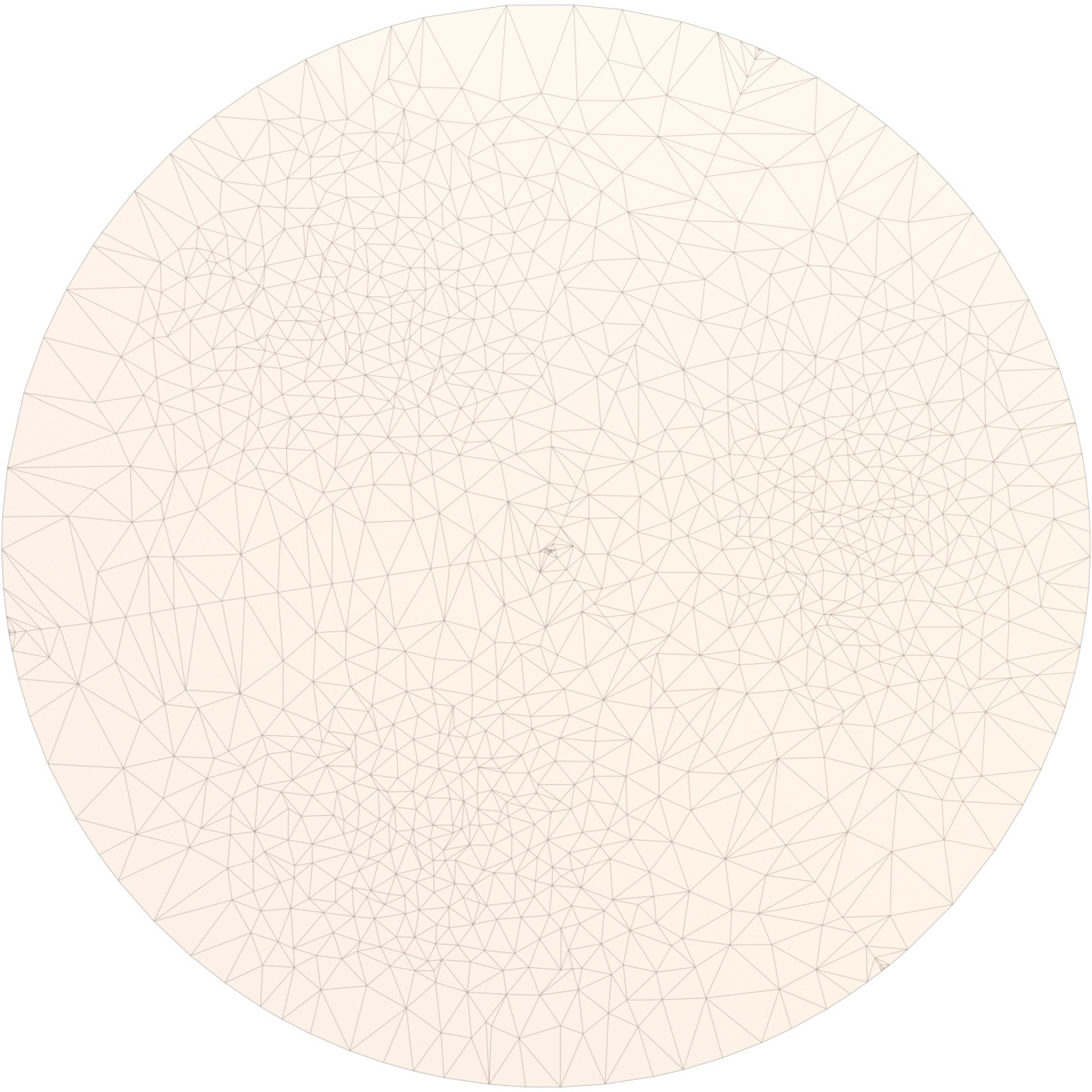}
		\caption{ }   
	\end{subfigure}   
	\begin{subfigure}[b]{0.15\textwidth}
		\includegraphics[width=\textwidth]{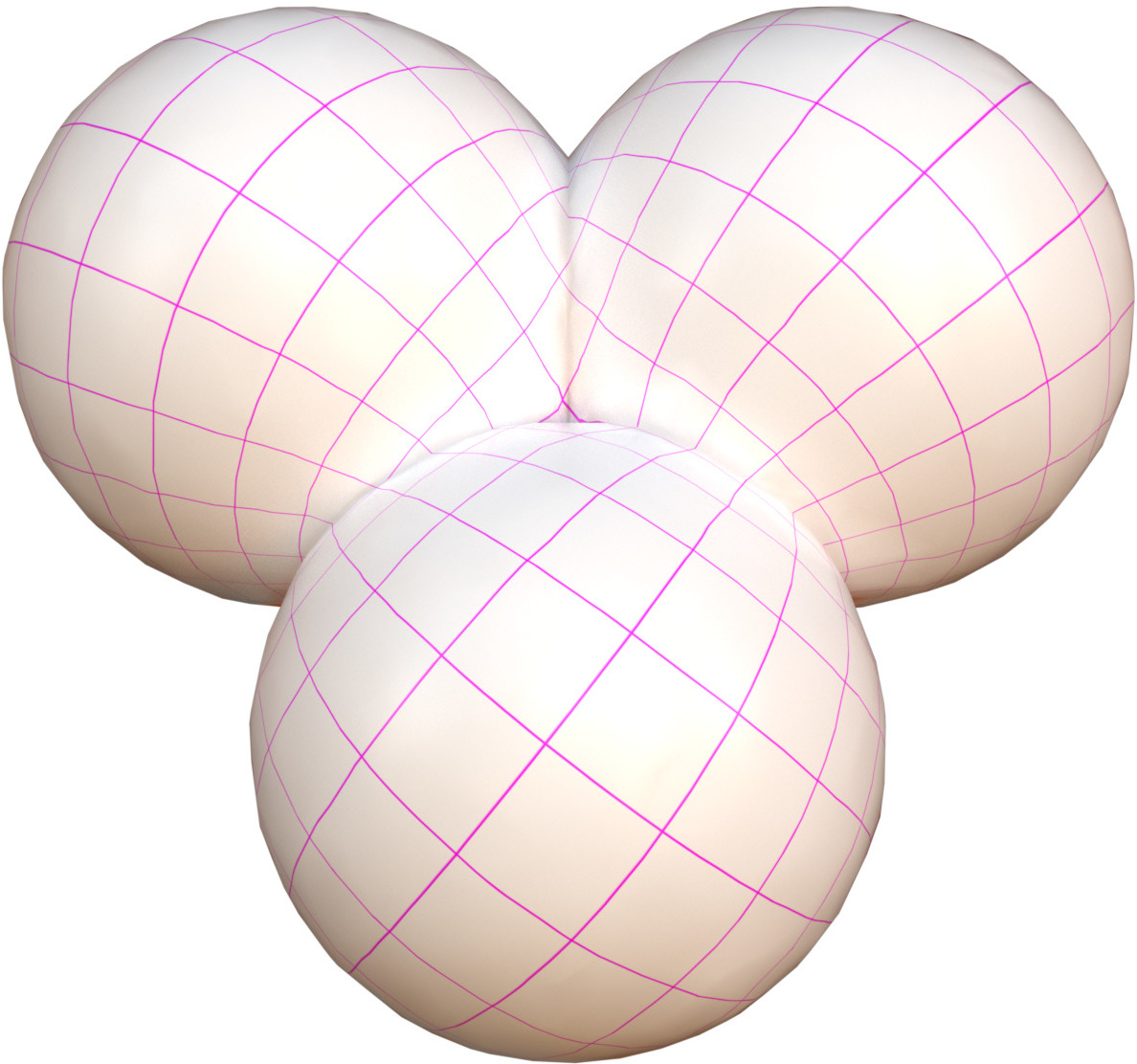}
		\caption{ }     
	\end{subfigure}     
	\caption{Calabi flow based parametrization with the circle boundary.}\label{fig:calabiCricle}
\end{figure} 

In figure \ref{fig:calabiCricle}, two Calabi flow results with circle boundaries are demonstrated.

\begin{figure} [th!] 
	\centering   
	\begin{subfigure}[b]{0.15\textwidth}
		\includegraphics[width=\textwidth]{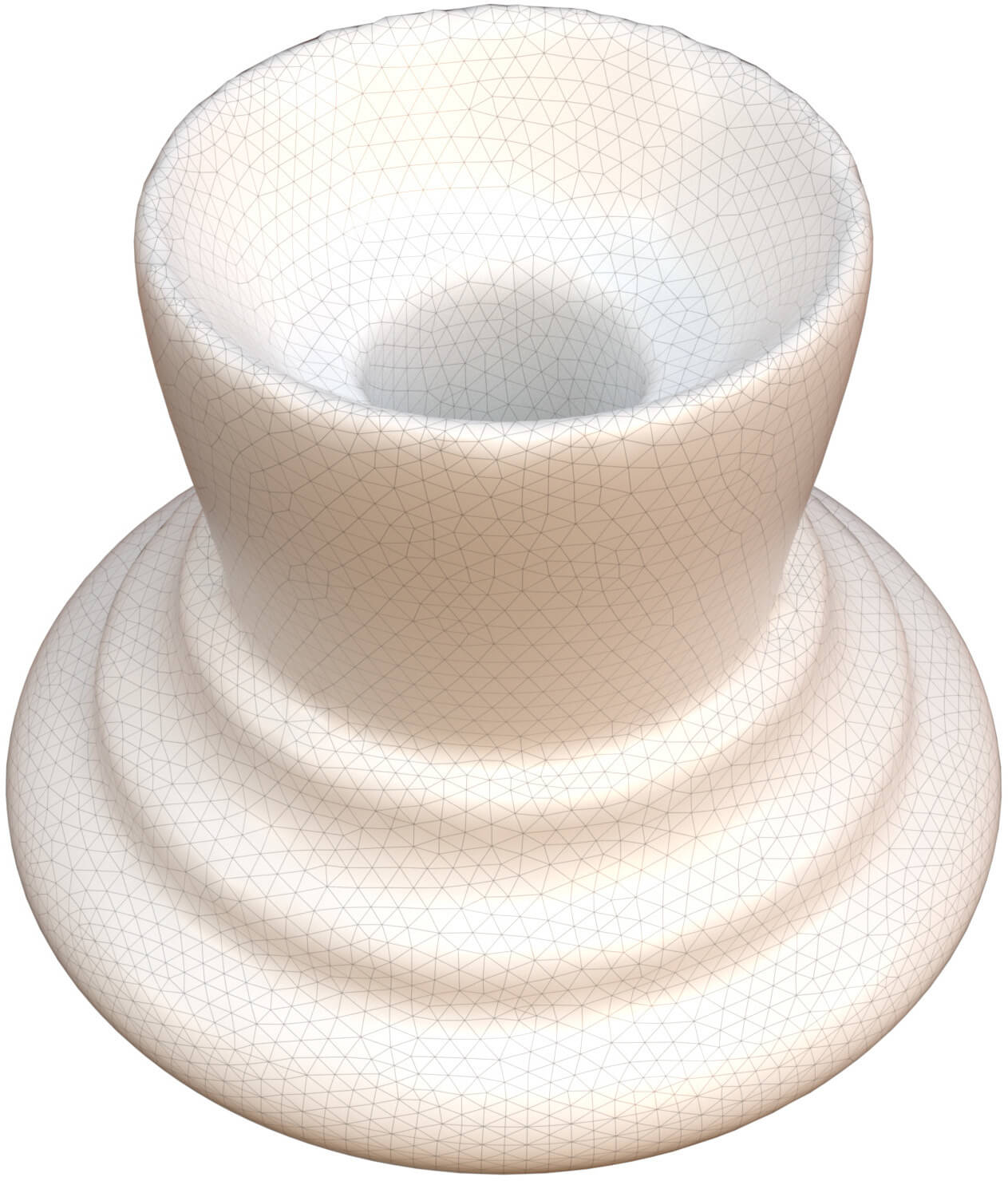}
		\caption{ }   \label{fig:bunny0}
	\end{subfigure}
	\begin{subfigure}[b]{0.15\textwidth}
		\includegraphics[width=\textwidth]{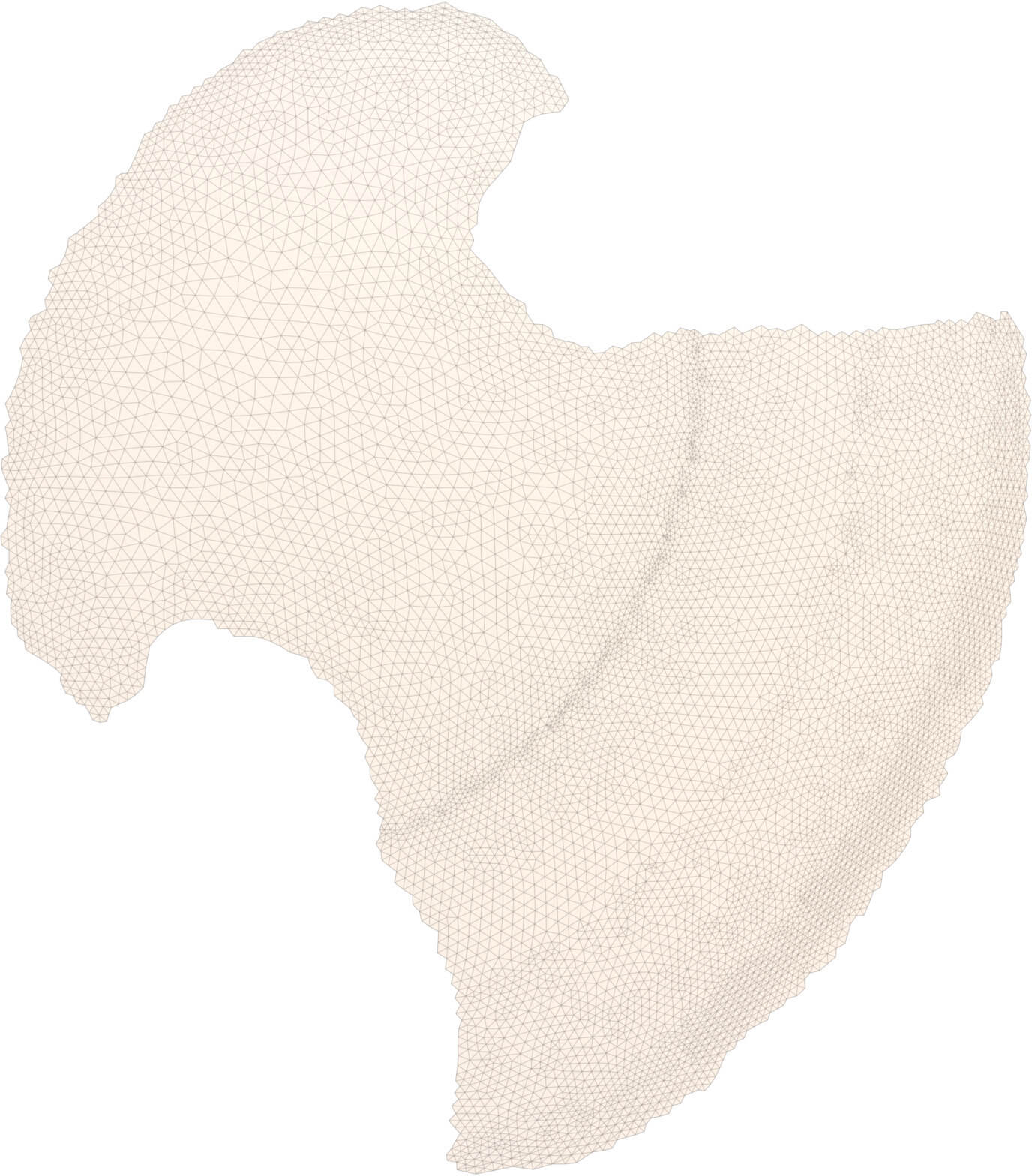}
		\caption{ }   
	\end{subfigure}   
	\begin{subfigure}[b]{0.15\textwidth}
		\includegraphics[width=\textwidth]{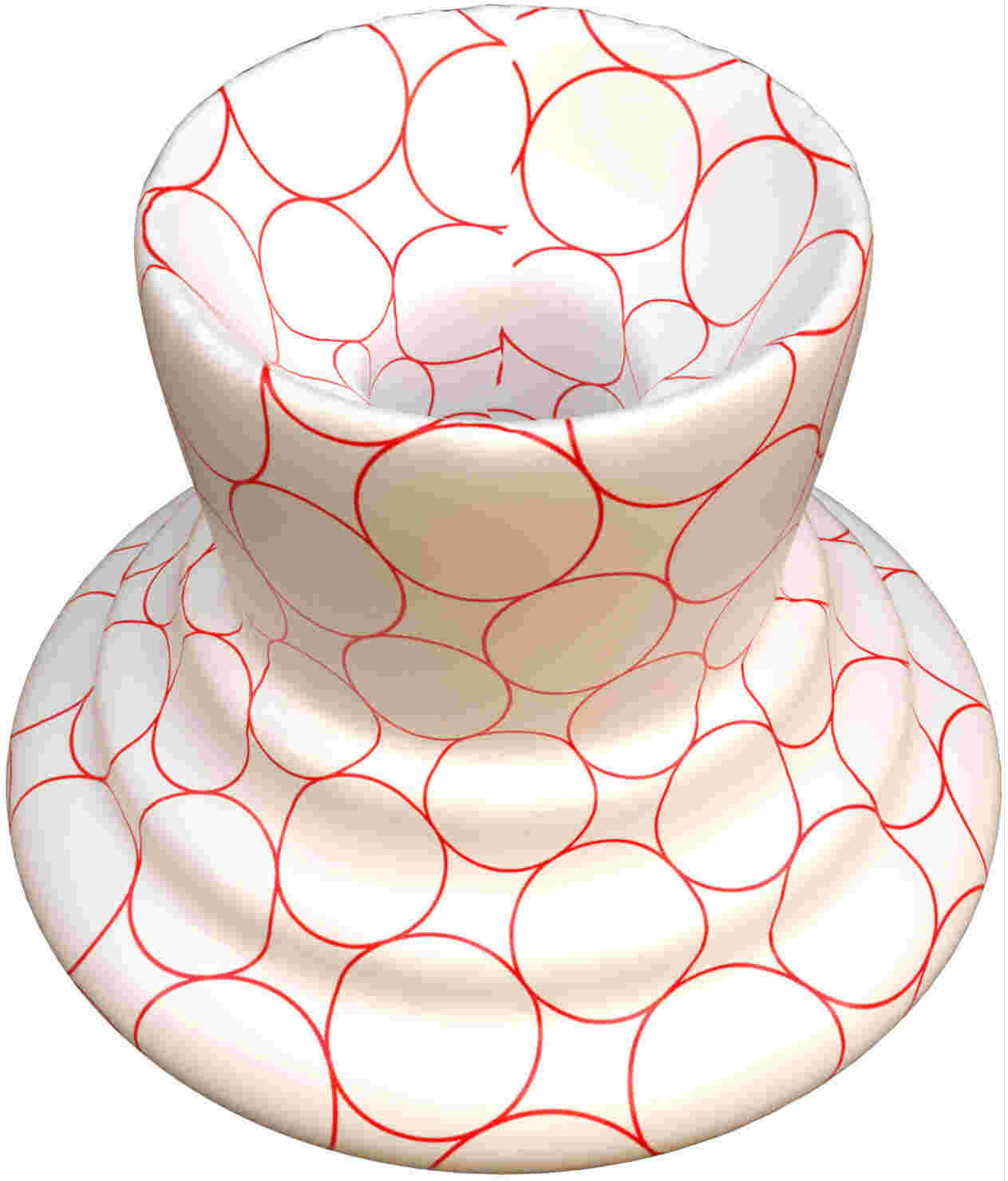}
		\caption{ }     
	\end{subfigure}  
	
	\begin{subfigure}[b]{0.13\textwidth}
		\includegraphics[width=\textwidth]{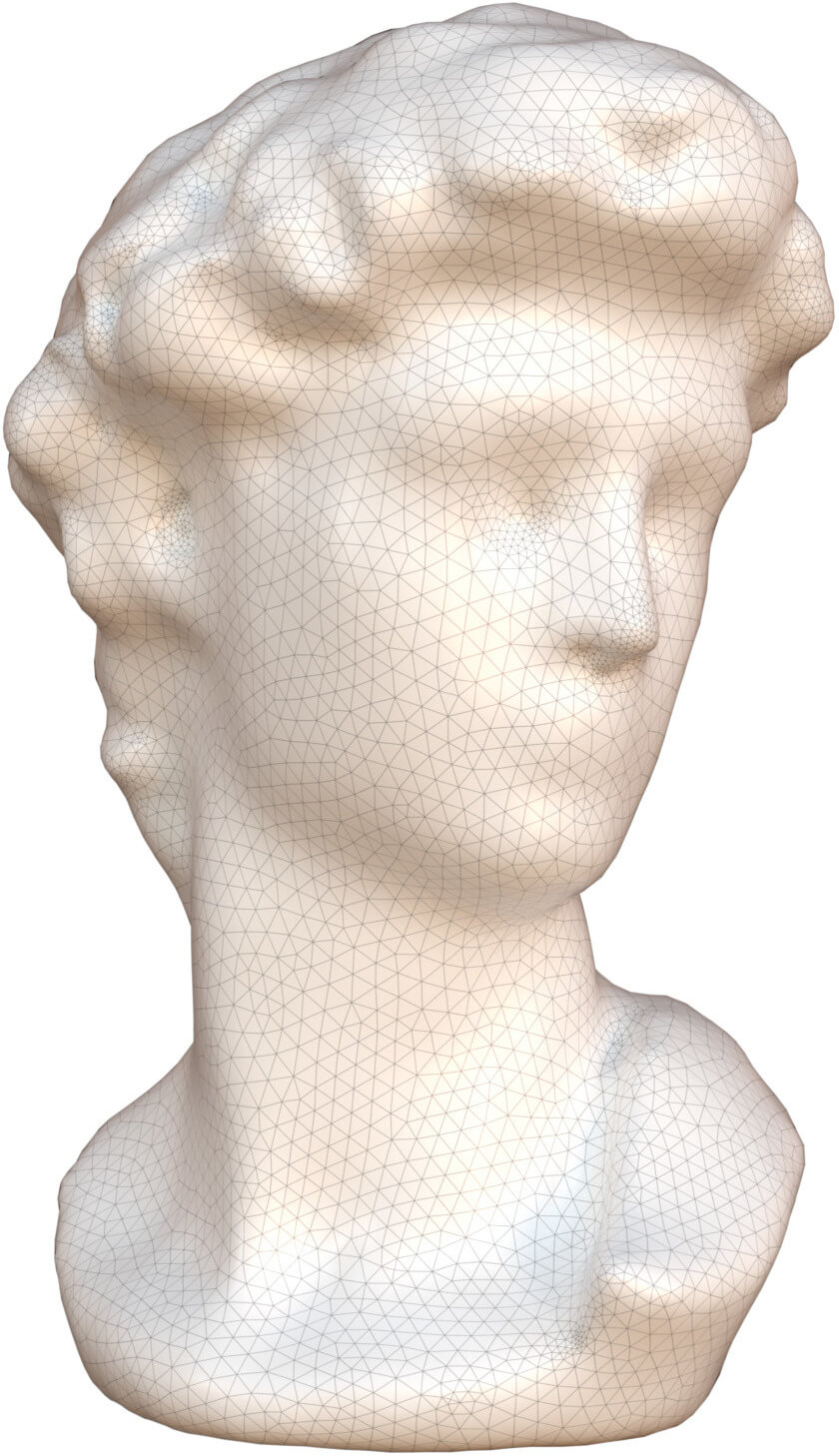}
		\caption{ }   
	\end{subfigure}
	\begin{subfigure}[b]{0.19\textwidth}
		\includegraphics[width=\textwidth]{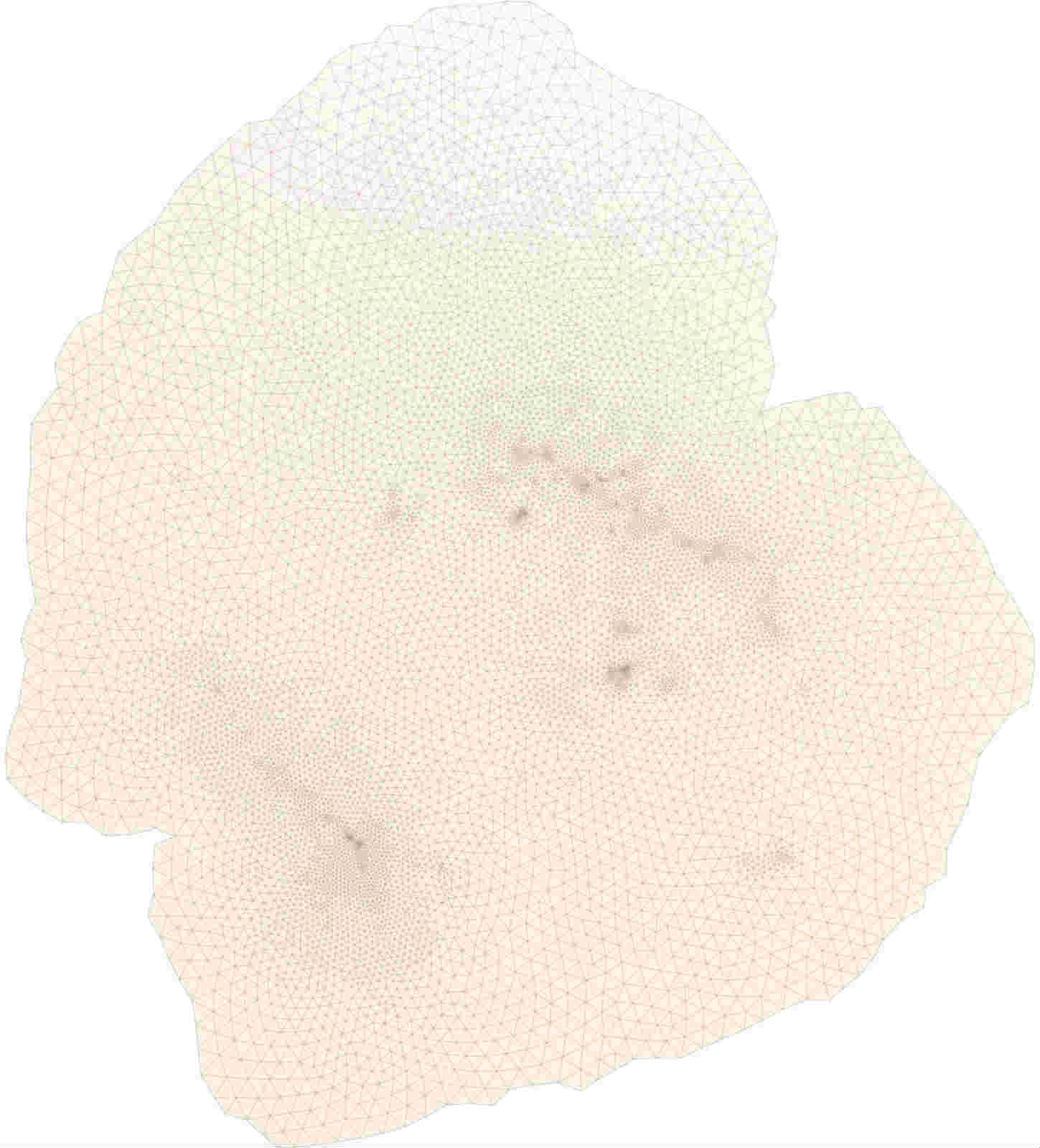}
		\caption{ }   
	\end{subfigure}   
	\begin{subfigure}[b]{0.13\textwidth}
		\includegraphics[width=\textwidth]{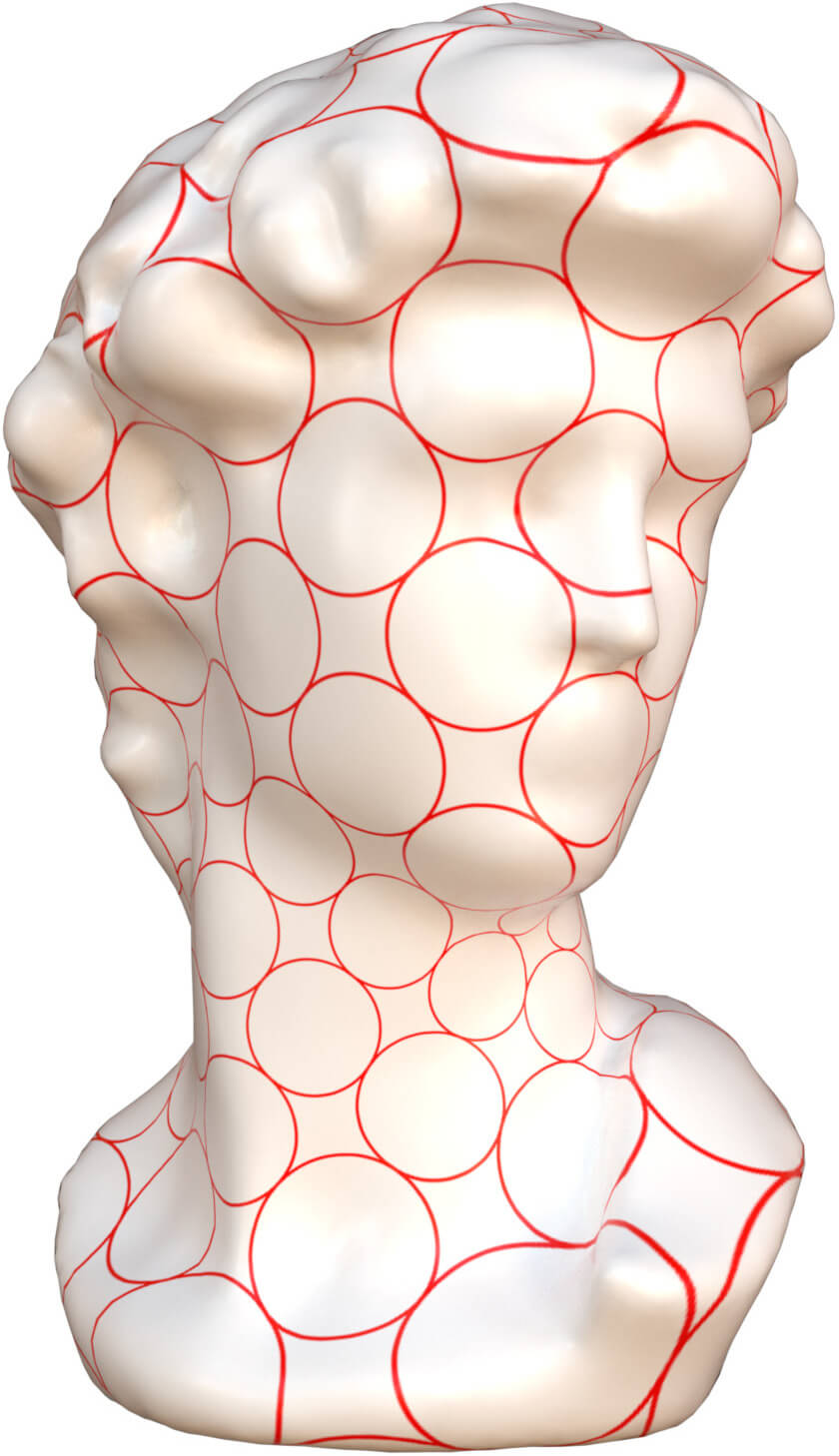}
		\caption{ }     
	\end{subfigure}     
	\caption{ Calabi flow based parameterizations with free boundary.}\label{fig:calabiFree}
\end{figure}

\textbf{Free Boundary.} In this setting, we do not set boundary target curvatures directly, instead we set $d\textbf{u}$ of boundary vertices in Alg. \ref{fix_k} to be zero. In this way the Calabi energy will continue to decrese, and this will lead to smaller area distortion. To be specific, we set all boundary $d\textbf{u}$ be zero, and only update the interior vertices. Some results are shown in the figure \ref{fig:calabiFree}.

\begin{figure*} [th!] 
	\centering   
		\begin{subfigure}[b]{0.075\textwidth}
		\includegraphics[width=\textwidth]{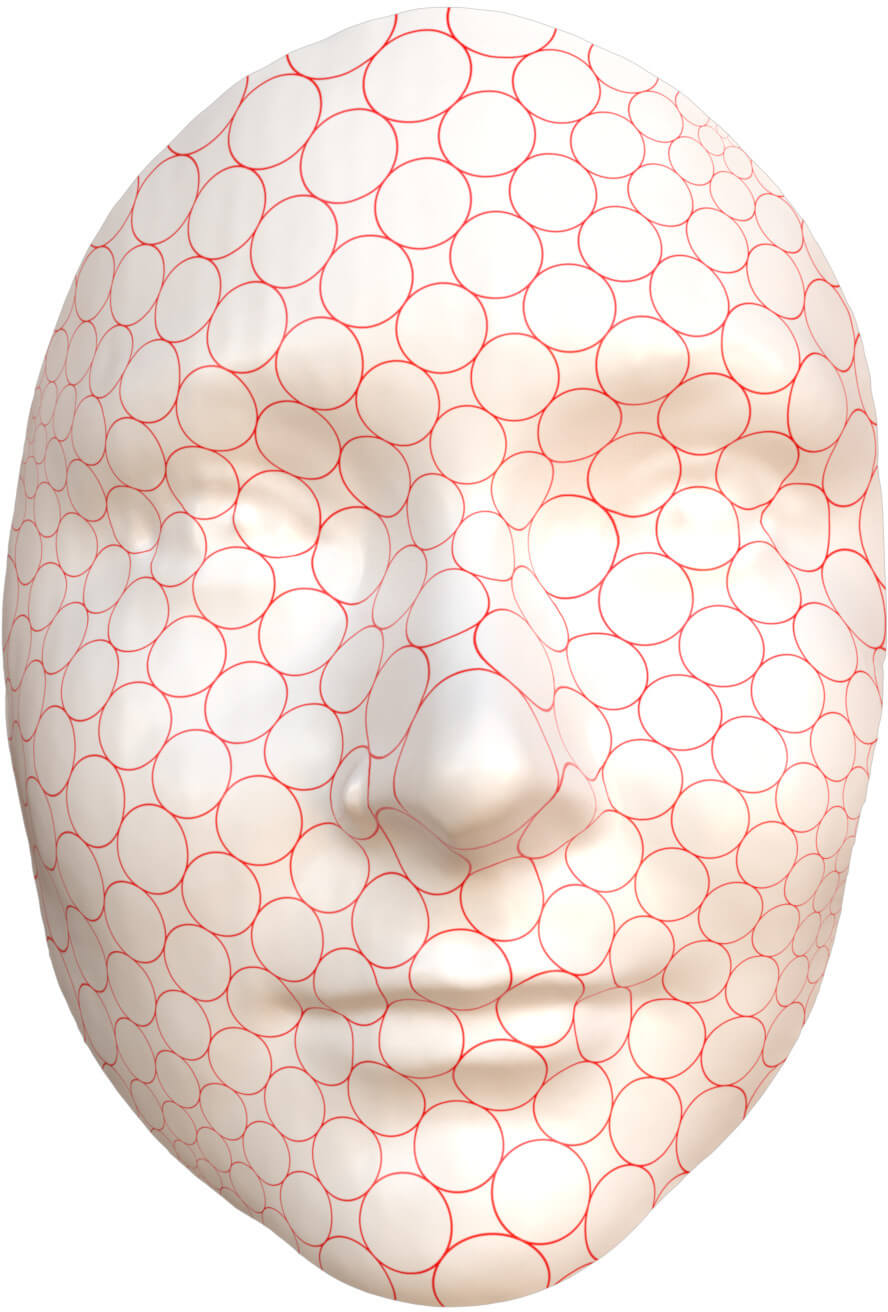}
		\caption*{ }   
	\end{subfigure}
	\begin{subfigure}[b]{0.055\textwidth}
		\includegraphics[width=\textwidth]{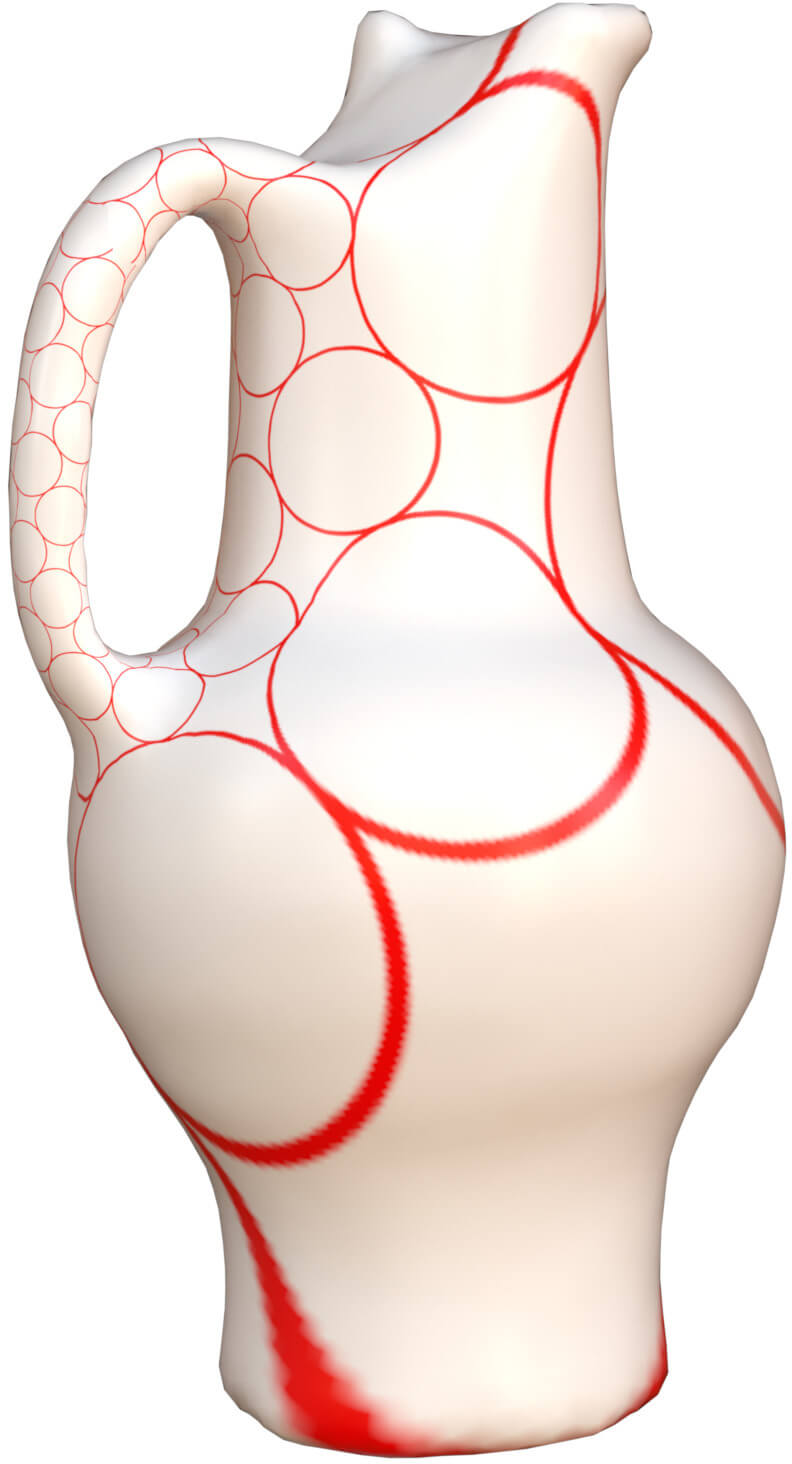}
		\caption*{ }     
	\end{subfigure}
	\begin{subfigure}[b]{0.17\textwidth}
		\includegraphics[width=\textwidth]{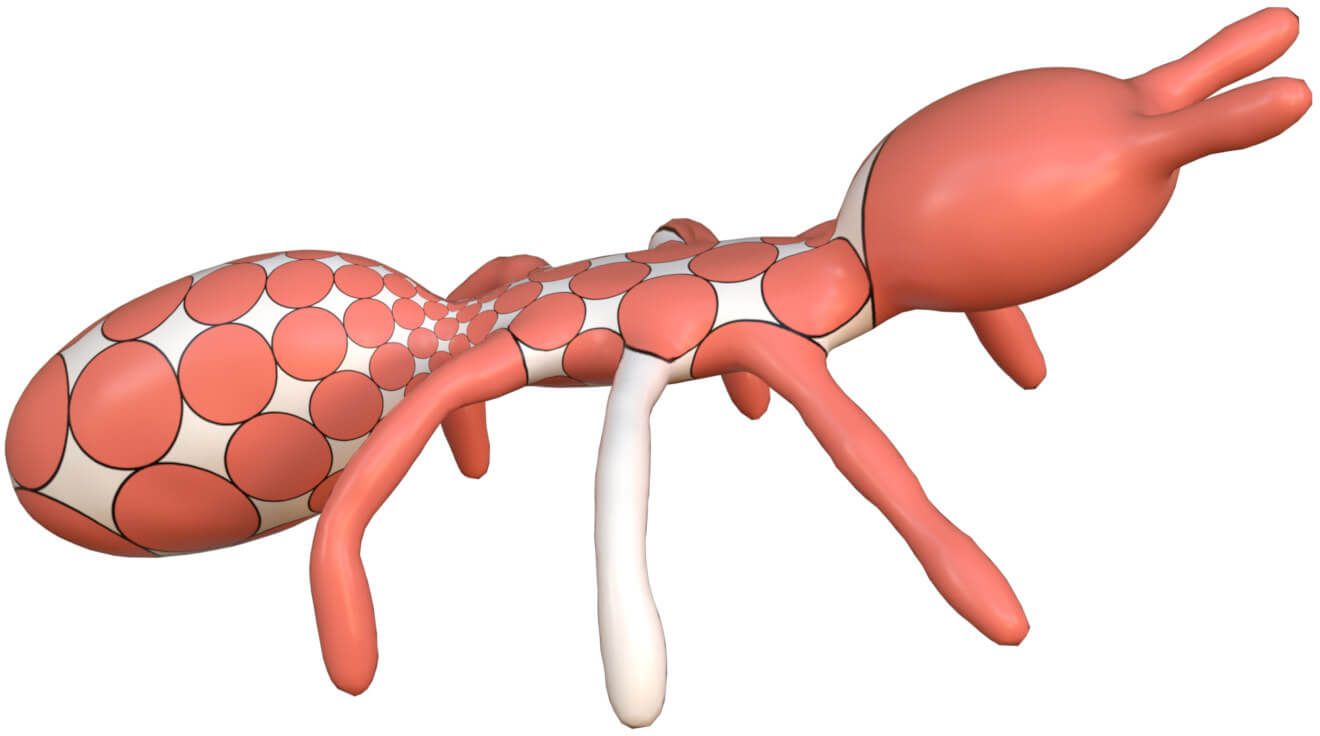}
		\caption*{ }   
	\end{subfigure}	
		\begin{subfigure}[b]{0.13\textwidth}
		\includegraphics[width=\textwidth]{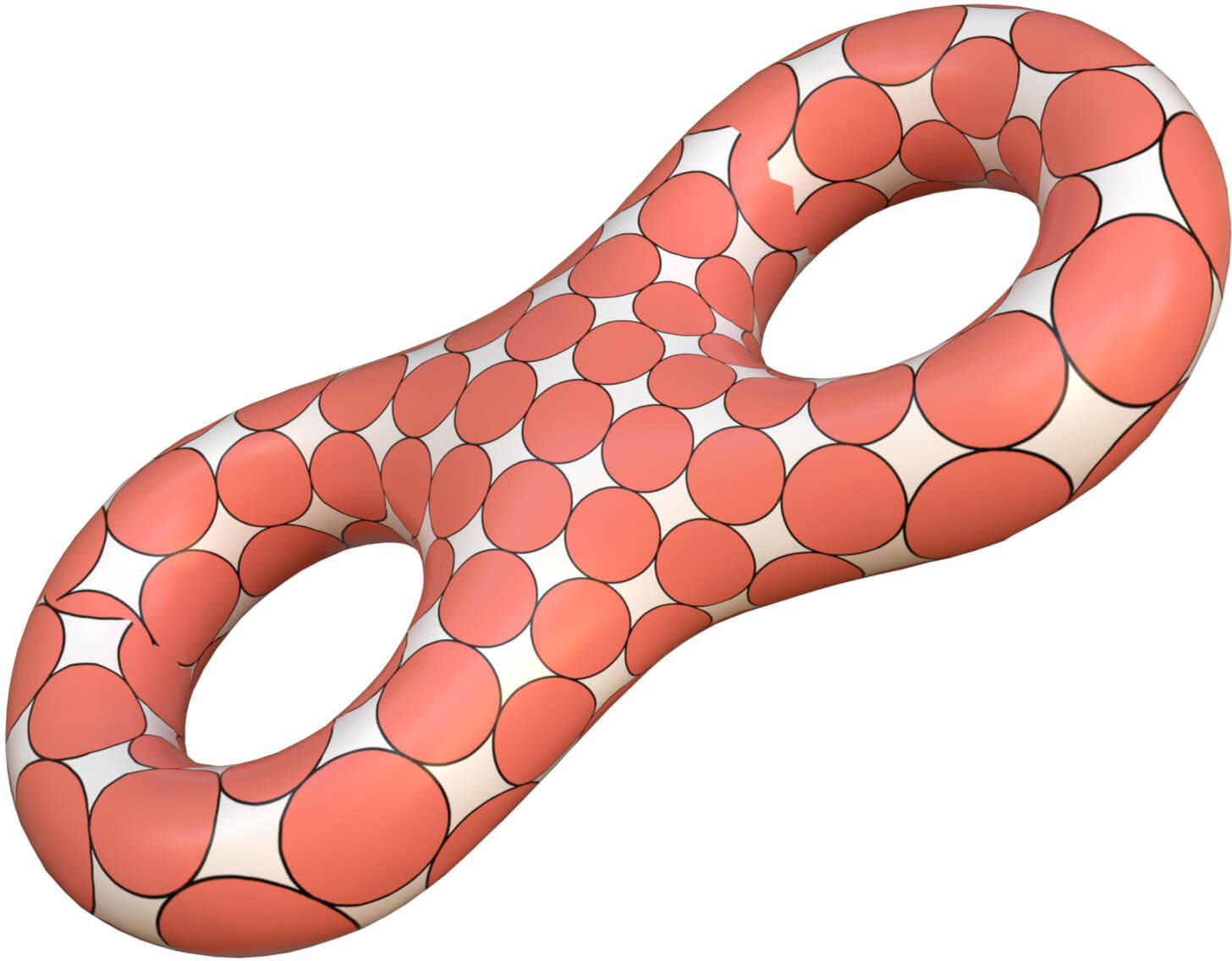}
		\caption*{ }       
	\end{subfigure}  
	\begin{subfigure}[b]{0.115\textwidth}
		\includegraphics[width=\textwidth]{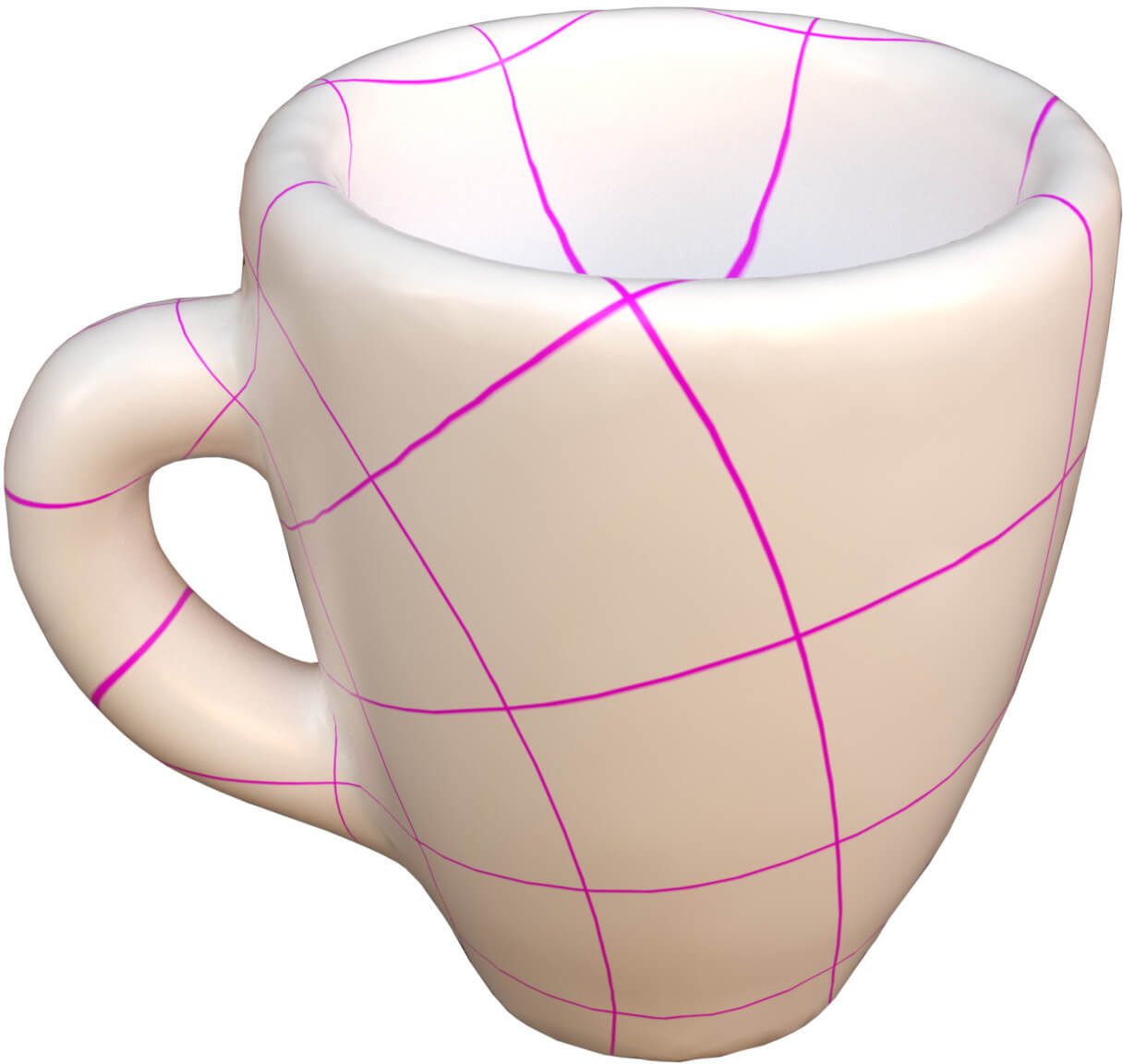}
		\caption*{ }   
	\end{subfigure}           
	\begin{subfigure}[b]{0.115\textwidth}
		\includegraphics[width=\textwidth]{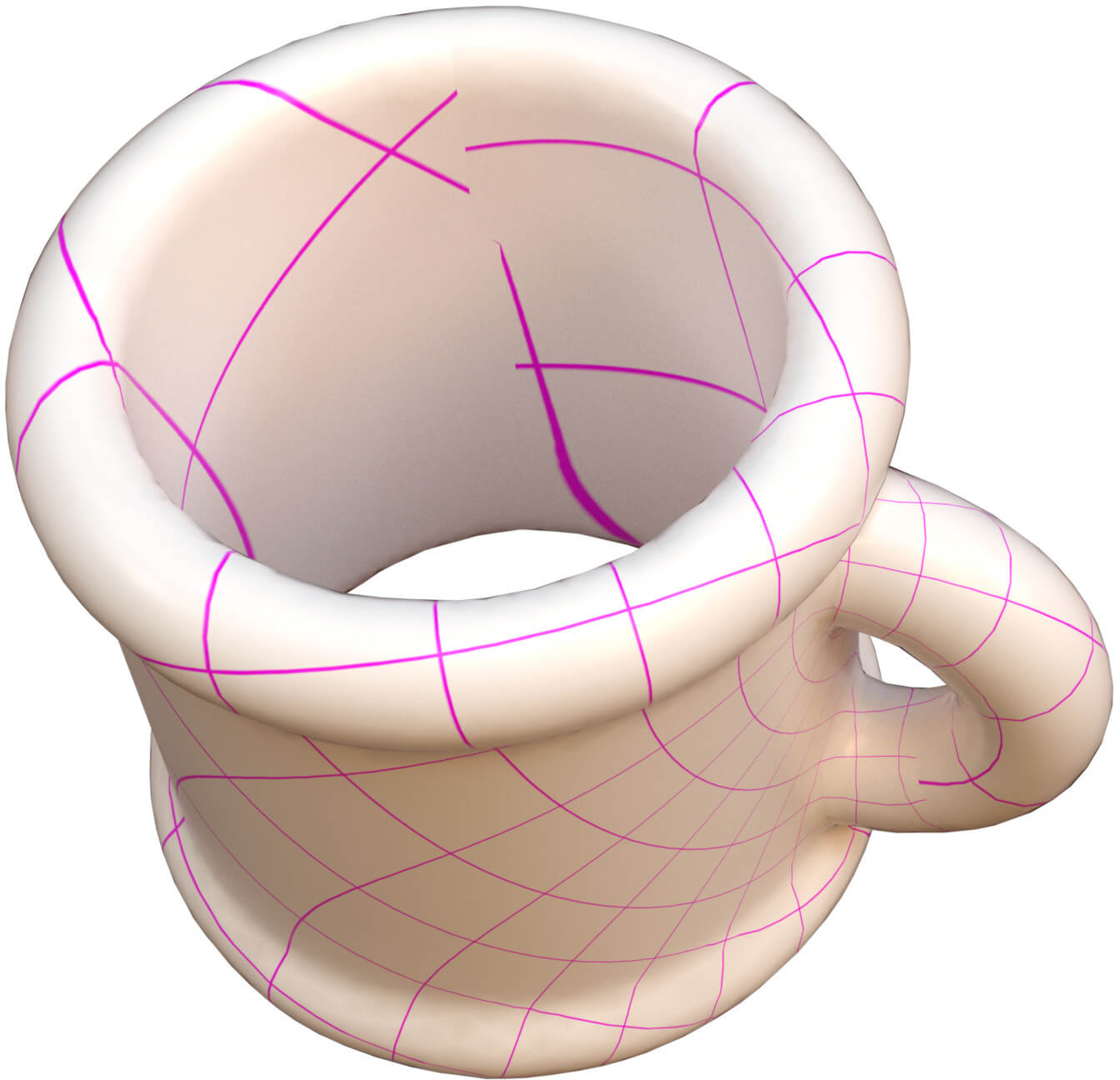}
		\caption*{ }         
	\end{subfigure}
	\begin{subfigure}[b]{0.09\textwidth}
		\includegraphics[width=\textwidth]{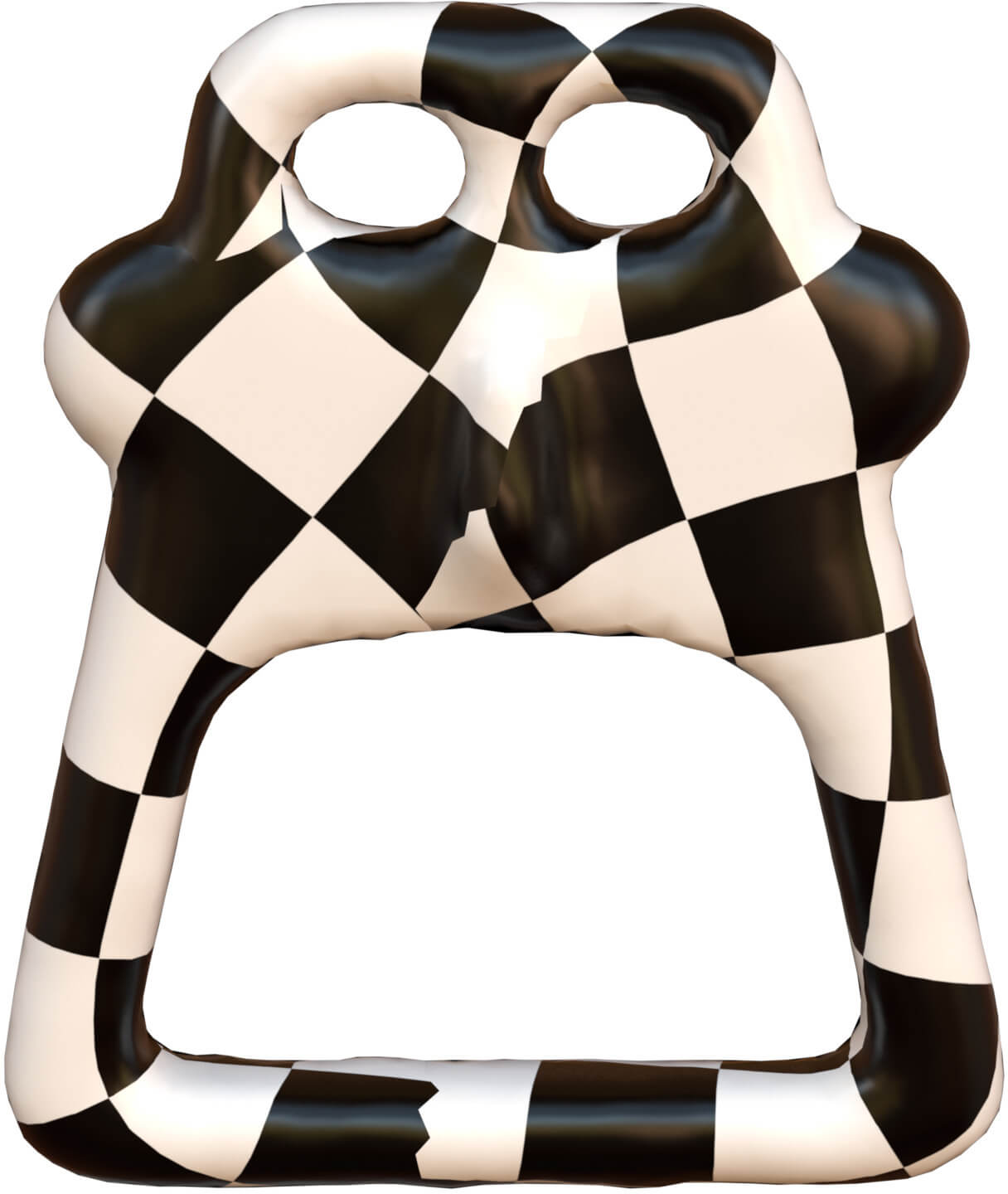}
		\caption*{ }        
	\end{subfigure}    
	\begin{subfigure}[b]{0.13\textwidth}
	\includegraphics[width=\textwidth]{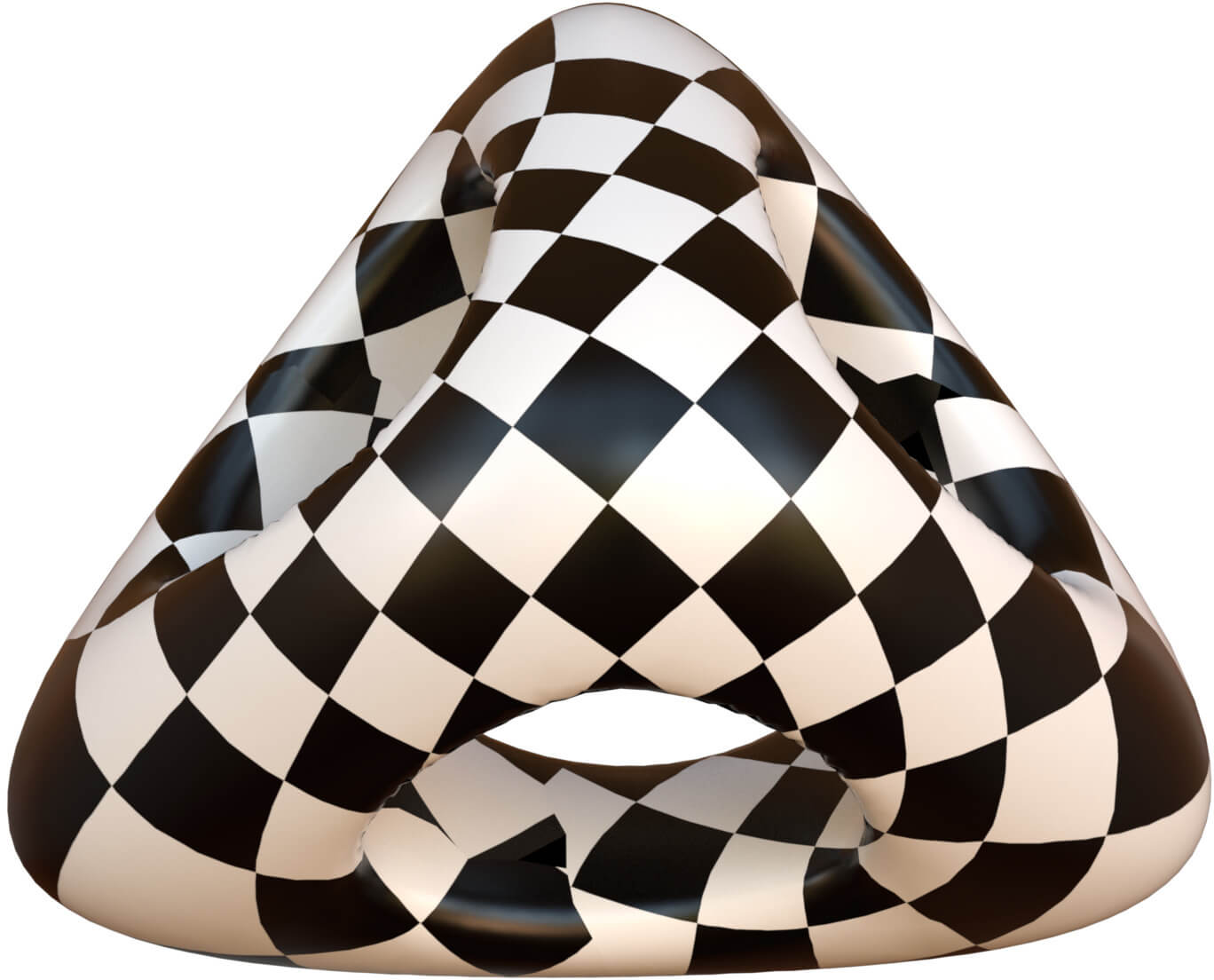}
	\caption*{ }        
    \end{subfigure} 
	
			\begin{subfigure}[b]{0.075\textwidth}
		\includegraphics[width=\textwidth]{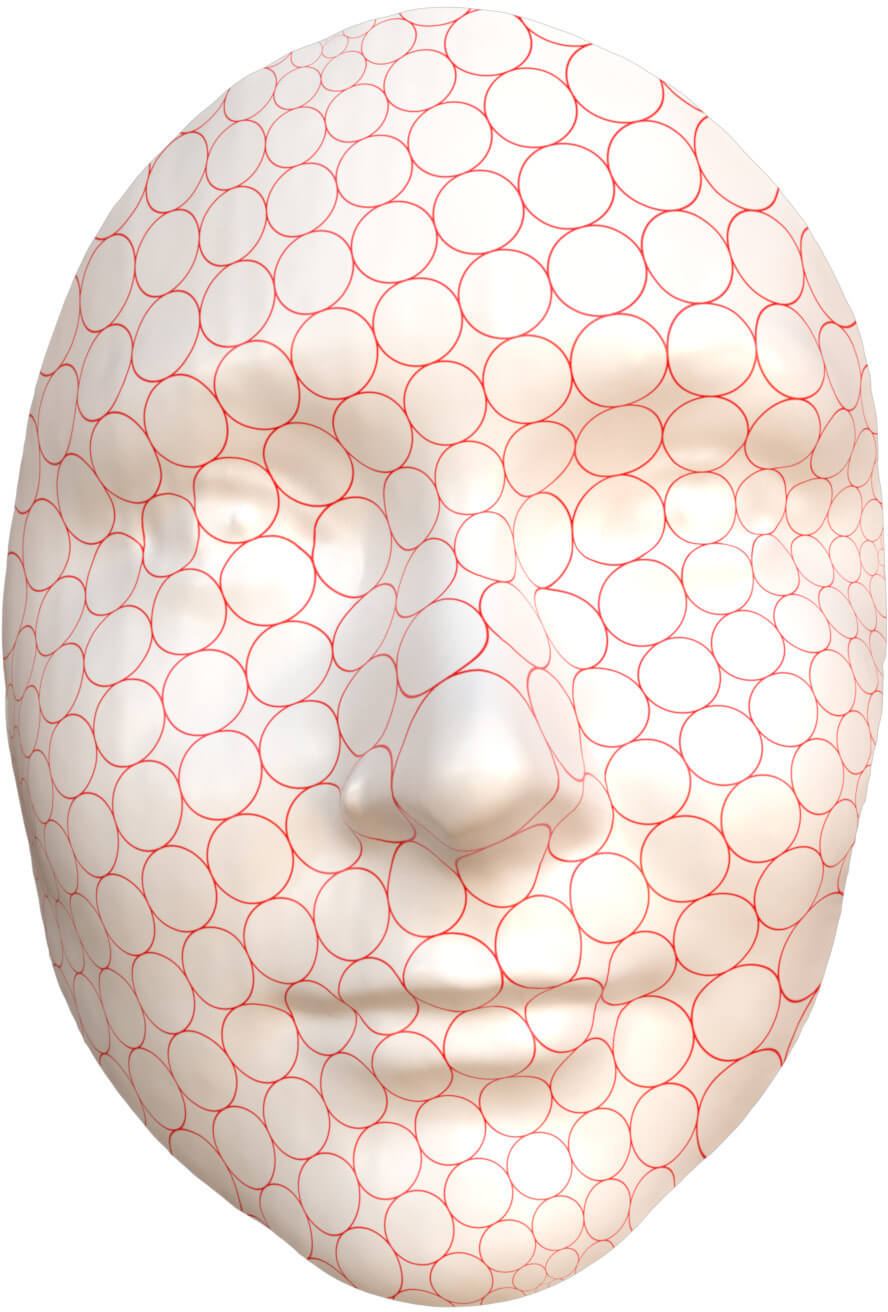}
		\caption*{ }   
	\end{subfigure}
	\begin{subfigure}[b]{0.055\textwidth}
		\includegraphics[width=\textwidth]{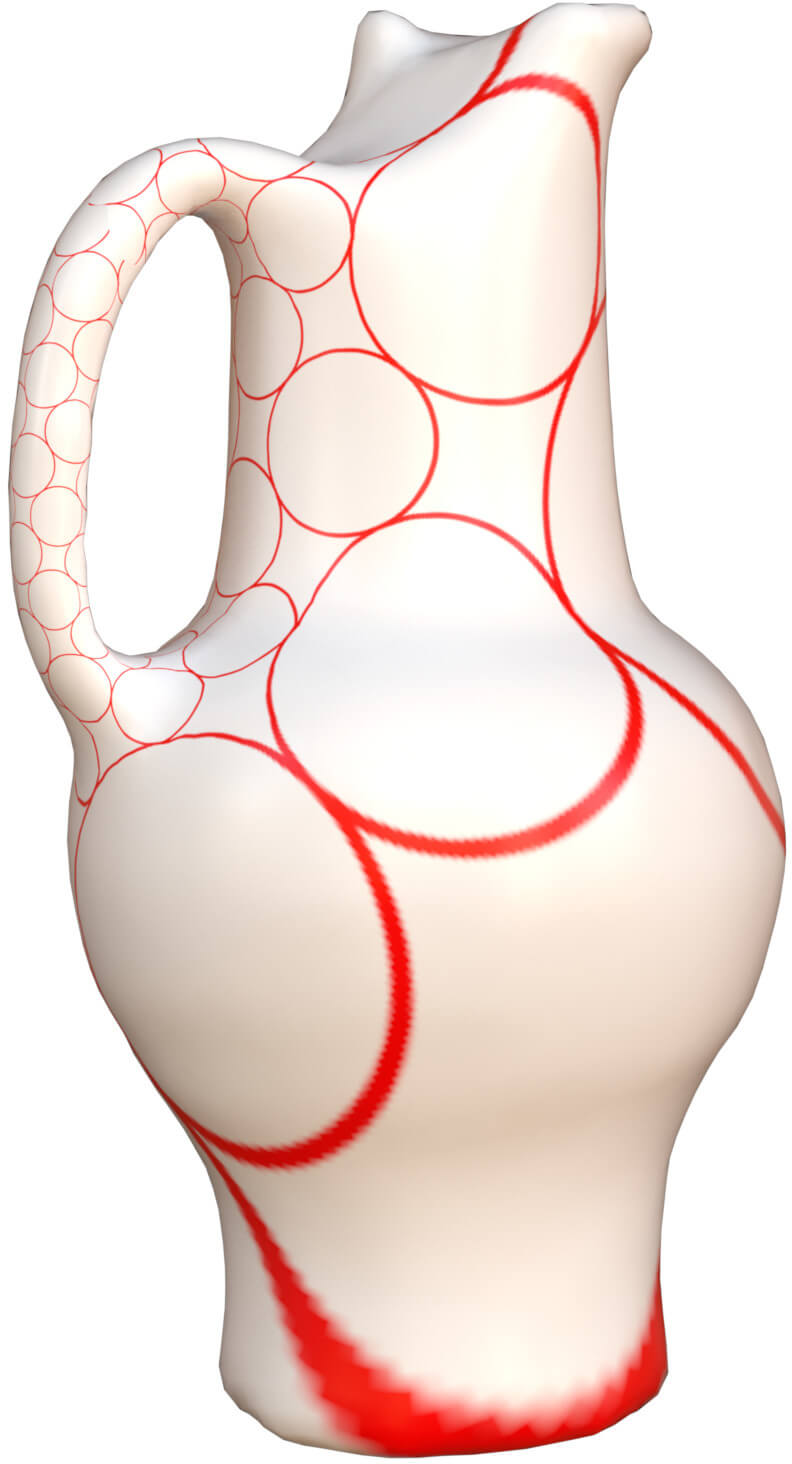}
		\caption*{ }     
	\end{subfigure}
	\begin{subfigure}[b]{0.17\textwidth}
		\includegraphics[width=\textwidth]{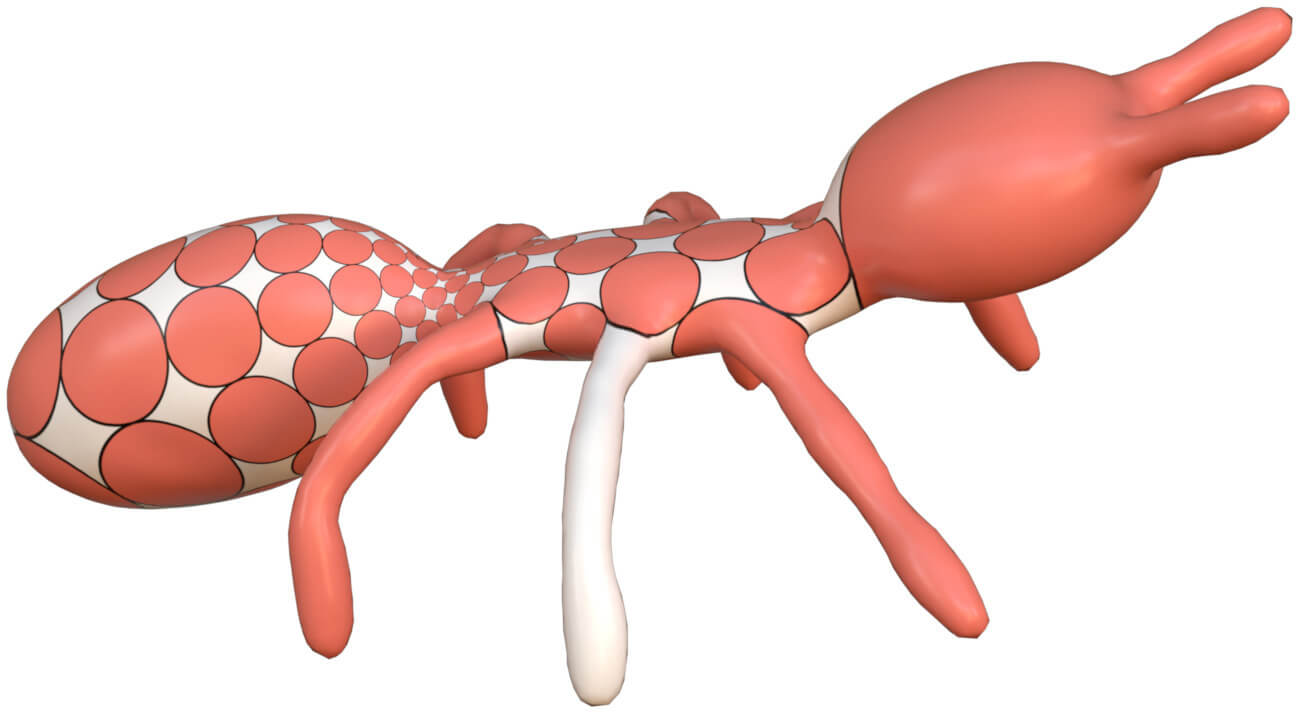}
		\caption*{ }   
	\end{subfigure}	
	\begin{subfigure}[b]{0.13\textwidth}
		\includegraphics[width=\textwidth]{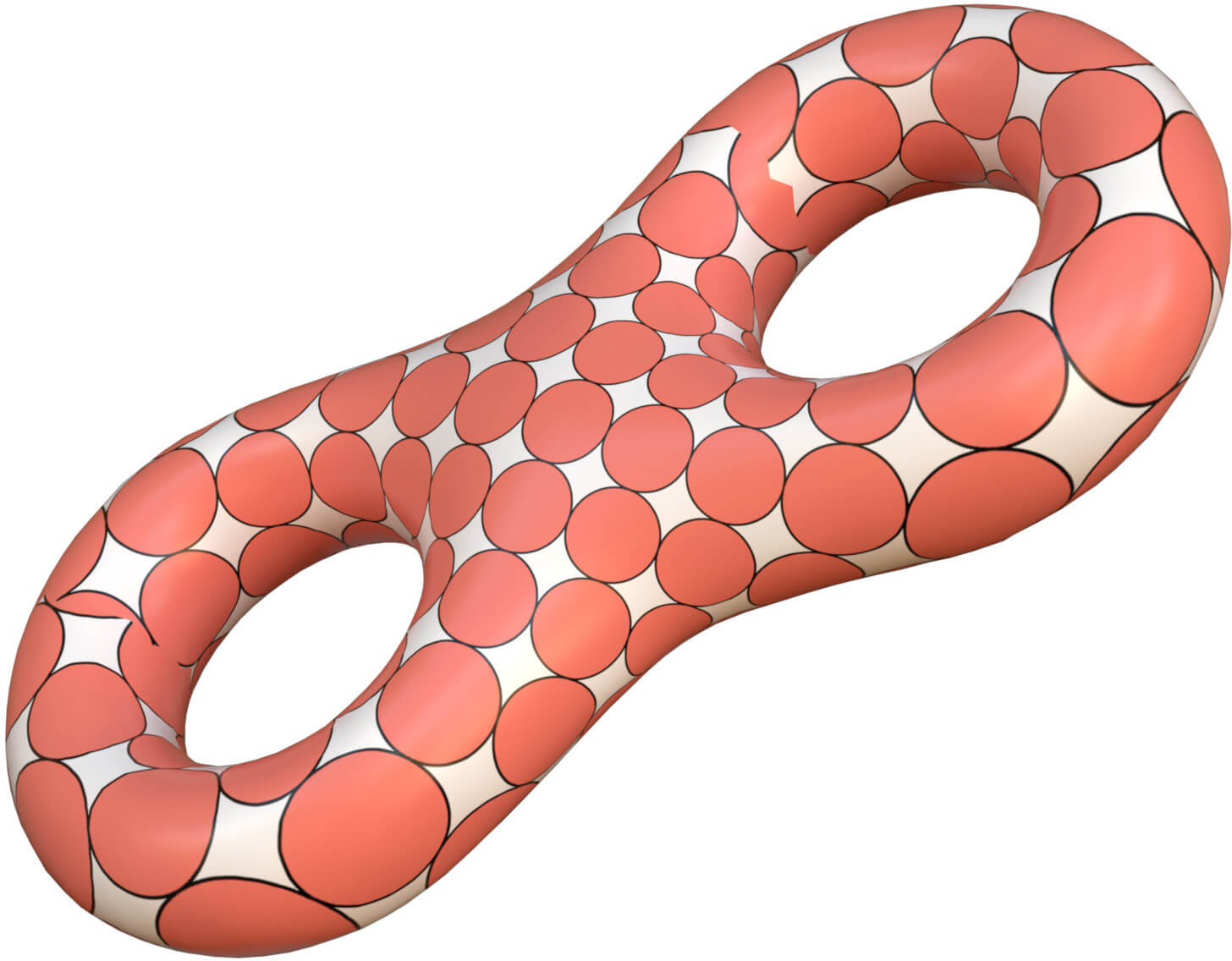}
		\caption*{ }       
	\end{subfigure}  
	\begin{subfigure}[b]{0.115\textwidth}
		\includegraphics[width=\textwidth]{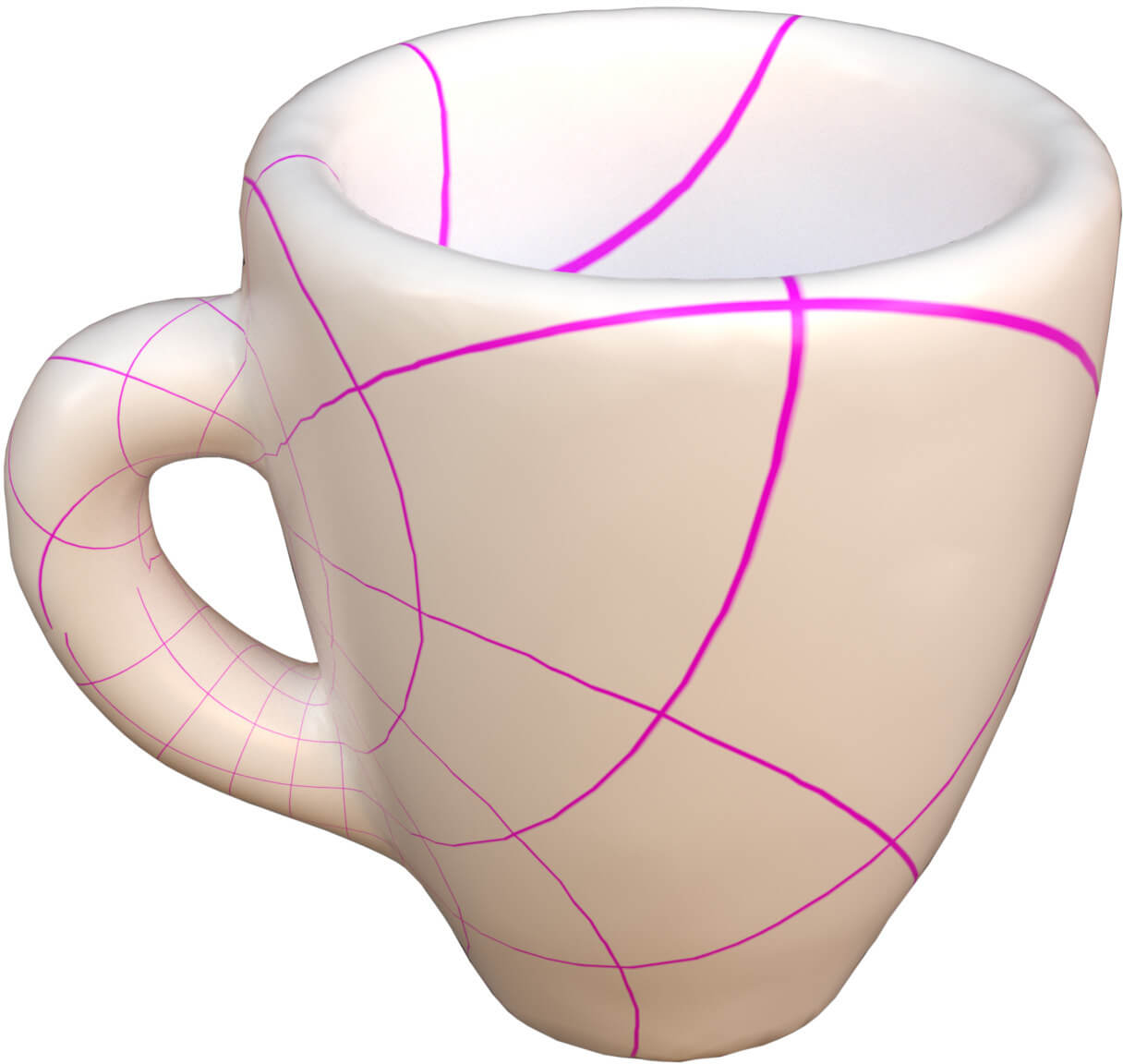}
		\caption*{ }   
	\end{subfigure}           
	\begin{subfigure}[b]{0.115\textwidth}
		\includegraphics[width=\textwidth]{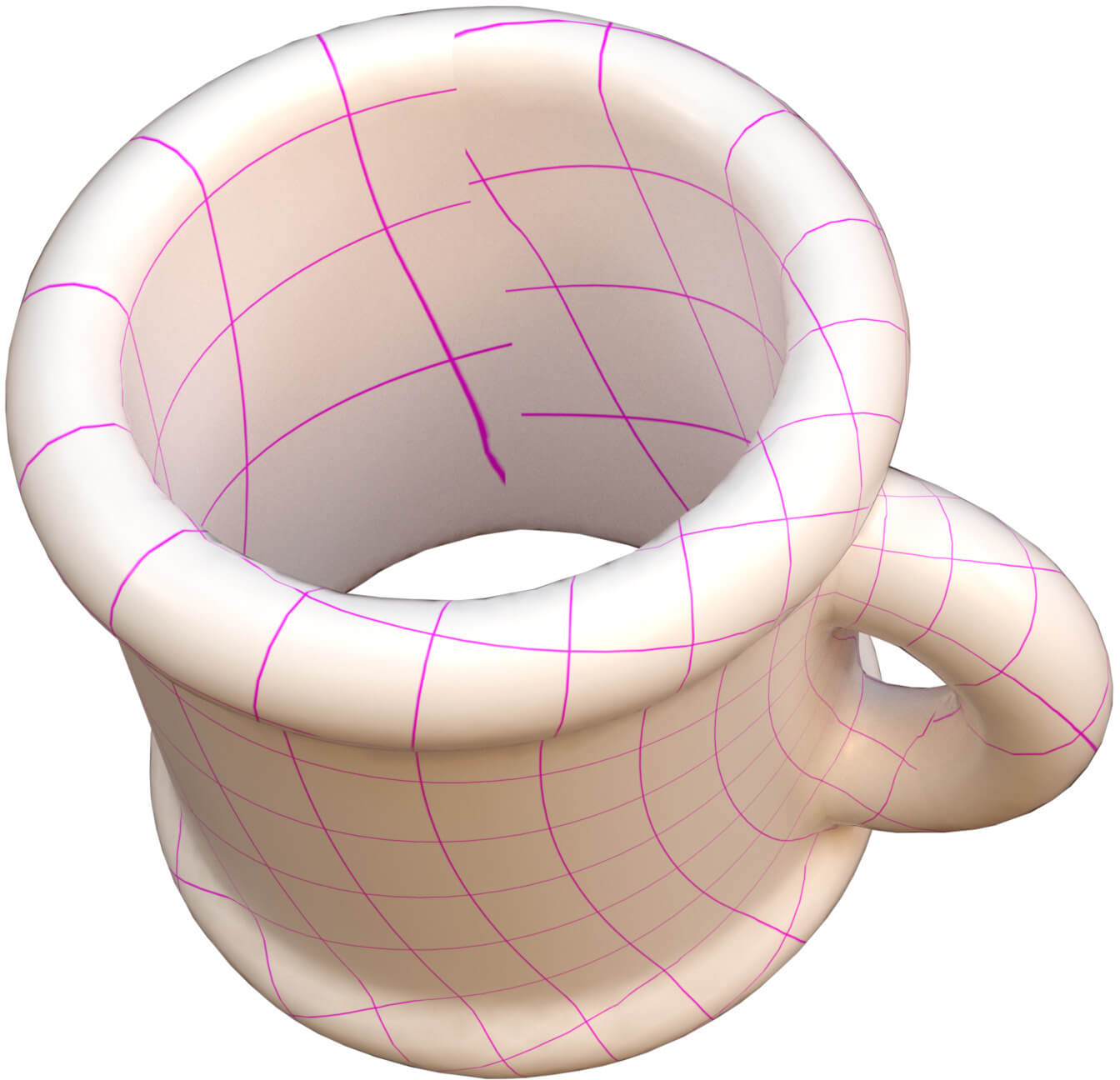}
		\caption*{ }         
	\end{subfigure}
	\begin{subfigure}[b]{0.09\textwidth}
		\includegraphics[width=\textwidth]{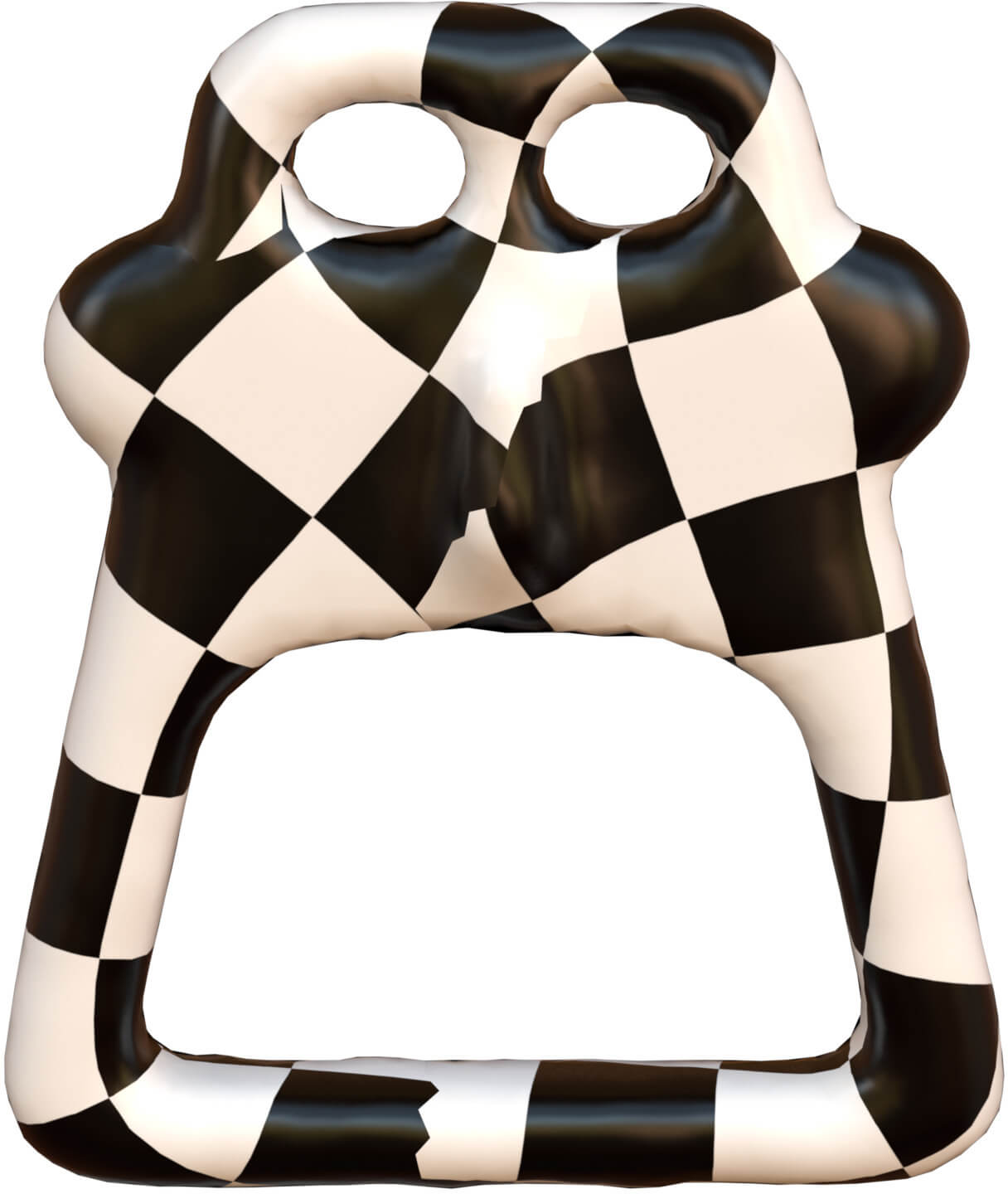}
		\caption*{ }        
	\end{subfigure}    
	\begin{subfigure}[b]{0.13\textwidth}
		\includegraphics[width=\textwidth]{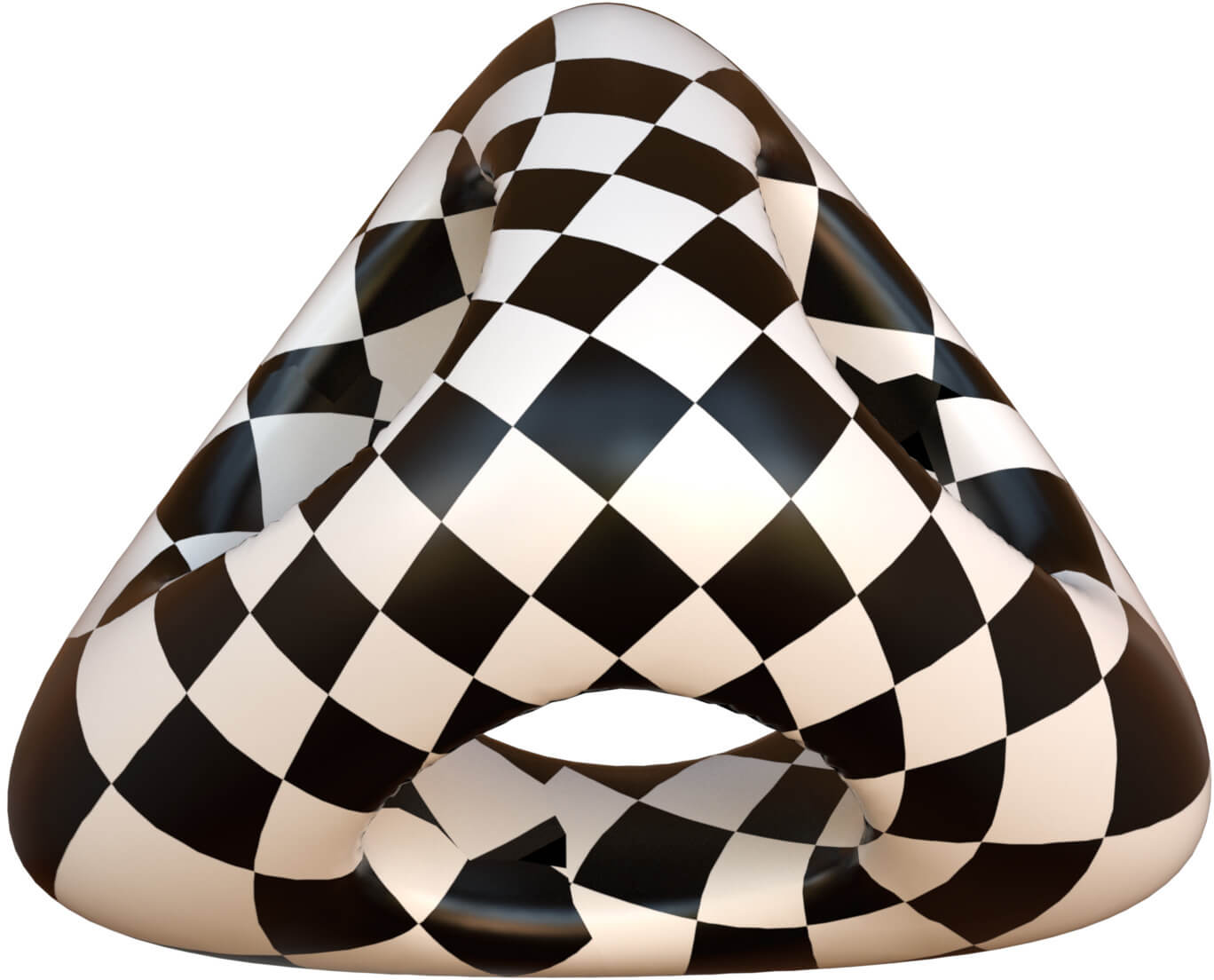}
		\caption*{ }        
	\end{subfigure} 
    
    		\begin{subfigure}[b]{0.075\textwidth}
    	\includegraphics[width=\textwidth]{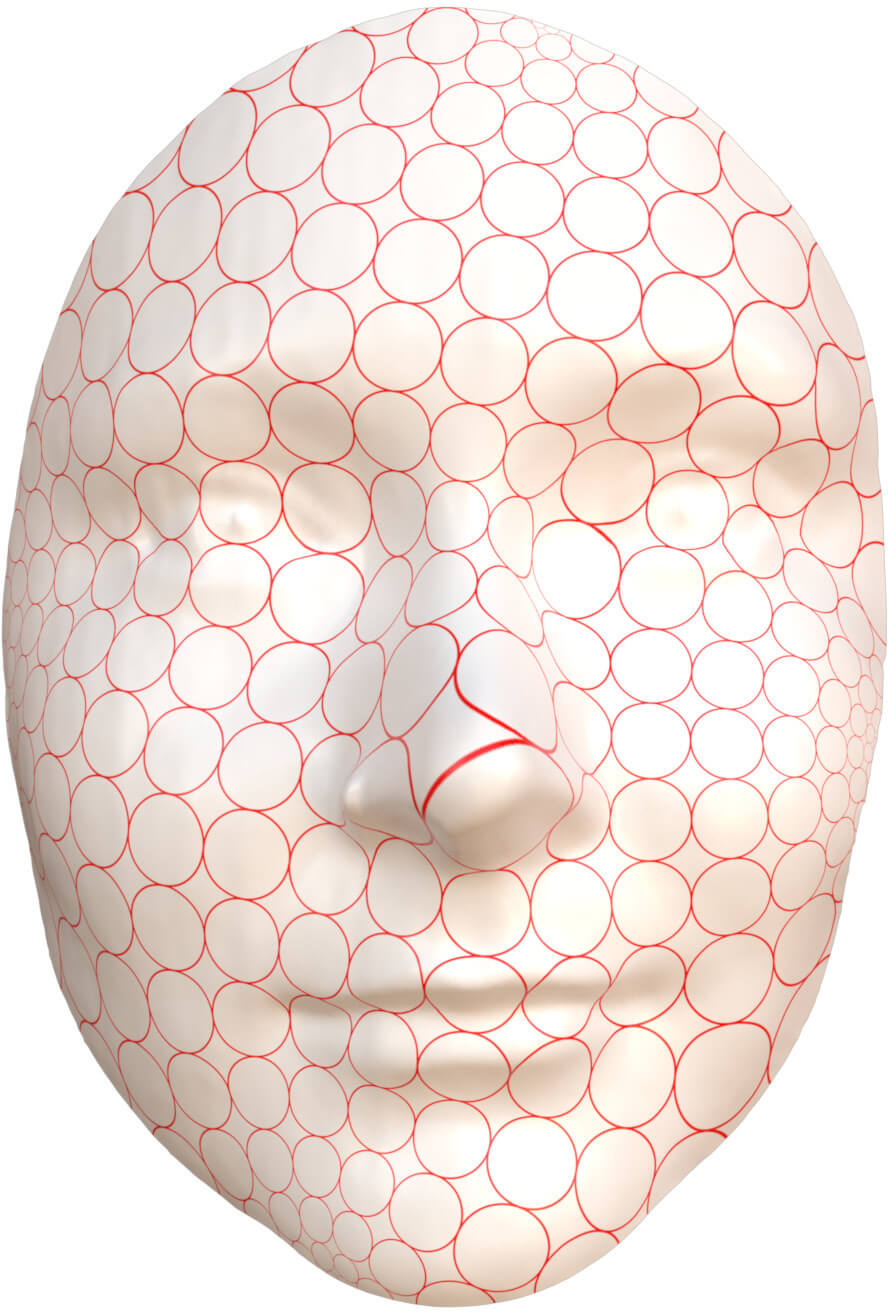}
    	\caption{ }   
    \end{subfigure}
    \begin{subfigure}[b]{0.055\textwidth}
    	\includegraphics[width=\textwidth]{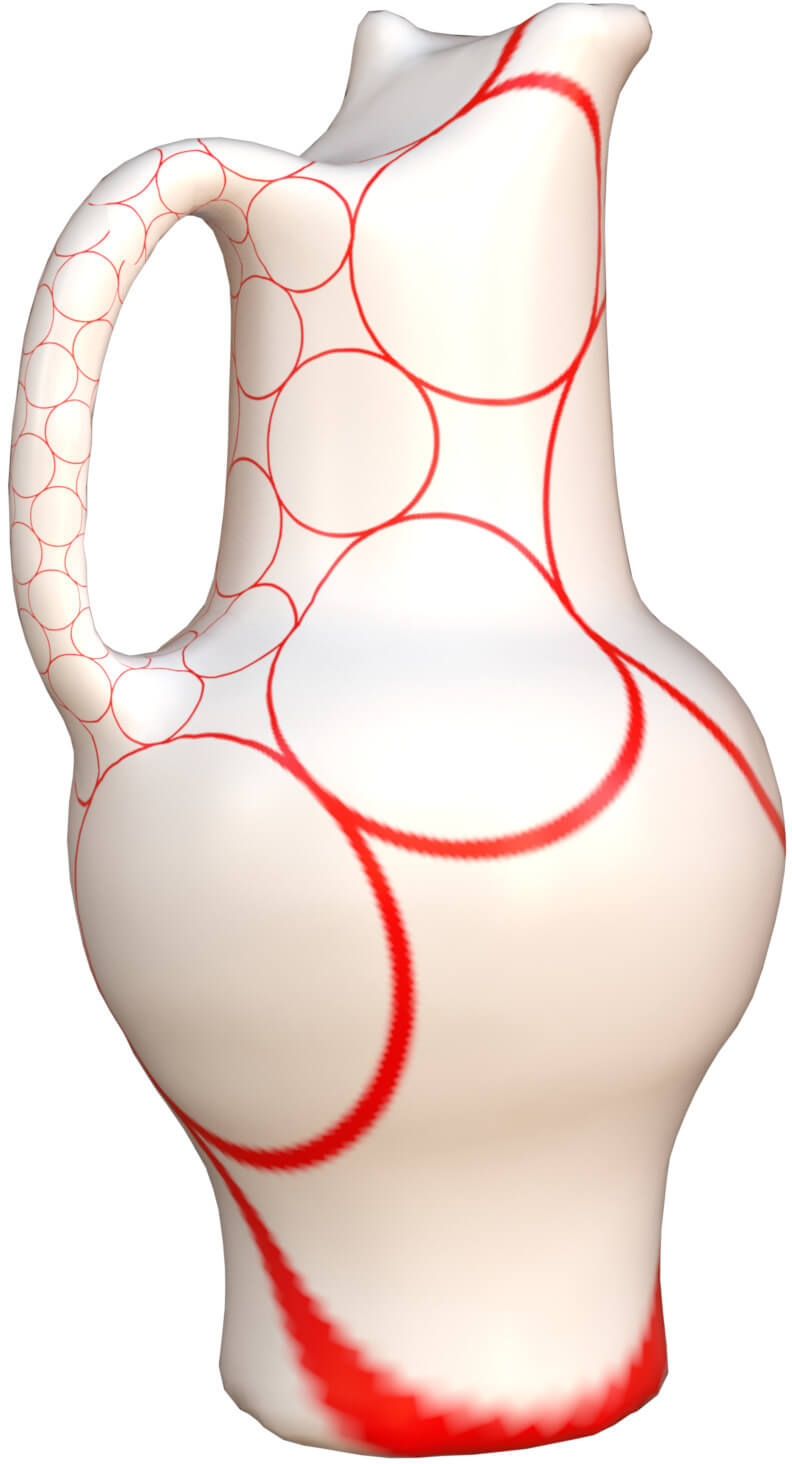}
    	\caption{ }     
    \end{subfigure}
    \begin{subfigure}[b]{0.17\textwidth}
    	\includegraphics[width=\textwidth]{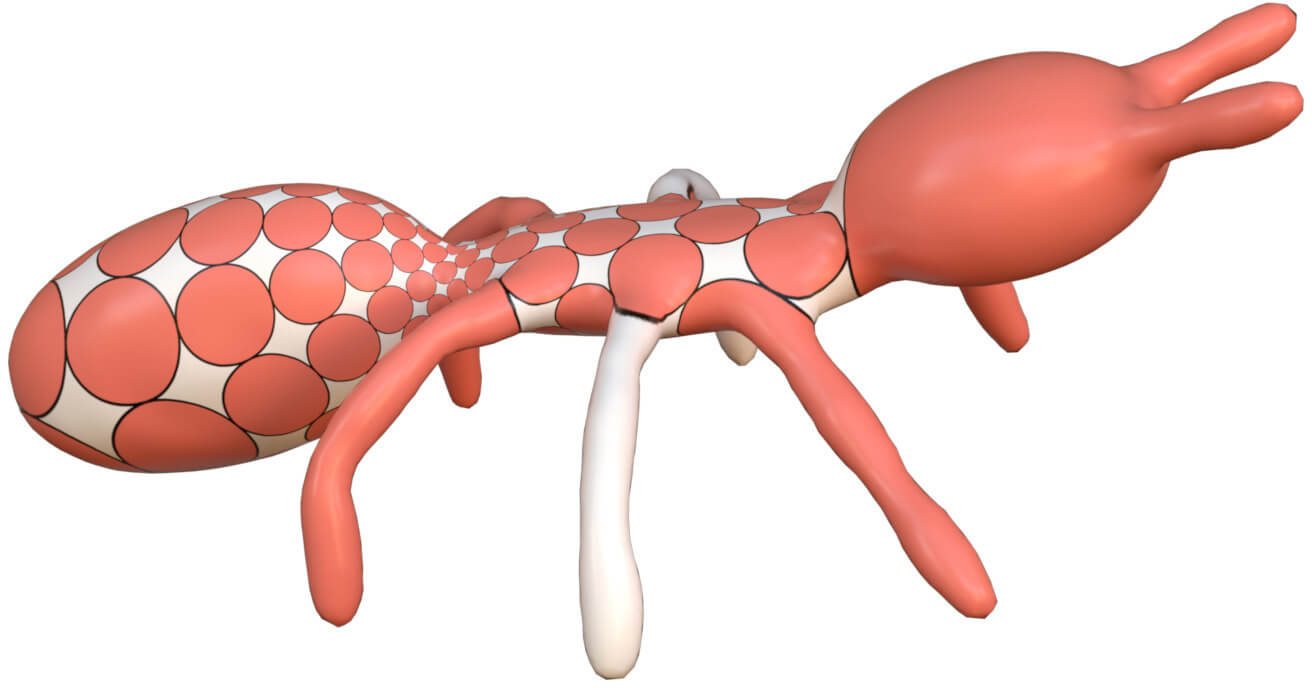}
    	\caption{ }   
    \end{subfigure}	
    \begin{subfigure}[b]{0.13\textwidth}
    	\includegraphics[width=\textwidth]{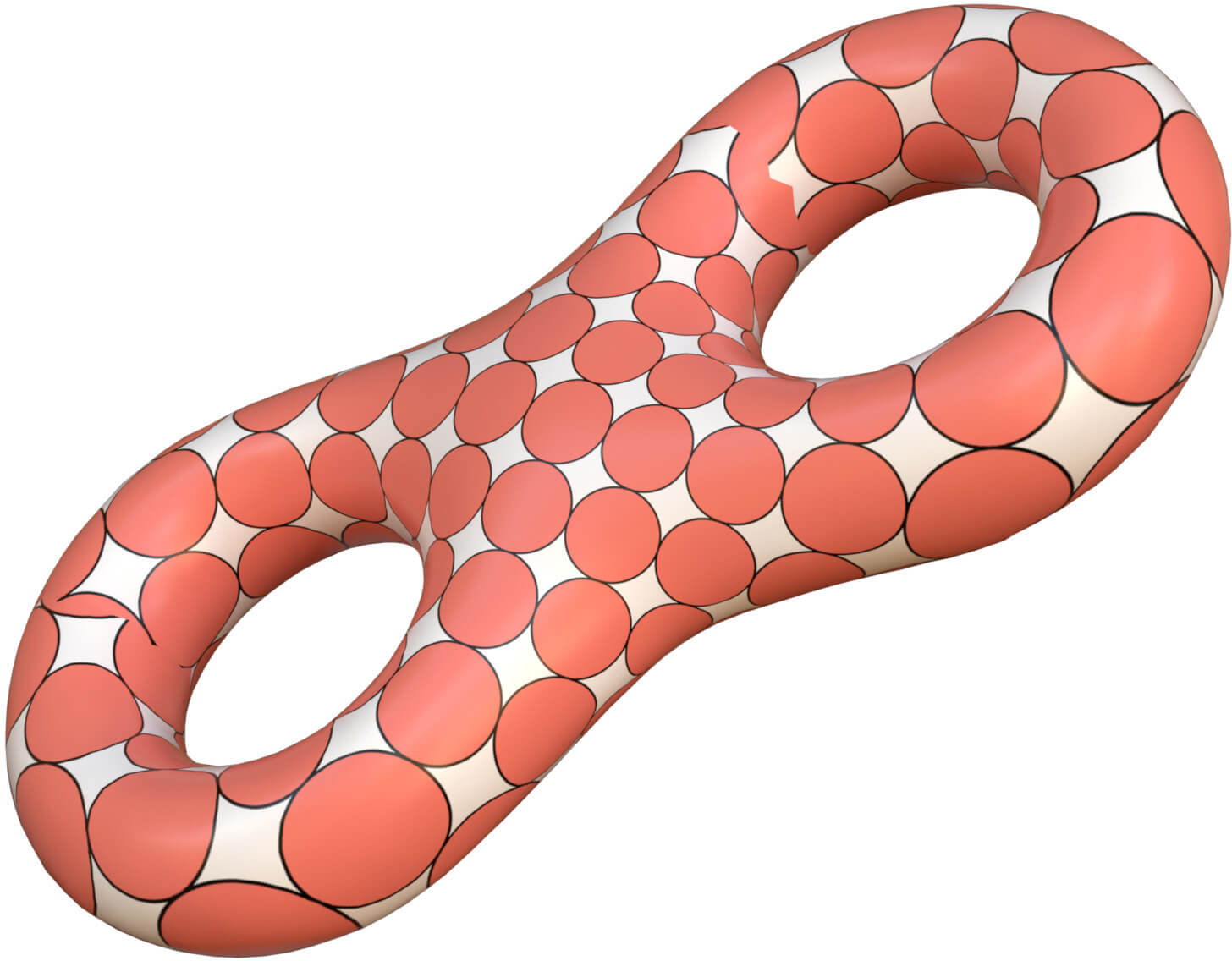}
    	\caption{ }       
    \end{subfigure}  
    \begin{subfigure}[b]{0.115\textwidth}
    	\includegraphics[width=\textwidth]{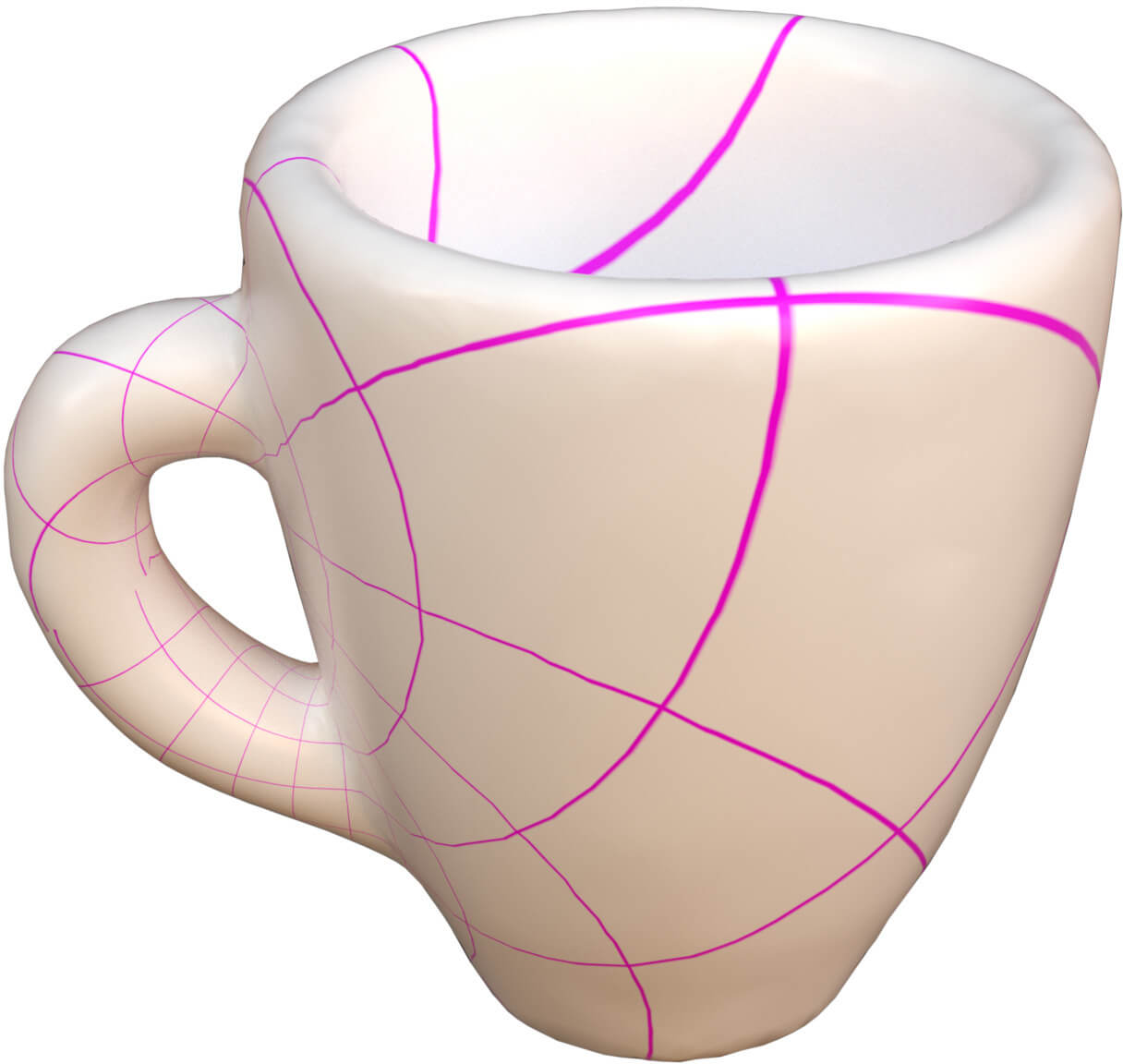}
    	\caption{ }   
    \end{subfigure}           
    \begin{subfigure}[b]{0.115\textwidth}
    	\includegraphics[width=\textwidth]{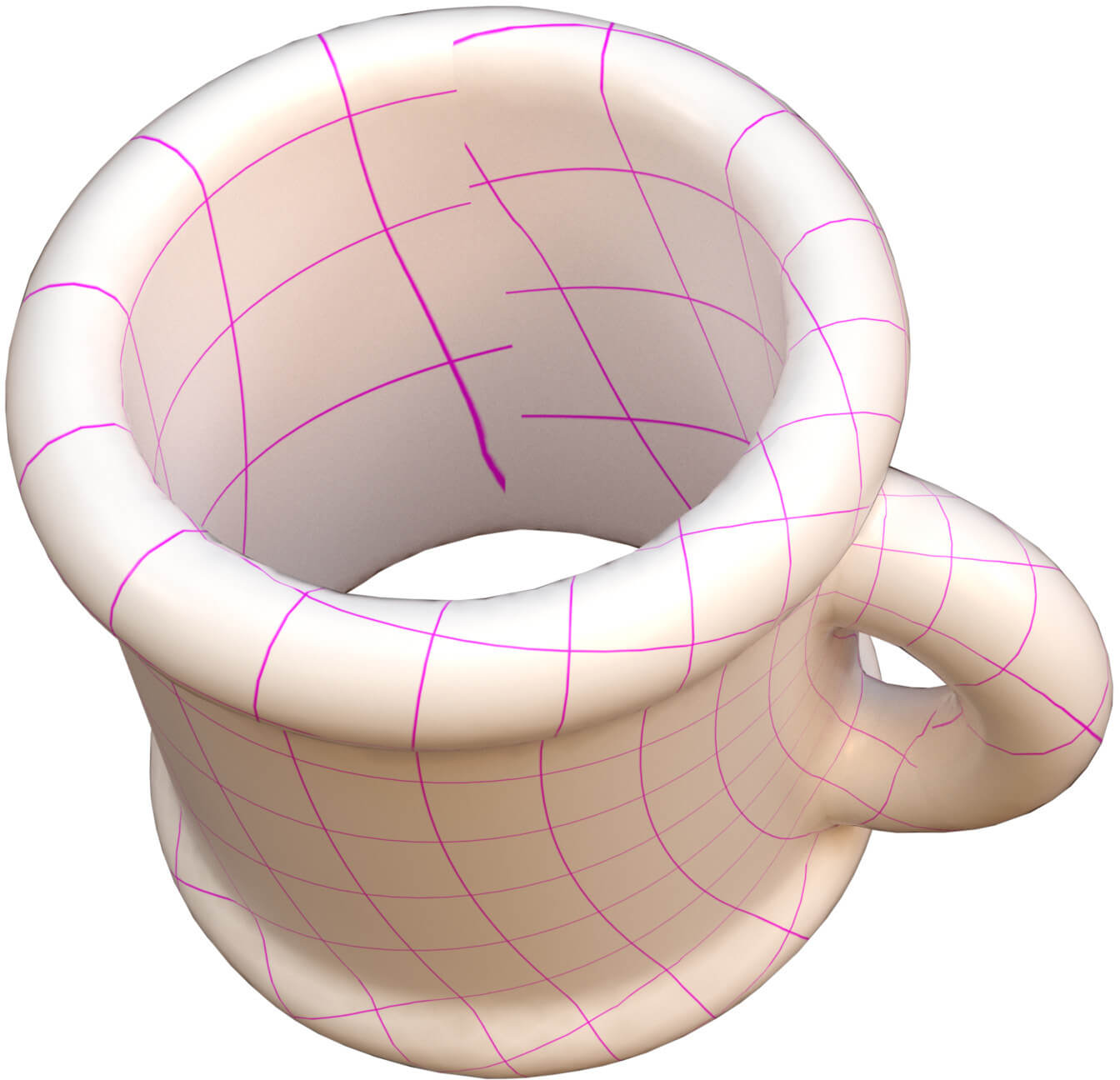}
    	\caption{ }         
    \end{subfigure}
    \begin{subfigure}[b]{0.09\textwidth}
    	\includegraphics[width=\textwidth]{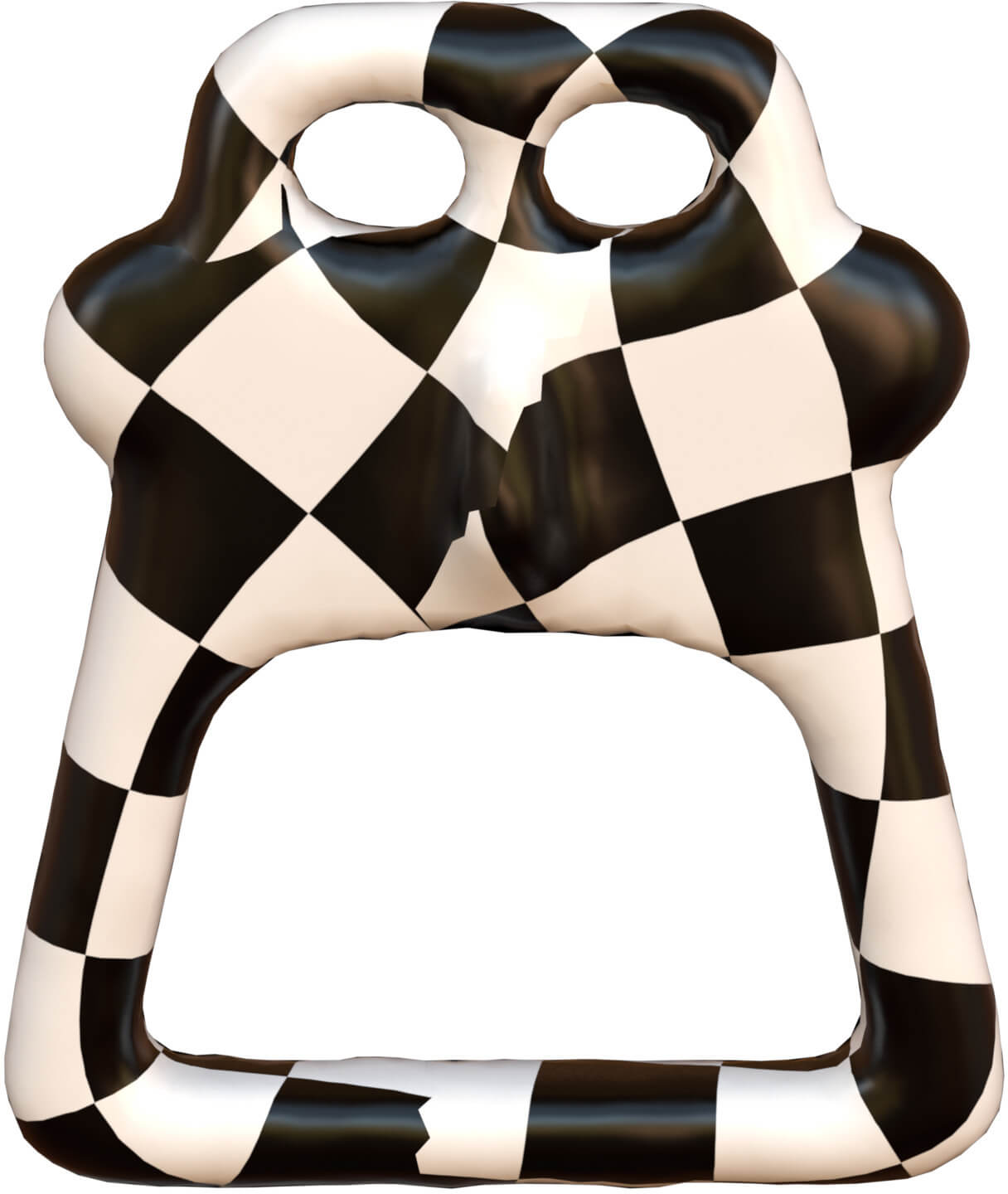}
    	\caption{ }        
    \end{subfigure}    
    \begin{subfigure}[b]{0.13\textwidth}
    	\includegraphics[width=\textwidth]{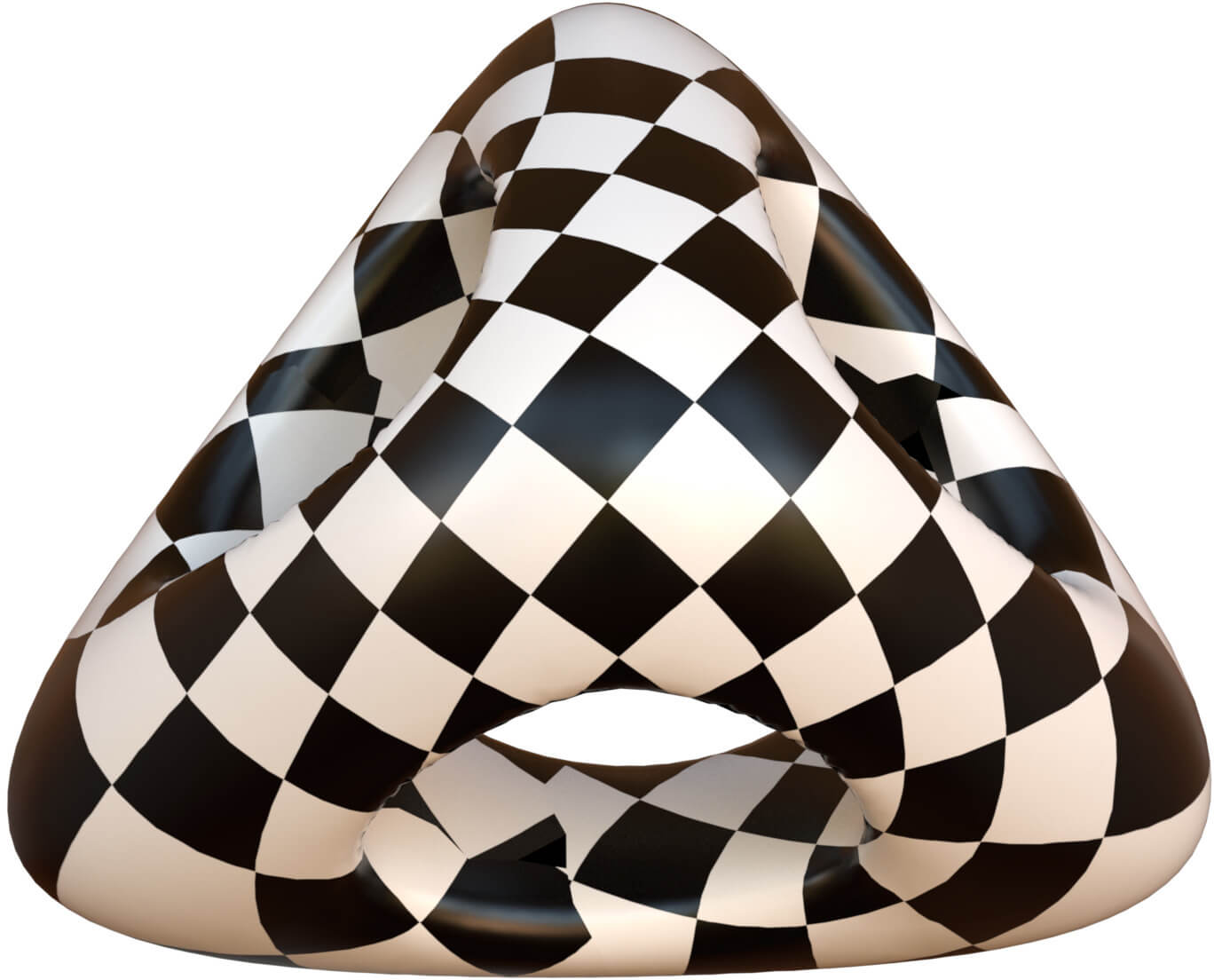}
    	\caption{ }        
    \end{subfigure} 
	\caption{The first row is with Calabi flow; the second is with Ricci flow; The third is with CETM.}\label{fig:calabicomparision}
\end{figure*}

We  run experiments on dozens of models and compare the conformities of our Calabi flow, Ricci flow and CETM algorithms.
In the figure \ref{fig:calabicomparision}, we show several meshes parametrized with Calabi flow, Ricci flow and CETM algorithms respectively. 
It is observed that three kinds of algorithms have almost the same commonality and all of them can preserve the angles perfectly.

\begin{figure}
	\begin{subfigure}[b]{0.15\textwidth}
		\includegraphics[width=\textwidth]{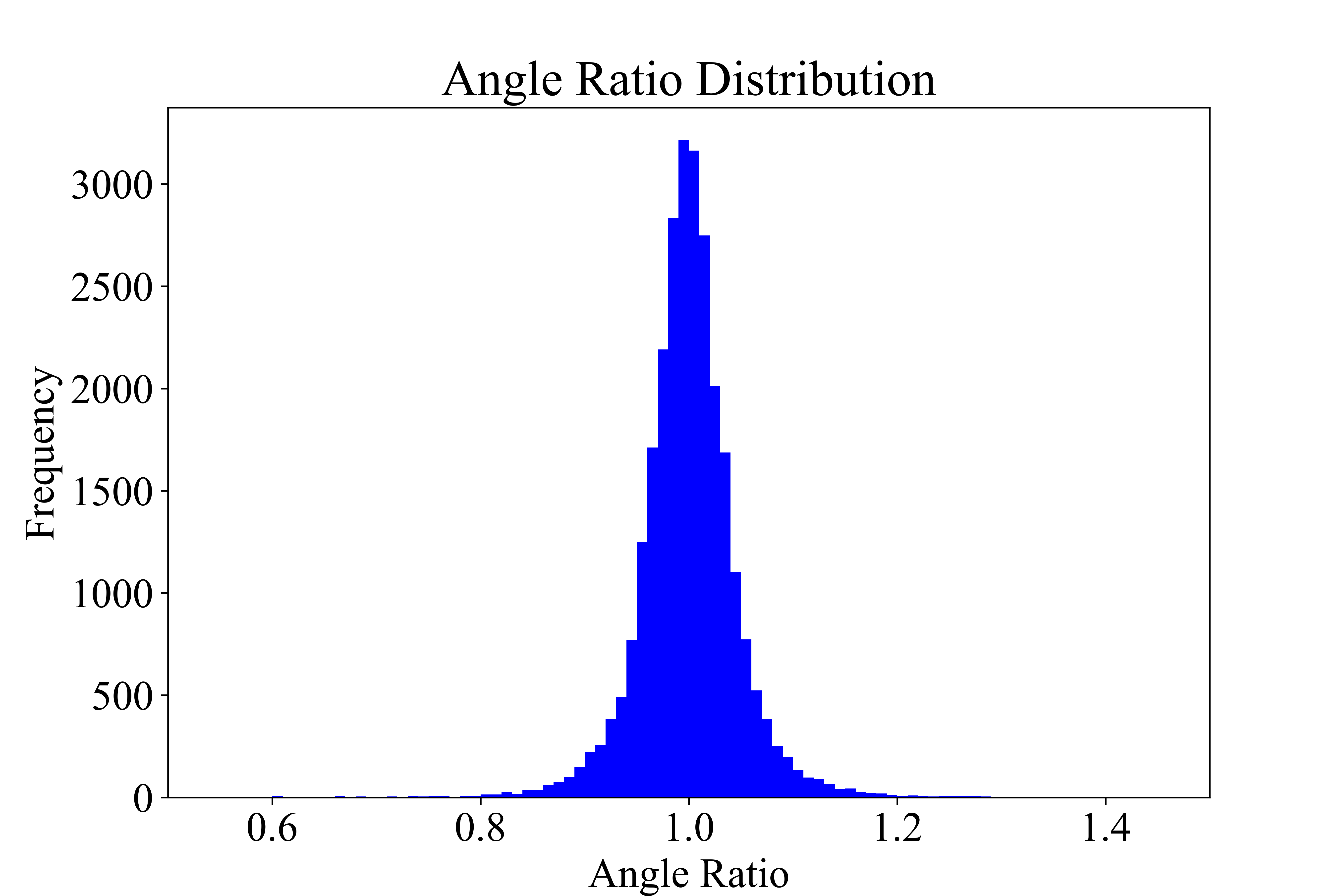}
		\caption{ }   
	\end{subfigure} 
	\begin{subfigure}[b]{0.15\textwidth}
		\includegraphics[width=\textwidth]{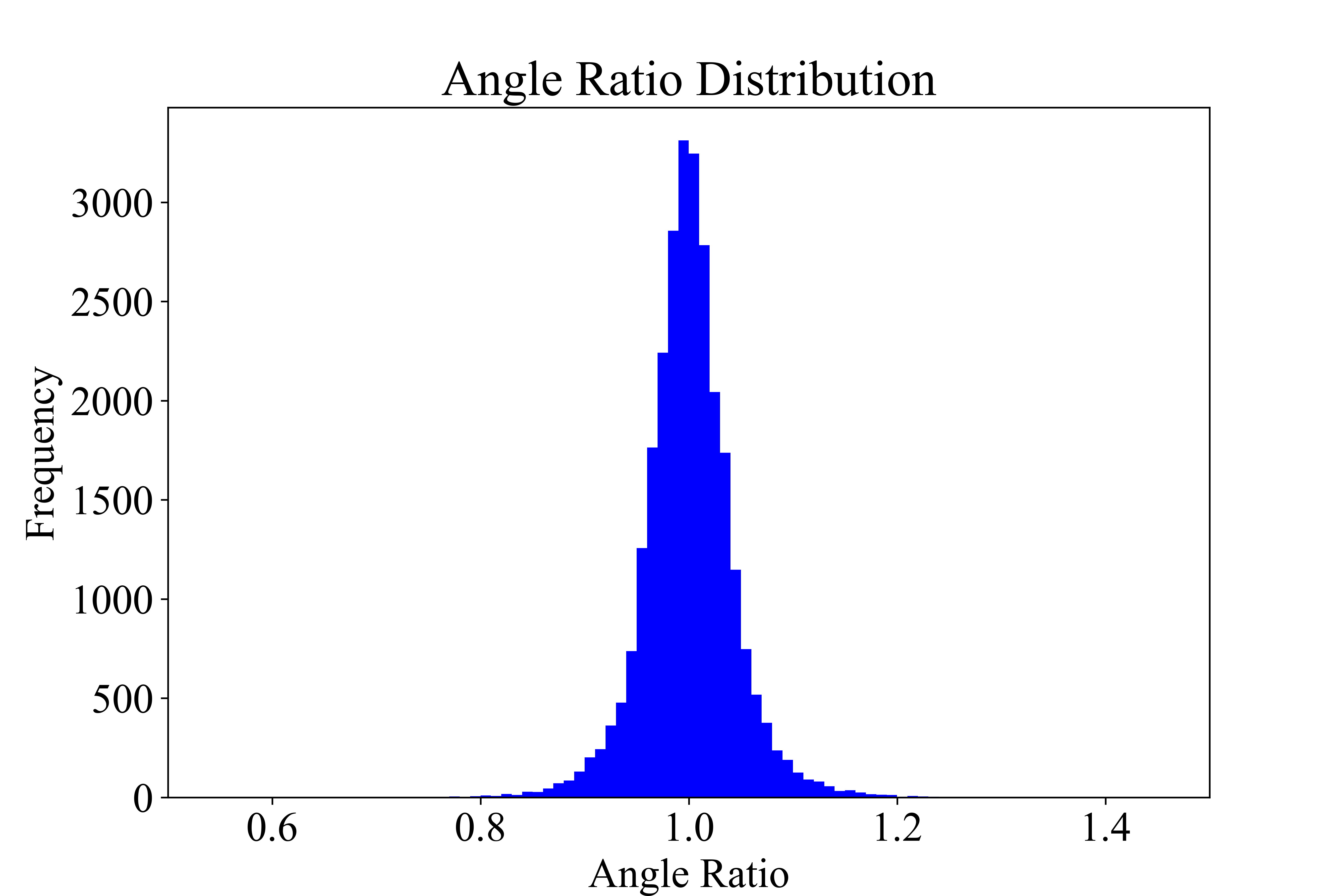}
		\caption{ }   
	\end{subfigure} 
	\begin{subfigure}[b]{0.15\textwidth}
		\includegraphics[width=\textwidth]{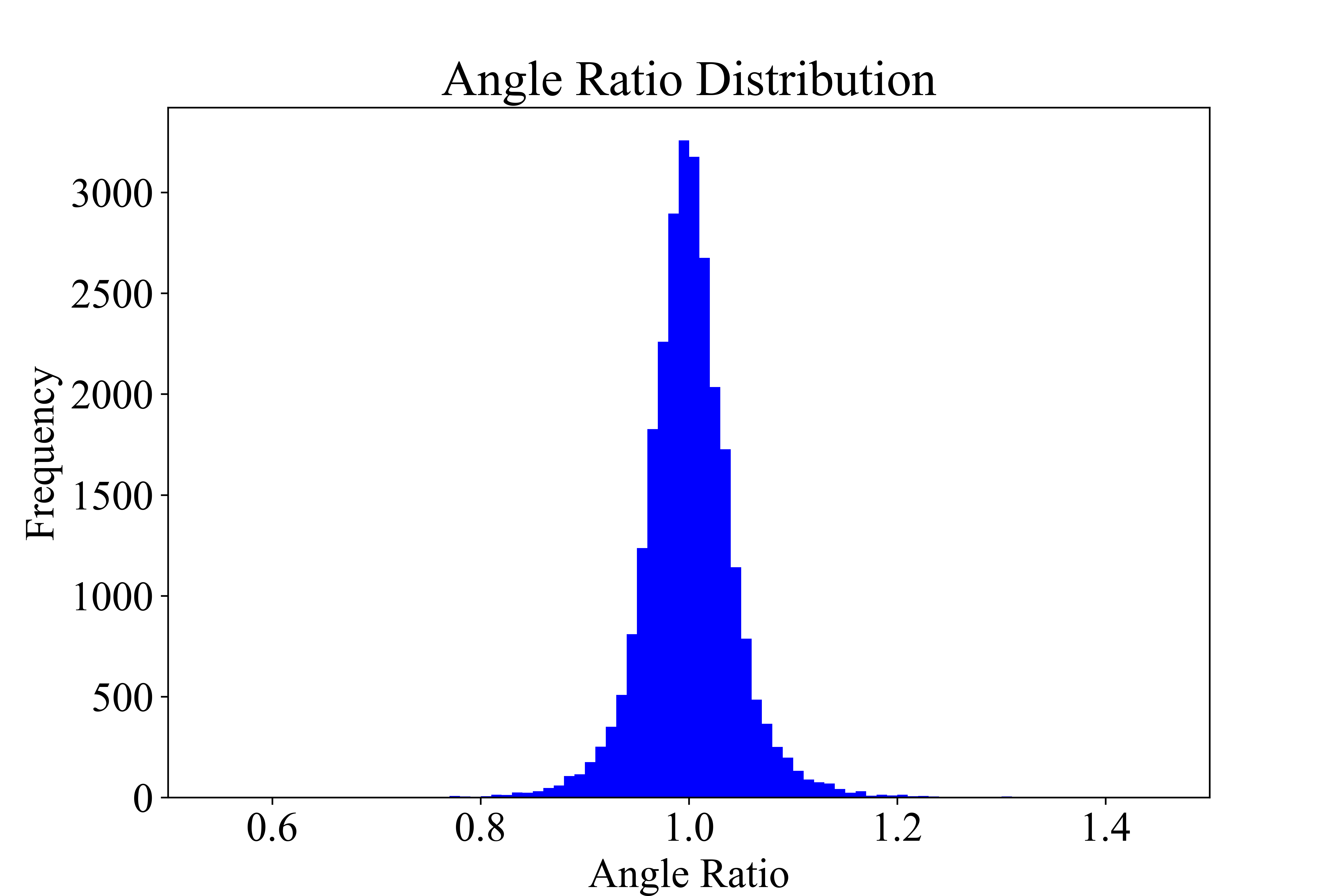}
		\caption{ }   
	\end{subfigure} 
	\caption{Angle ratio distributions of the kitten model   by Calabi flow (a), Ricci flow (b) and CETM (c) respectively.}
	\label{fig:conformitytest}
\end{figure}

\begin{figure}
		\centering
		\includegraphics[width=0.5\textwidth]{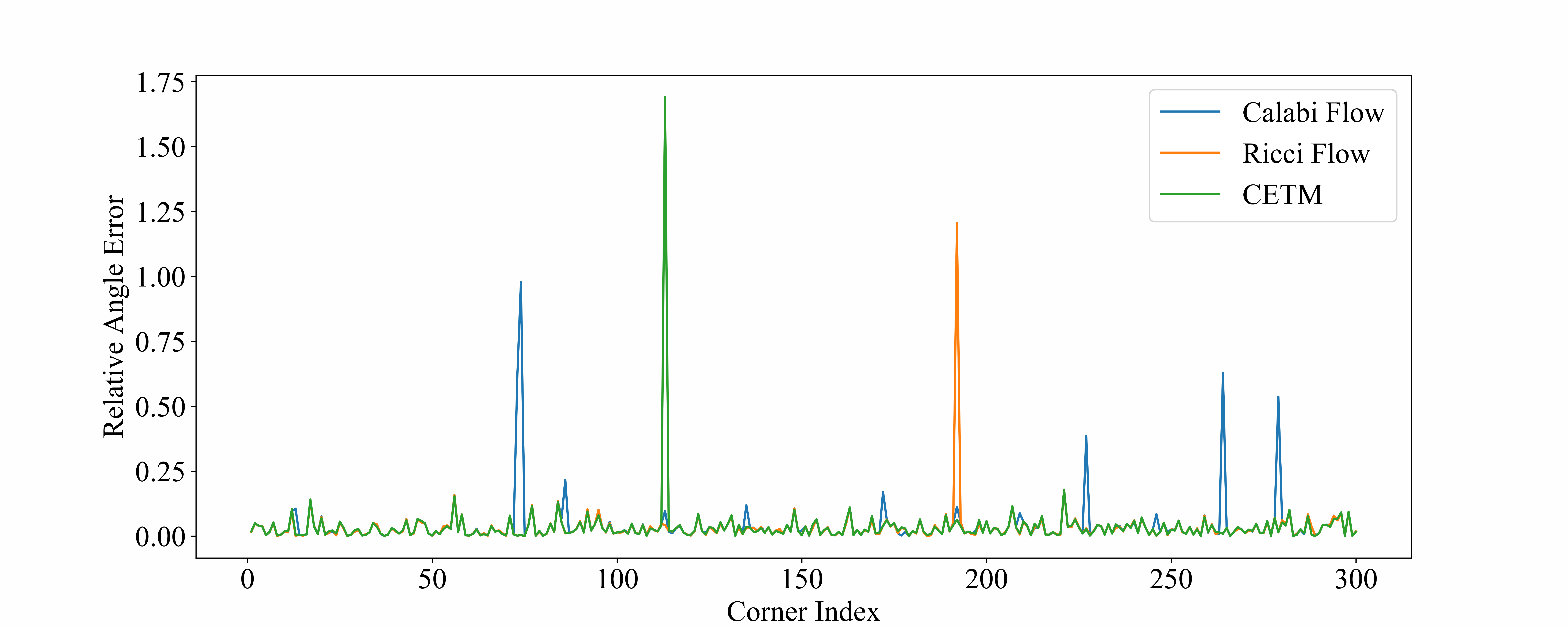}
	\caption{Angle relative errors on randomly selected 300 faces of kitten model.}
	\label{fig:conformitytest2}
\end{figure}

\begin{figure} 	
	\centering
		\includegraphics[width = 0.35\textwidth]{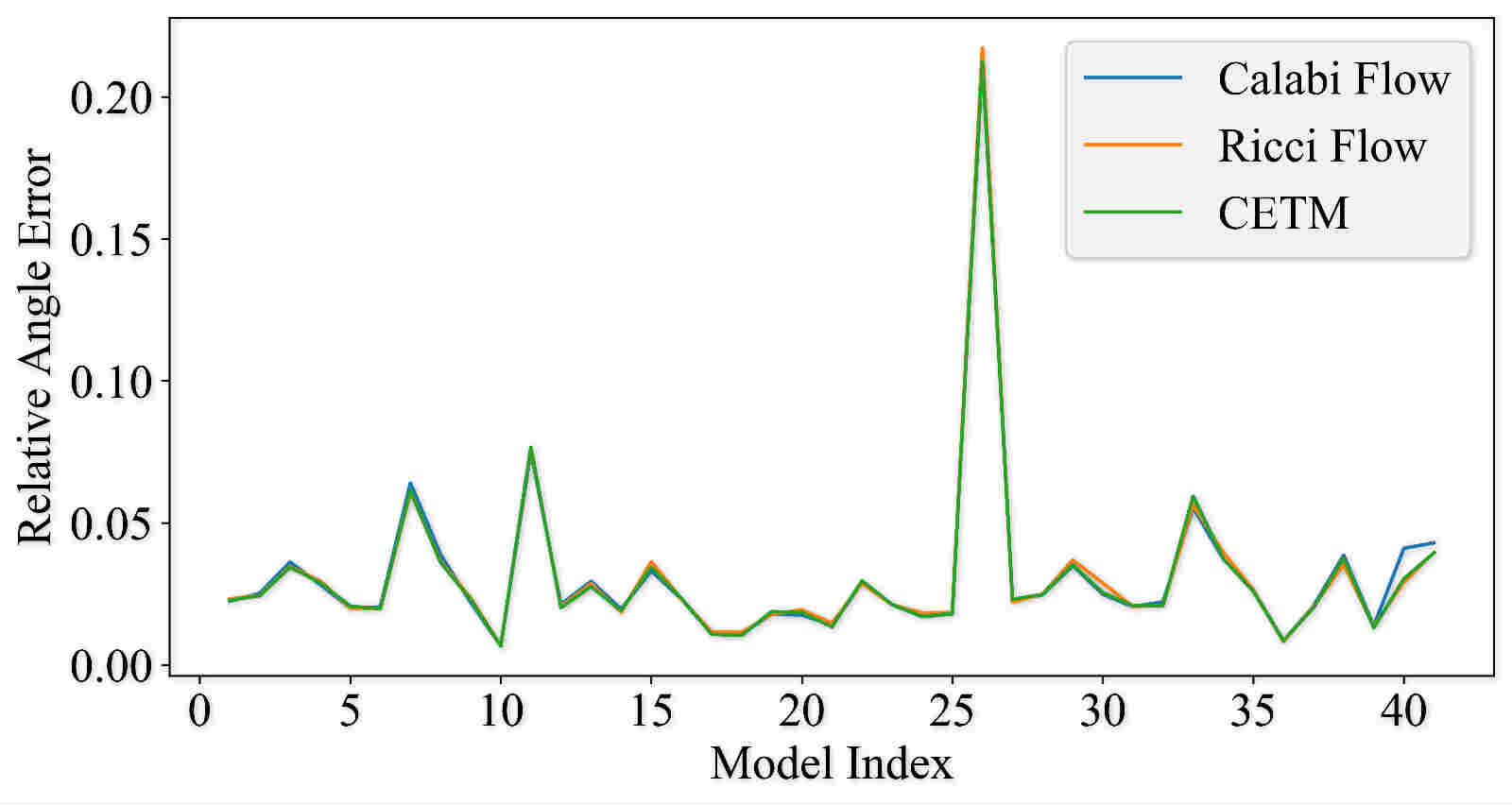}
	\caption{Mean angle relative errors on all models.}
	\label{fig:conformitytest3}
\end{figure}

We calculate  the statistic of angle ratios of the original kitten model and its parameterization by three kinds of algorithms. In figure \ref{fig:conformitytest}, it is shown that three methods has almost the same distribution of angle ratios.  We choose 300 corners of the kitten model randomly. The relative angle errors are shown in figure \ref{fig:conformitytest2}. It is observed that there is some local conformal difference of three algorithms.  Finally we exhibit the mean relative angle errors of all our test models in figure \ref{fig:conformitytest3}. 
Based on these experiments, we conclude that our Calabi flow method has the same conformity with Ricci flow and CETM.

In our experiments, as same as Ricci flow and CETM, Calabi flow may  fail on the meshes with very narrow triangles or with large curvatures. It means that the convergence of these flow based methods depends on the mesh quality.

\section{Conclusions}

In this paper, we present a conformal mesh parameterization based on discrete Calabi flow. This flow is different from previous well-known Ricci flow and CETM.  
Currently Calabi flow works on the Euclidean background, it is also possible to explore Calabi flow in the hyperbolic background \cite{ge20162}. 

Calabi energy has a clear geometric meaning, it expressed the difference between current Gaussian curvature and target Gaussian curvature. It is potential to exploit this feature in mesh smoothing and other related operations.

The conformal flow tool, such as Ricci flow, has already been used in the applications of brain analysis \cite{Wang2007BrainSC}, shape analysis \cite{Gu2007RicciFF}, network routing \cite{Sarkar2009GreedyRW,Gao2016DiscreteRF}, shape modeling \cite{Patan2014AnIT}, shape registration \cite{Gao2013HighRC}, face matching \cite{Zeng20083DFM}, geometric structures \cite{Jin2007ComputingGG}, and so on.  In the future, we will explore the applicability of  Calabi flow in these kinds of applications.

\section*{Acknowledgements}

We would like to thank  anonymous reviewers for their insightful feedbacks,  valuable
comments and suggestions. We also want to thank Ke Xu for the rendering.  This work is
supported in part by the AAA. 
 Some pictures are rendered by Mitsuba \cite{Mitsuba}.  
All 3D models are from the
AIM@SHAPE shape repository and Thingi10K repository. Thanks MeshDGP \cite{Hui:MeshDGP:2016} framework for the implementation reference.

	%-------------------------------------------------------------------------

\bibliographystyle{eg-alpha-doi}
\bibliography{calabiflow}

\newcommand{\etalchar}[1]{$^{#1}$}
\begin{thebibliography}{\uppercase{SKPSH13}}

\bibitem[AL15]{Aigerman2015OrbifoldTE}
\textsc{Aigerman N., Lipman Y.}:
\newblock Orbifold tutte embeddings.
\newblock \emph{ACM Trans. Graph. 34} (2015), 190.

\bibitem[AL16]{Aigerman2016HyperbolicOT}
\textsc{Aigerman N., Lipman Y.}:
\newblock Hyperbolic orbifold tutte embeddings.
\newblock \emph{ACM Trans. Graph. 35} (2016), 217.

\bibitem[APL14]{Aigerman2014LiftedBF}
\textsc{Aigerman N., Poranne R., Lipman Y.}:
\newblock Lifted bijections for low distortion surface mappings.
\newblock \emph{ACM Trans. Graph. 33} (2014), 69:1--69:12.

\bibitem[BCGB08]{ben2008conformal}
\textsc{Ben-Chen M., Gotsman C., Bunin G.}:
\newblock Conformal flattening by curvature prescription and metric scaling.
\newblock In \emph{Computer Graphics Forum} (2008), vol.~27, Wiley Online
  Library, pp.~449--458.

\bibitem[BS04]{bowers2004uniformizing}
\textsc{Bowers P.~L., Stephenson K.}:
\newblock \emph{Uniformizing Dessins and Bely? Maps Via Circle Packing},
  vol.~170.
\newblock American Mathematical Soc., 2004.

\bibitem[Cal82]{calabi1982kahler}
\textsc{Calabi E.}:
\newblock Extremal k\"ahler metrics.
\newblock \emph{Annals of Mathematics Studies} (1982), 259--290.

\bibitem[CBK15]{Campen2015QuantizedGP}
\textsc{Campen M., Bommes D., Kobbelt L.}:
\newblock Quantized global parametrization.
\newblock \emph{ACM Trans. Graph. 34} (2015), 192.

\bibitem[CD06]{cohen2006designing}
\textsc{Cohen Y. T. P. A.~D., Desbrun S.~M.}:
\newblock Designing quadrangulations with discrete harmonic forms.
\newblock In \emph{Eurographics symposium on geometry processing} (2006),
  pp.~1--10.

\bibitem[CH08]{chen2008calabi}
\textsc{Chen X., He W.}:
\newblock On the calabi flow.
\newblock \emph{American Journal of Mathematics 130}, 2 (2008), 539--570.

\bibitem[Che55]{chern1995isothermal}
\textsc{Chern S.-S.}:
\newblock An elementary proof of the existence of isothermal parameters on a
  surface.
\newblock \emph{Proceedings of the American Mathematical Society 6}, 5 (1955),
  771--782.

\bibitem[CL{\etalchar{*}}03]{chow2003combinatorial}
\textsc{Chow B., Luo F., et~al.}:
\newblock Combinatorial ricci flows on surfaces.
\newblock \emph{Journal of Differential Geometry 63}, 1 (2003), 97--129.

\bibitem[DMA02a]{desbrun2002intrinsic}
\textsc{Desbrun M., Meyer M., Alliez P.}:
\newblock Intrinsic parameterizations of surface meshes.
\newblock In \emph{Computer Graphics Forum} (2002), vol.~21, Wiley Online
  Library, pp.~209--218.

\bibitem[DMA02b]{Desbrun2002IntrinsicPO}
\textsc{Desbrun M., Meyer M., Alliez P.}:
\newblock Intrinsic parameterizations of surface meshes.
\newblock \emph{Comput. Graph. Forum 21} (2002), 209--218.

\bibitem[FLG15a]{fu2015computing}
\textsc{Fu X.-M., Liu Y., Guo B.}:
\newblock Computing locally injective mappings by advanced mips.
\newblock \emph{ACM Transactions on Graphics (TOG) 34}, 4 (2015), 71.

\bibitem[FLG15b]{Fu2015ComputingLI}
\textsc{Fu X.-M., Liu Y., Guo B.}:
\newblock Computing locally injective mappings by advanced mips.
\newblock \emph{ACM Trans. Graph. 34} (2015), 71.

\bibitem[Flo03]{Floater2003OnetoonePL}
\textsc{Floater M.~S.}:
\newblock One-to-one piecewise linear mappings over triangulations.
\newblock \emph{Math. Comput. 72} (2003), 685--696.

\bibitem[Ge17]{ge2012combinatorial}
\textsc{Ge H.}:
\newblock Combinatorial calabi flows on surfaces.
\newblock \emph{Transactions of AMS} (2017).

\bibitem[GGL16]{Gao2016DiscreteRF}
\textsc{Gao J., Gu X., Luo F.}:
\newblock Discrete ricci flow for geometric routing.
\newblock In \emph{Encyclopedia of Algorithms} (2016).

\bibitem[GGT06]{Gortler2006DiscreteOO}
\textsc{Gortler S.~J., Gotsman C., Thurston D.}:
\newblock Discrete one-forms on meshes and applications to 3d mesh
  parameterization.
\newblock \emph{Computer Aided Geometric Design 23} (2006), 83--112.

\bibitem[GJ17a]{ge2017deformation2}
\textsc{Ge H., Jiang W.}:
\newblock On the deformation of inversive distance circle packings, ii.
\newblock \emph{Journal of Functional Analysis 272}, 9 (2017), 3573--3595.

\bibitem[GJ17b]{ge2017deformation3}
\textsc{Ge H., Jiang W.}:
\newblock On the deformation of inversive distance circle packings, iii.
\newblock \emph{Journal of Functional Analysis 272}, 9 (2017), 3596--3609.

\bibitem[Gli05a]{glickenstein2005combinatorial}
\textsc{Glickenstein D.}:
\newblock A combinatorial yamabe flow in three dimensions.
\newblock \emph{Topology 44}, 4 (2005), 791--808.

\bibitem[Gli05b]{glickenstein2005geometric}
\textsc{Glickenstein D.}:
\newblock Geometric triangulations and discrete laplacians on manifolds.
\newblock \emph{arXiv preprint math/0508188} (2005).

\bibitem[GLSY13]{gu2013variational}
\textsc{Gu X., Luo F., Sun J., Yau S.-T.}:
\newblock Variational principles for minkowski type problems, discrete optimal
  transport, and discrete monge-ampere equations.
\newblock \emph{arXiv preprint arXiv:1302.5472} (2013).

\bibitem[GSZ{\etalchar{*}}13]{Gao2013HighRC}
\textsc{Gao M., Shi R., Zhang S., Zeng W., Qian Z., Gu X., Metaxas D.~N., Axel
  L.}:
\newblock High resolution cardiac shape registration using ricci flow.
\newblock In \emph{ISBI} (2013).

\bibitem[GWK{\etalchar{*}}07]{Gu2007RicciFF}
\textsc{Gu X., Wang S., Kim J., Zeng Y., Wang Y., Qin H., Samaras D.}:
\newblock Ricci flow for 3d shape analysis.
\newblock In \emph{ICCV} (2007).

\bibitem[GX16]{ge20162}
\textsc{Ge H., Xu X.}:
\newblock 2-dimensional combinatorial calabi flow in hyperbolic background
  geometry.
\newblock \emph{Differential Geometry and its Applications 47} (2016), 86--98.

\bibitem[GY03]{gu2003global}
\textsc{Gu X., Yau S.-T.}:
\newblock Global conformal surface parameterization.
\newblock In \emph{Proceedings of the 2003 Eurographics/ACM SIGGRAPH symposium
  on Geometry processing} (2003), Eurographics Association, pp.~127--137.

\bibitem[HG00]{hormann2000mips}
\textsc{Hormann K., Greiner G.}:
\newblock \emph{MIPS: An efficient global parametrization method}.
\newblock Tech. rep., DTIC Document, 2000.

\bibitem[Jak10]{Mitsuba}
\textsc{Jakob W.}:
\newblock Mitsuba renderer, 2010.
\newblock http://www.mitsuba-renderer.org.

\bibitem[JKG07]{Jin2007DiscreteSR}
\textsc{Jin M., Kim J., Gu X.}:
\newblock Discrete surface ricci flow: Theory and applications.
\newblock In \emph{IMA Conference on the Mathematics of Surfaces} (2007).

\bibitem[JKLG08a]{Jin2008DiscreteSR}
\textsc{Jin M., Kim J., Luo F., Gu X.}:
\newblock Discrete surface ricci flow.
\newblock \emph{IEEE Trans. Vis. Comput. Graph. 14} (2008), 1030--1043.

\bibitem[JKLG08b]{Jin2008VariationalMO}
\textsc{Jin M., Kim J., Luo F., Gu X.}:
\newblock Variational method on discrete ricci flow.
\newblock In \emph{IWCIA Special Track on Applications} (2008).

\bibitem[JLG06]{jin2006computing}
\textsc{Jin M., Luo F., Gu X.}:
\newblock Computing surface hyperbolic structure and real projective structure.
\newblock In \emph{Proceedings of the 2006 ACM symposium on Solid and physical
  modeling} (2006), ACM, pp.~105--116.

\bibitem[JLG07]{Jin2007ComputingGG}
\textsc{Jin M., Luo F., Gu X.}:
\newblock Computing general geometric structures on surfaces using ricci flow.
\newblock \emph{Computer-Aided Design 39} (2007), 663--675.

\bibitem[KSS06]{kharevych2006discrete}
\textsc{Kharevych L., Springborn B., Schr{\"o}der P.}:
\newblock Discrete conformal mappings via circle patterns.
\newblock \emph{ACM Transactions on Graphics (TOG) 25}, 2 (2006), 412--438.

\bibitem[Lee03]{lee2003manifolds}
\textsc{Lee J.~M.}:
\newblock \emph{Introduction to Smooth Manifolds}.
\newblock Springer, 2003.

\bibitem[Lip12]{Lipman2012BoundedDM}
\textsc{Lipman Y.}:
\newblock Bounded distortion mapping spaces for triangular meshes.
\newblock \emph{ACM Trans. Graph. 31} (2012), 108:1--108:13.

\bibitem[LPRM02]{levy2002least}
\textsc{L{\'e}vy B., Petitjean S., Ray N., Maillot J.}:
\newblock Least squares conformal maps for automatic texture atlas generation.
\newblock In \emph{Acm transactions on graphics (tog)} (2002), vol.~21, ACM,
  pp.~362--371.

\bibitem[Luo03]{luo2003yamabe}
\textsc{Luo F.}:
\newblock Combinatorial yamabe flow on surfaces.
\newblock \emph{Communications in Contemporary Mathematics} (2003).

\bibitem[LZX{\etalchar{*}}08]{liu2008local}
\textsc{Liu L., Zhang L., Xu Y., Gotsman C., Gortler S.~J.}:
\newblock A local/global approach to mesh parameterization.
\newblock In \emph{Computer Graphics Forum} (2008), vol.~27, Wiley Online
  Library, pp.~1495--1504.

\bibitem[MPZ14]{myles2014robust}
\textsc{Myles A., Pietroni N., Zorin D.}:
\newblock Robust field-aligned global parametrization.
\newblock \emph{ACM Transactions on Graphics (TOG) 33}, 4 (2014), 135.

\bibitem[MZ12]{myles2012global}
\textsc{Myles A., Zorin D.}:
\newblock Global parametrization by incremental flattening.
\newblock \emph{ACM Transactions on Graphics (TOG) 31}, 4 (2012), 109.

\bibitem[MZ13]{myles2013controlled}
\textsc{Myles A., Zorin D.}:
\newblock Controlled-distortion constrained global parametrization.
\newblock \emph{ACM Transactions on Graphics (TOG) 32}, 4 (2013), 105.

\bibitem[Pet06]{riemanngeometry}
\textsc{Petersen P.}:
\newblock \emph{Riemannian geometry}.
\newblock Springer-Verlag New York, 2006.

\bibitem[PLG14]{Patan2014AnIT}
\textsc{Patan{\`e} G., Li X., Gu X.}:
\newblock An introduction to ricci flow and volumetric approximation with
  applications to shape modeling.
\newblock In \emph{SIGGRAPH ASIA Courses} (2014).

\bibitem[SCQ{\etalchar{*}}16]{su2016area}
\textsc{Su K., Cui L., Qian K., Lei N., Zhang J., Zhang M., Gu X.~D.}:
\newblock Area-preserving mesh parameterization for poly-annulus surfaces based
  on optimal mass transportation.
\newblock \emph{Computer Aided Geometric Design 46} (2016), 76--91.

\bibitem[SHL{\etalchar{*}}07]{Hormann2007MeshPT}
\textsc{Sheffer A., Hormann K., Levy B., Desbrun M., Zhou K., Praun E., Hoppe
  H.}:
\newblock Mesh parameterization: Theory and practice.

\bibitem[SKPSH13]{schuller2013locally}
\textsc{Sch{\"u}ller C., Kavan L., Panozzo D., Sorkine-Hornung O.}:
\newblock Locally injective mappings.
\newblock In \emph{Computer Graphics Forum} (2013), vol.~32, Wiley Online
  Library, pp.~125--135.

\bibitem[SLMB05]{Sheffer2005ABFFA}
\textsc{Sheffer A., L{\'e}vy B., Mogilnitsky M., Bogomyakov A.}:
\newblock Abf++: fast and robust angle based flattening.
\newblock \emph{ACM Trans. Graph. 24} (2005), 311--330.

\bibitem[SPR06]{Sheffer2006MeshPM}
\textsc{Sheffer A., Praun E., Rose K.}:
\newblock Mesh parameterization methods and their applications.
\newblock \emph{Foundations and Trends in Computer Graphics and Vision 2}
  (2006).

\bibitem[SS15]{Smith2015BijectivePW}
\textsc{Smith J., Schaefer S.}:
\newblock Bijective parameterization with free boundaries.
\newblock \emph{ACM Trans. Graph. 34} (2015), 70.

\bibitem[SSP08]{springborn2008conformal}
\textsc{Springborn B., Schr{\"o}der P., Pinkall U.}:
\newblock Conformal equivalence of triangle meshes.
\newblock In \emph{ACM Transactions on Graphics (TOG)} (2008), vol.~27, ACM,
  p.~77.

\bibitem[Ste05]{stephenson2005introduction}
\textsc{Stephenson K.}:
\newblock \emph{Introduction to circle packing: The theory of discrete analytic
  functions}.
\newblock Cambridge University Press, 2005.

\bibitem[SYG{\etalchar{*}}09]{Sarkar2009GreedyRW}
\textsc{Sarkar R., Yin X., Gao J., Luo F., Gu X.}:
\newblock Greedy routing with guaranteed delivery using ricci flows.
\newblock In \emph{IPSN} (2009).

\bibitem[SZS{\etalchar{*}}13a]{shi2013hyperbolic}
\textsc{Shi R., Zeng W., Su Z., Damasio H., Lu Z., Wang Y., Yau S.-T., Gu X.}:
\newblock Hyperbolic harmonic mapping for constrained brain surface
  registration.
\newblock In \emph{Proceedings of the IEEE Conference on Computer Vision and
  Pattern Recognition} (2013), pp.~2531--2538.

\bibitem[SZS{\etalchar{*}}13b]{su2013area}
\textsc{Su Z., Zeng W., Shi R., Wang Y., Sun J., Gu X.}:
\newblock Area preserving brain mapping.
\newblock In \emph{Proceedings of the IEEE Conference on Computer Vision and
  Pattern Recognition} (2013), pp.~2235--2242.

\bibitem[Thu76]{thurston}
\textsc{Thurston W.}:
\newblock \emph{Geometry and topology of three-manifolds}.
\newblock Princeton Lecture Notes, 1976.

\bibitem[WGC{\etalchar{*}}07]{Wang2007BrainSC}
\textsc{Wang Y., Gu X., Chan T.~F., Thompson P.~M., Yau S.-T.}:
\newblock Brain surface conformal parameterization with the ricci flow.
\newblock In \emph{ISBI} (2007).

\bibitem[WZ14]{Weber2014LocallyIP}
\textsc{Weber O., Zorin D.}:
\newblock Locally injective parametrization with arbitrary fixed boundaries.
\newblock \emph{ACM Trans. Graph. 33} (2014), 75:1--75:12.

\bibitem[YGL{\etalchar{*}}09]{Yang2009GeneralizedDR}
\textsc{Yang Y.-L., Guo R., Luo F., Hu S.-M., Gu X.}:
\newblock Generalized discrete ricci flow.
\newblock \emph{Comput. Graph. Forum 28} (2009), 2005--2014.

\bibitem[YLZG17]{yu2017surface}
\textsc{Yu X., Lei N., Zheng X., Gu X.}:
\newblock Surface parameterization based on polar factorization.
\newblock \emph{Journal of Computational and Applied Mathematics} (2017).

\bibitem[ZGZ{\etalchar{*}}14]{Zhang2014TheUD}
\textsc{Zhang M., Guo R., Zeng W., Luo F., Yau S.-T., Gu X.}:
\newblock The unified discrete surface ricci flow.
\newblock \emph{Graphical Models 76} (2014), 321--339.

\bibitem[Zha16]{Hui:MeshDGP:2016}
\textsc{Zhao H.}:
\newblock {MeshDGP}: A {C Sharp} mesh processing framework, 2016.
\newblock http://meshdgp.github.io/.

\bibitem[ZSG{\etalchar{*}}13]{zhao2013area}
\textsc{Zhao X., Su Z., Gu X.~D., Kaufman A., Sun J., Gao J., Luo F.}:
\newblock Area-preservation mapping using optimal mass transport.
\newblock \emph{IEEE transactions on visualization and computer graphics 19},
  12 (2013), 2838--2847.

\bibitem[ZYZ{\etalchar{*}}08]{Zeng20083DFM}
\textsc{Zeng W., Yin X., Zeng Y., Lai Y.-K., Gu X., Samaras D.}:
\newblock 3d face matching and registration based on hyperbolic ricci flow.
\newblock In \emph{CVPR Workshops} (2008).

\bibitem[ZZG{\etalchar{*}}15]{Zhang2015SurveyOD}
\textsc{Zhang M., Zeng W., Guo R., Luo F., Gu X.~D.}:
\newblock Survey on discrete surface ricci flow.
\newblock \emph{J. Comput. Sci. Technol. 30} (2015), 598--613.

\end{thebibliography}
	
	%-------------------------------------------------------------------------
\newpage

\end{document}